\documentclass[aps,prx,reprint,longbibliography,superscriptaddress]{revtex4-1}
\usepackage{amsmath,amssymb,psfrag,bm}
\usepackage[dvipsnames]{xcolor}
\usepackage{pifont}
\usepackage{float}
\usepackage[caption=false]{subfig}
\usepackage{hyperref}
\usepackage{multirow}
\usepackage{color,soul}
\usepackage{footnote}
\usepackage{cancel}
\usepackage{epstopdf}
\usepackage{pstool}
\usepackage{etoolbox}
\usepackage{changes}
\usepackage[titletoc]{appendix}
\usepackage{float}

\begin{document}

\title{Uniform-Velocity Spacetime Crystals}

\author{Zo\'e-Lise Deck-L\'eger}
\affiliation{Dpt. of Electrical Engineering,
Polytechnique Montr\'eal, Montr\'eal, QC H3T 1J4, Canada}

\author{Nima Chamanara}
\affiliation{Dpt. of Electrical Engineering,
Polytechnique Montr\'eal, Montr\'eal, QC H3T 1J4, Canada}

\author{Maksim Skorobogatiy}
\affiliation{Dpt. of Physics Engineering,
Polytechnique Montr\'eal, Montr\'eal, QC H3T 1J4, Canada}

\author{M\'ario G. Silveirinha}
\affiliation{Dpt. of Electrical Engineering, University of Lisbon, 1049-001 Lisboa, Portugal}

\author{Christophe Caloz}
\affiliation{Dpt. of Electrical Engineering,
Polytechnique Montr\'eal, Montr\'eal, QC H3T 1J4, Canada}

\setcounter{tocdepth}{1}

\newcommand{\ud}{\,\mathrm{d}}
\newcommand{\D}{\,\partial}
\newcommand{\rect}{\,\mathrm{rect}}
\newcommand{\sinc}{\,\mathrm{sinc}}
\newcommand{\bu}{\,\noindent $\bullet$}

\begin{abstract}
We perform a comprehensive analysis of uniform-velocity bilayer spacetime crystals, combining concepts of conventional photonic crystallography and special relativity. Given that a spacetime crystal consists of a sequence of spacetime discontinuities, we do this by solving the following sequence of problems: 1)~the spacetime interface, 2)~the double spacetime interface, or spacetime slab, 3)~the unbounded crystal, and 4)~the truncated crystal. For these problems, we present the following respective new results: 1)~an extension of the Stokes principle to spacetime interfaces, 2)~an interference-based analysis of the interference phenomenology, 3)~a quick linear approximation of the dispersion diagrams, a description of simultaneous wavenumber and frequency bandgaps, and 4)~the explanation of the effects of different types of spacetime crystal truncations, and the corresponding scattering coefficients. This work may constitute the foundation for a virtually unlimited number of novel canonical spacetime media and metamaterial problems.
\end{abstract}

\maketitle
%

\section{Introduction}\label{sec:intro}

Crystals are structures that are periodic, either at the atomic level (e.g. electronic crystals)~\cite{ashcroft1976introduction} or at the supermolecular level of artificial scatterers (e.g. photonic crystals)~\cite{joannopoulos2011photonic}. Their periodicity confers them a dispersion in the form of a bandgap structure, which leads to spatio-temporal filtering. They have a myriad of applications, such as for instance
X-ray imaging~\cite{saleh2007} and Bragg gratings~\cite{bragg1913reflection} in the electromagnetic case.

In conventional crystals, the periodicity is purely spatial: the structure is composed of a 1D, 2D or 3D lattice of molecules or artificial particles, which may be seen as a variation of medium parameters in space. Medium variation may also occur in time, and a periodic time variation leads then to time crystals. Time crystals were first studied in the 1960s under the form of temporally modulated 1D structures, whose controlled instabilities led to parametric amplifiers~\cite{holberg1966parametric,fante1971transmission} and whose asymmetric frequency transitions led to nonreciprocal devices~\cite{kamal1960parametric,baldwin1961nonreciprocal}. Such time crystals have recently experienced a regain of interest~\cite{halevi2009,vila2017bloch,koutserimpas2018electromagnetic}, and more generally time-varying structures~\cite{kalluri2010electromagnetics,shlivinski2018beyond,mirmoosa2019time}. Moreover, the concept of spontaneous time crystals, which are time crystals produced by a non-periodic stimulus, was introduced as a new state of matter in~\cite{wilczek2012classical}.

Generally, medium variation may occur in both space and time, and a periodic spacetime variation leads to spacetime crystals, supporting extremely rich and largely unexplored physics. Spacetime crystals may be moving-medium structures~\cite{skorobogatiy2000rigid,wang2013optical} or modulated-medium structures, the former involving the motion of matter and the latter involving the motion of a perturbation. The two share much of the same physics, including Doppler-like frequency transitions~\cite{doppler1842}, amplification~\cite{ramasastry1967wave,tsai1967wave}, and Bradley-aberration deflections~\cite{Bradley_1729}. Modulated structures are easier to realize, more practical, and more diverse, as they may be superluminal, have multiple periodicities, and accelerate without the application of any force~\cite{caloz2019spacetime}. For these reasons, the paper
focuses on modulated structures, while comparing them with moving structures whenever appropriate.

Spacetime crystals may generally include the three dimensions of space and the single dimension of time, and are hence (3+1)-D. In this paper, the quantitative developments are restricted to a single dimension of space, corresponding to (1+1)-D crystals, but the qualitative concepts essentially apply to (1,2,3+1)-D spacetime crystals.

\begin{figure}[h]
\centering
\includegraphics[width=1\columnwidth]{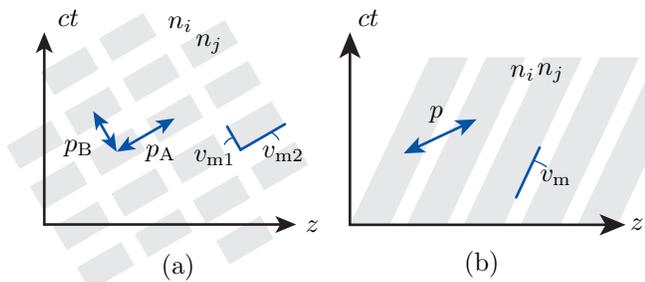}{
\psfrag{a}[c][c]{(a)}
\psfrag{b}[c][c]{(b)}
\psfrag{A}[c][c]{$p_\text{A}$}
\psfrag{B}[c][c]{$p_\text{B}$}
\psfrag{p}[c][c]{$p$}
\psfrag{z}[c][c]{$z$}
\psfrag{t}[c][c]{$ct$}
\psfrag{Z}[c][c]{$z'$}
\psfrag{T}[c][c]{$ct'$}
\psfrag{v}[c][c]{$v_\text{m}$}
\psfrag{1}[c][c][1]{$n_i$}
\psfrag{2}[c][c][1]{$n_j$}
\psfrag{i}[c][c][1]{$v_{\text{m}2}$}
\psfrag{0}[c][c][1]{$v_{\text{m}1}$}
}
\caption{Representation of two canonical spacetime crystals. Here, the variable $z$ is intended to represent the hyperspace, i.e. the three dimensions of space. The white and gray regions correspond to media with refractive index $n_i$ and $n_j$, respectively. (a)~Double-periodicity ($p_A$ and $p_B$) structure, associated with multiple velocities (here 2 velocities, $v_{\text{m}1}$ and $v_{\text{m}2}$), and hence acceleration. (b)~Single-periodicity ($p$) structure, associated with a unique and uniform velocity, $v_\text{m}$.}
\label{fig:1+1D}
\end{figure}

Figure~\ref{fig:1+1D} represents two canonical spacetime crystals in spacetime diagrams, where $z$ and $t$ denote space and time, and $c$ the speed of light in free space. For simplicity, throughout the paper we consider crystals composed of just two homogeneous media with refractive indices $n_i$ and $n_j$ separated by sharp, step discontinuities. The first crystal type, represented in Fig.~\ref{fig:1+1D}(a), is a multiple-periodicity structure (2 periods in the (1+1)-D rectangular-scatterer case). Such multiple-periodicity, which may involve up to 4 periods in the (3+1)-D case, requires the presence of different slopes in the spacetime diagram, and hence multiple velocities, which may come either in discrete values (e.g. 2 velocities in the figure) or in continuous values (e.g. as obtained by rounding the corners of the patches in the figure). Some of the features of such crystals were studied in~\cite{lurie2006wave,lurie2009mathematicali,milton2017field,mattei2017field,mattei2018field2} using space-time ray tracing.

The second crystal type, represented in Fig.~\ref{fig:1+1D}(b), is a single-periodicity structure, corresponding to a single and uniform velocity. Such uniform-velocity modulated spacetime crystals have been investigated under the form of traveling-wave amplifiers~\cite{pierce1959use,tien1958parametric,cullen1958travelling}, and, more recently, under various forms for magnetless nonreciprocity~\cite{fan2009isolation,bahabad2010quasi,taravati2017nrmodulated,chamanara2017optical,caloz2018nonreciprocity,caloz2018editorial}. The present paper pertains to this type of spacetime crystal because of its greater simplicity, and because this fundamental problem already exhibits many unreported and interesting physics.

Specifically, this work studies the scattering of waves from uniform-velocity spacetime crystals [Fig.~\ref{fig:1+1D}(b)] by generalizing standard crystallography and leveraging special relativity. It retrieves both the dispersion diagrams of~\cite{cassedy1963dispersion,cassedy1967} and the transfer-matrix results of~\cite{biancalana2007dynamics}, and presents a number of new results, including the generalization of the Stokes principle for spacetime interfaces, the derivation of the interference condition in spacetime slabs, and a complete description of the dispersion diagrams of spacetime crystals, with the identification and explanation of dispersion-slope asymmetry, the demonstration of simultaneous frequency and wavenumber bandgaps and the analytical derivation of the bandgap positions. The calculation of the scattered-wave amplitudes for finite crystals is also provided.

The paper is organized as follows. Section~\ref{sec:preliminaries} presents preliminary concepts and principles, including the description of a spacetime interface, the introduction of generalized spacetime diagrams, the distinction of three velocity regimes in spacetime structures, our strategy to compute the  fields scattered by spacetime structures, a brief recall on Lorentz transformations, the proper selection of Lorentz frames for subluminal and superluminal structures, and a discussion on the duality existing between space and time in spacetime systems. Section~\ref{sec:interface} deals with a single interface of a spacetime crystal, deriving its scattering coefficients, its frequency transitions, and generalizing the Stokes relations. Section~\ref{sec:ST_slab} solves the double-interface problem of a spacetime slab, describing its scattering phenomenology, deriving its frequency transitions, phase shifts and scattering coefficients, and graphically illustrates the gap interference condition. Section~\ref{sec:unb_cryst} studies an unbounded bilayer crystal. It develops a linear approximation of the dispersion diagram, finds the position of the centers of the bandgaps, derives the transfer matrix for a unit cell of a spacetime crystal, subsequently computes the dispersion relation and explains its spacetime specifics in terms of the Brillouin zone, the complex bandgaps, and the bandgap edges. Section~\ref{sec:truncated} uses the previously derived unit-cell transfer matrix to compute the scattered amplitudes across a finite spacetime crystal. Finally, conclusions are provided in Sec.~\ref{sec:concl}.

\section{Preliminaries}\label{sec:preliminaries}
\subsection{Interface Perspective}

Since a crystal electromagnetically consists of a succession of medium discontinuities, we shall first study spacetime interfaces corresponding to such discontinuities. These interfaces should be understood as modulated perturbations moving at a velocity that constitutes the velocity of the crystal. A possible analogy would be that of a chain of standing domino tiles, sufficiently closely spaced to topple their neighbours upon falling, so that a chain reaction can be launched by knocking down the first tile. In such a reaction, one clearly sees the ``interface'' between the fallen and standing parts of the chain propagating along the structure at a specific velocity, with the dominoes themselves on either side of the interface either standing or fallen not moving in the direction of propagation. The velocity of such an interface along the chain would be limited by the time it takes for one tile to topple the next. We could nevertheless imagine the situation where the tiles could be knocked down by independent external triggers distributed along the domino chain. If these triggers occurred successively at vanishingly small time intervals, the interface velocity would be superluminal, and it would even be infinite if the triggers were all simultaneous.

\subsection{Spacetime Diagram}~\label{sec:interface_st_diag}

\begin{figure}[h]
\includegraphics[width=1\columnwidth]{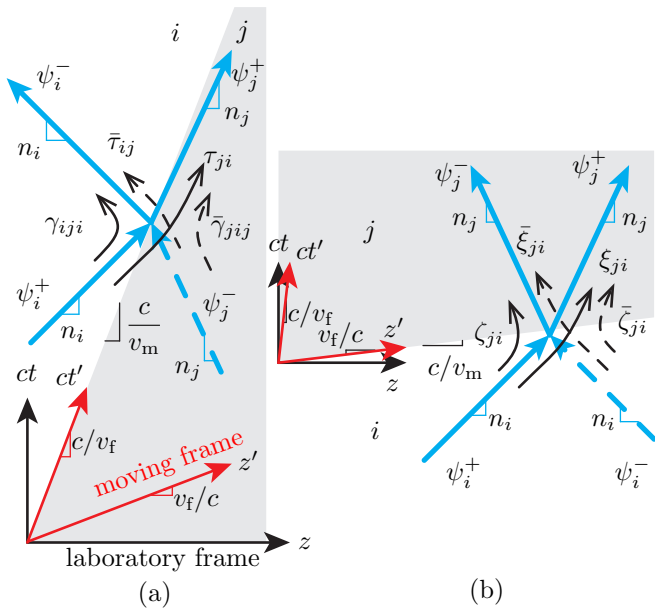}{
\psfrag{z}[c][c]{$z$}
\psfrag{t}[c][c]{$ct$}
\psfrag{Z}[c][c]{$z'$}
\psfrag{T}[c][c]{$ct'$}
\psfrag{e}[c][c]{$c/v_\text{f}$}
\psfrag{E}[c][c]{$v_\text{f}/c$}
\psfrag{v}[c][c]{$\dfrac{c}{v_\text{m}}$}
\psfrag{V}[c][c]{$c/v_\text{m}$}
\psfrag{a}[c][c]{(a)}
\psfrag{b}[c][c]{(b)}
\psfrag{c}[c][c]{(c)}
\psfrag{d}[c][c]{(d)}
\psfrag{1}[c][c]{$i$}
\psfrag{2}[c][c]{$j$}
\psfrag{i}[c][c][1]{$n_i$}
\psfrag{j}[c][c][1]{$n_j$}
\psfrag{A}[c][c][1]{$\psi_i^+$}
\psfrag{B}[c][c][1]{$\psi_i^-$}
\psfrag{C}[c][c][1]{$\psi_j^+$}
\psfrag{D}[c][c][1]{$\psi_j^-$}
\psfrag{g}[r][r]{$\gamma_{iji}$}
\psfrag{G}[c][c]{$\bar{\gamma}_{jij}$}
\psfrag{p}[c][c]{$\bar{\tau}_{ij}$}
\psfrag{P}[c][c]{$\tau_{ji}$}
\psfrag{X}[c][c]{$\bar{\zeta}_{ji}$}
\psfrag{x}[c][c]{$\zeta_{ji}$}
\psfrag{y}[c][c]{$\bar{\xi}_{ji}$}
\psfrag{Y}[c][c]{$\xi_{ji}$}
\psfrag{M}[c][b][1][20]{\textcolor{red}{moving frame}}
\psfrag{m}[c][c]{laboratory frame}
}
\centering
\caption{Scattering at a spacetime crystal interface. The white and gray regions correspond to media $i$ and $j$, with refractive indices $n_i$ and $n_j$ ($n_i<n_j$). The blue arrows represent wave trajectories (referring to a specific phase point of the waveform), while the black arrows represent scattering coefficients. The dashed lines correspond to incidence from the right. The laboratory and moving frames are superimposed with common origin.  (a)~Subluminal modulation velocity ($v_\text{m}<c$), with moving frame having the same velocity as the interface ($v_\text{f}/c=v_\text{m}/c$). (b)~Superluminal modulation velocity ($v_\text{m}>c$), with moving frame having the inverse velocity of the interface ($v_\text{f}/c=c/v_\text{m}$).}\label{fig:interface}
\end{figure}

Subluminal ($v_\text{m}<c$) and superluminal ($v_\text{m}>c$) modulation regimes give rise to distinct physics~\cite{cassedy1963dispersion,cassedy1967,caloz2019spacetime} and require distinct treatments. Figures~\ref{fig:interface}(a) and~\ref{fig:interface}(b) depict the scattering from a subluminal interface and from a superluminal interface in spacetime diagrams, with corresponding spacetime interface trajectories, wave trajectories, scattering coefficients and relevant spacetime coordinate frames.

Given our assumption that the transitions between the homogeneous media $i$ and $j$ forming the crystal are step transitions, the spacetime interface problem is mathematically described by the refractive index function
\begin{subequations}
\begin{equation}\label{eq:ref_index_def}
n(z-v_\text{m}t)=n_i+(n_j-n_i)\theta(z-v_\text{m}t)
\end{equation}
for Fig.~\ref{fig:interface}(a) and
\begin{equation}\label{eq:ref_index_def_sup}
n(t-z/v_\text{m})=n_i+(n_j-n_i)\theta(t-z/v_\text{m})
\end{equation}
\end{subequations}
for Fig.~\ref{fig:interface}(b), where $\theta(\cdot)$ is the Heaviside step function and $v_\text{m}$ is the velocity of the interface. Note that setting \mbox{$v_\text{m}=0$} in \eqref{eq:ref_index_def} reduces the interface to a purely spatial one, while setting $v_\text{m}=\infty$ in \eqref{eq:ref_index_def_sup} reduces the interface to a purely temporal one. Purely spatial and purely temporal problems have been widely investigated, and our results throughout the paper may be verified to correspond to those in the literature at these limit cases. Also note that~\eqref{eq:ref_index_def} and~\eqref{eq:ref_index_def_sup} are different from each other, because we have chosen here to have the incident wave always in medium $i$, for symmetry with respect to the luminal axis $ct=z$; the other option would have been to use~\eqref{eq:ref_index_def} for all regimes, with an increasing $v_\text{m}$ corresponding to a rotation of the interface in Fig.~\ref{fig:interface}(a). In that case, the incidence media would be different for the subluminal or superluminal regimes. The symmetry between subluminal and superluminal problems will be further discussed in Sec.~\ref{sec:st_duality}.

To solve scattering problems involving constant velocities, we will use the Lorentz transformations between the laboratory frame and the moving frame, moving at the uniform velocity $v_\text{f}$. These frames are drawn in Fig.~\ref{fig:interface}, with coordinates axes ($z,ct$) and ($z',ct'$), respectively. In the subluminal case [Fig.~\ref{fig:interface}(a)], the interface appears stationary in the moving frame, i.e., the $ct'$ axis is parallel to the interface. In the superluminal case [Fig.~\ref{fig:interface}(b)], the interface appears purely temporal in the moving frame, i.e. the $z'$ axis is parallel to the interface. The justification for this choice will be provided in Sec.~\ref{sec:resolution_strategy}.

The trajectories of the interfaces in the spacetime diagrams of Fig.~\ref{fig:interface} have a slope that is inversely proportional to their velocity, i.e., equal to $\partial(ct)/\partial{z_\text{traj.}(t)}=[\partial z_\text{traj.}(t)/\partial{(ct)}]^{-1}=c/\text{velocity}$. The slope is therefore steeper for the subluminal case than for the superluminal case. Since the media at either side of the interface are homogeneous, the waves propagate in them at uniform velocities $v_i=c/n_i$ and $v_j=c/n_j$, that correspond to straight trajectories with slopes $c/v_i=n_i$ and $c/v_j=n_j$. Thus the slopes of the wave trajectories are steeper in the slower, or denser, medium $j$.

The waves in medium $p$ in the $\pm z$ directions are denoted $\psi_p^\pm$, with $\psi$ representing any of the transverse components of the electromagnetic fields $\mathbf{E}$, $\mathbf{D}$, $\mathbf{H}$ and $\mathbf{B}$~\cite{jackson1999classical}. The scattering problem for incidence from the left is called the forward problem and for incidence from the right the backward problem. The scattering coefficients for the forward problem are $\gamma$ and $\tau$ for the subluminal regime and $\zeta$ and $\xi$ for the superluminal regime, and the scattering coefficients for the backward problem are denoted by barred oversets ($\bar{\gamma},\bar{\tau}$ and $\bar{\zeta},\bar{\xi}$), that we shall simply refer to as ``barred" for conciseness. All these coefficients will be detailed in Sec.~\ref{sec:scat_coef}.
Note that the incident wave can propagate in the same direction as the interface or medium, i.e., $\text{sgn}(v_\text{inc})=\text{sgn}(v_\text{m})$, as in Fig.~\ref{fig:interface}, or in the direction opposite to the interface or medium, i.e., $\text{sgn}(v_\text{inc})=-\text{sgn}(v_\text{m})$. These two cases, which we refer here to as codirectional and contradirectional, respectively, are important to distinguish because they involve drastically different physics, as clearly apparent in the Doppler phenomenon.

\subsection{Space-Time Duality}~\label{sec:st_duality}

\begin{figure}[h!]
\centering
\includegraphics[width=1\columnwidth]{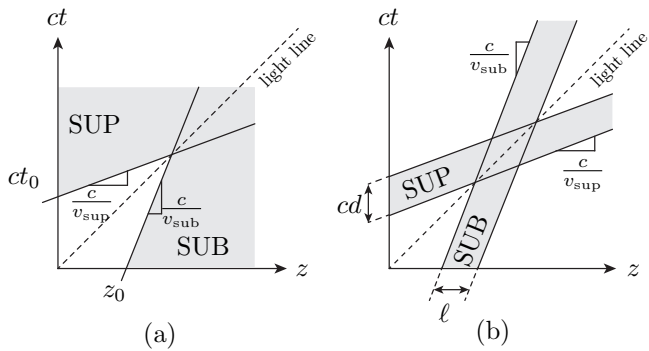}{
\psfrag{a}[c][c]{(a)}
\psfrag{b}[c][c]{(b)}
\psfrag{p}[c][c]{SUB}
\psfrag{P}[c][c]{SUP}
\psfrag{s}[c][c][1][68]{SUB}
\psfrag{S}[c][c][1][21]{SUP}
\psfrag{i}[c][c][0.7][45]{light line}
\psfrag{m}[c][c]{$\ell$}
\psfrag{o}[c][c]{$cd$}
\psfrag{L}[c][c]{$z_0$}
\psfrag{M}[c][c]{$z_0$}
\psfrag{O}[c][c]{$ct_0$}
\psfrag{z}[c][c]{$z$}
\psfrag{t}[c][c]{$ct$}
\psfrag{Z}[c][c]{$z'$}
\psfrag{T}[c][c]{$ct'$}
\psfrag{v}[c][c]{$\frac{c}{v_\text{sub}}$}
\psfrag{V}[c][c]{$\frac{c}{v_\text{sup}}$}
\psfrag{w}[c][c]{$\frac{1}{\beta}$}
}
\caption{Spacetime-inversion symmetry of subluminal (SUB) and superluminal (SUP) structures. (a)~Interfaces. (b)~Slabs. }
\label{fig:duality}
\end{figure}

Subluminal and superluminal structures are symmetric around the light line ($z=ct$) in the spacetime diagram, as shown in Fig.~\ref{fig:duality} for the cases of the interface and the slab. This symmetry is due to the fact that changing from a subluminal to a superluminal problem reverts to exchanging space and time: $v_\text{sub}/c=\D z/\D (ct)\leftrightarrow\D (ct)/\D z =c/v_\text{sub}=v_\text{sup}/c$. The subluminal and superluminal structures are thus related through a spacetime inversion. This geometrical symmetry manifests a fundamental duality between the subluminal and superluminal problems.

This structural duality is completed  by the spacetime duality of Maxwell equations. Consider for instance the electric field in any of the (homogeneous) medium on either side of the discontinuities,
\begin{equation}\label{eq:wave_equation}
\frac{\D^2 E_x}{\D z^2}-n^2(z-v_\text{m}t)\frac{\D^2 E_x}{\D(ct)^2}=0.
\end{equation}
Interchanging space and time in this equation, and also applying $n\rightarrow1/n$, since $n=c/v=\D(ct)/\D z$ and $v_\text{m}/c\rightarrow c/v_\text{m}$, indeed transforms the equation into itself. Thus, the wave equation is also symmetric under spacetime inversion.

Therefore, subluminal and superluminal spacetime electromagnetic problems are the dual of each other. Table~\ref{tab:duality} lists the variables related to the spacetime duality. The first four row substitutions are the ones just discussed for the problem of an interface. The fifth row refers to the duality between the length $\ell$ of a subluminal slab and the duration $cd$ of a superluminal slab, as illustrated in Fig.~\ref{fig:duality}(b). The sixth and sevenths rows correspond to the duality of plane wave solutions in spacetime scattering problems.

\bgroup
\def\arraystretch{2}
\begin{table}[h]  \centering
\label{tab:duality}
\begin{tabular}{ l @{\hskip 0.2cm}c@{\hskip 1cm} l @{\hskip 0.2cm}c }
 \toprule
  \multicolumn{2}{l}{Subluminal regime} & \multicolumn{2}{l }{Superluminal regime}\\\colrule
  space &$z$ & time & $ct$ \\
  time &$ct$ & space& $z$ \\
  modulation velocity &$\dfrac{v_\text{m}}{c}$ & inverse velocity &$\dfrac{c}{v_\text{m}}$  \\[3pt]
  refractive index &$n$ & inverse index &$\dfrac{1}{n}$ \\
  length & $\ell$ & duration& $cd$  \\
   \hline
  wavelength & $\lambda$ & period& $cT$  \\
  frequency&$\dfrac{\omega}{c}$ & wavenumber &$k$ \\[3pt]
  \botrule
\end{tabular}
\caption{Duality transformations between the subluminal and superluminal regimes.}
\label{tab:duality}
\end{table}
\egroup

Note that the duality between space and time is incomplete, because one can go backwards in space but not in time. As a consequence, we shall see that the scattering coefficients are not related to each other, although the phase quantities will always be.

\subsection{Velocity Regimes}\label{sec:velocity_regimes}

In the preceding sections, we distinguished subluminal and superluminal regimes, but omitted to discuss a third velocity regime which we will describe here. To be more precise, we should refer to a modulated system as strictly subluminal when the modulation is slower than the slowest wave involved, i.e., here, $v_\text{m}<v_j$, with $v_j<v_i$, and strictly superluminal when it is faster than the fastest wave involved, i.e., here, $v_\text{m}>v_i$. The velocity range between these two regimes, $v_j\leq v_\text{m} \leq v_i$, corresponds to a third regime, which we call here the interluminal regime. This regime involves yet different physics and requires yet a different treatment than the strictly sub- and superluminal regimes: the incident wave in medium $i$ scatters a single wave in the codirectional case (as in Fig.~\ref{fig:interface}) and three waves in the contradirectional case~\cite{ostrovskii1975,shui2014one,deck2018interluminal}.

Let us first consider the codirectional case. In the forward problem of the strictly subluminal regime [Fig.~\ref{fig:interface}(a)], the incident wave, $\psi_i^+$, is scattered into a reflected wave, $\psi_i^-$, and a transmitted wave, $\psi_j^+$. Progressively increasing $v_\text{m}$, and therefore decreasing the slope of the interface, brings the interface closer and closer to the trajectory of the transmitted wave. At the point where the interface propagates at the same velocity as the wave transmitted in medium $j$, i.e., $v_\text{m}=v_j$, a shock wave is produced from the accumulation of the transmitted wave at the interface. This is the upper limit of the strictly subluminal regime, and the lower limit of the interluminal regime. Beyond this limit ($v_\text{m}>v_j$, but still with $v_\text{m}<v_i$), the incident wave cannot penetrate any more into medium $j$ because it would otherwise be overtaken by the interface. Thus, scattering reduces here to a single reflected wave. Further increasing the velocity leads to the situation where the interface propagates at the same velocity as the incident wave, i.e., $v_\text{m}=v_i$. At this point, the incident wave is barely catching up with the interface, and so reflection tends to zero. Beyond this limit, we enter the strictly superluminal regime, where the incident wave cannot catch up with the interface unless it was launched earlier than it, which is the case in Fig.~\ref{fig:interface}(b). In this regime, the scattered waves are not the conventional reflected and refracted waves; they are instead later forward and later backward waves, and both propagate in medium $j$.

For the contradirectional case, we perform a similar experiment: starting with a strictly subluminal contradirectional problem, by setting a negative slope for the interface in [Fig.~\ref{fig:interface}(a)], and increasing the velocity, we arrive to a point where the velocity of the modulation is equal to the velocity of the wave in the media, i.e. $v_\text{m}=-v_j$, and so a wave will form, propagating along the interface. The scattering will therefore include three waves: a reflected wave, a later backward wave and a later forward wave. For a further increased velocity, the interface catches up with the reflected wave, i.e. $v_\text{m}=-v_i$, leading to another shock wave from the accumulation of the reflected wave. Beyond this limit, we enter the strictly superluminal regime.

In this paper, we shall focus on the strictly subluminal and strictly superluminal regimes, which we will call subluminal or superluminal for brevity.

\subsection{Recall of Special Relativity Concepts}\label{sec:Lorentz}

We noted in Sec.~\ref{sec:interface_st_diag} that we will use Lorentz transformations between the laboratory and moving frames to solve spacetime problems. We present here a brief recall Lorentz tranformation tools required for this.

\subsubsection{Lorentz Transformations and Lorentz Frames}\label{sec:Lorentz_st}

Intervals of space and time in frames moving at a relative velocity $v_\text{f}$ are related through the Lorentz transformations~\cite{lorentz1937electromagnetic}
\begin{equation}\label{eq:Lorentz_ct_z_p}
z'=\gamma\left(z-\frac{v_\text{f}}{c} ct\right), \qquad ct'=\gamma\left(ct-\frac{v_\text{f}}{c} z\right),
\end{equation}
where
\begin{equation}\label{eq:gamma_def}
\gamma=(1-v_\text{f}^2/c^2)^{-1/2}
\end{equation}
is the Lorentz factor, where the primed and unprimed quantities refer to quantities measured in the moving frame and in the laboratory frame, respectively.
The inverse relations are found by exchanging the primed and unprimed quantities and replacing $v_\text{f}$ by $-v_\text{f}$ in~\eqref{eq:Lorentz_ct_z_p}:
\begin{equation}\label{eq:Lorentz_ct_z}
z=\gamma\left(z'+\frac{v_\text{f}}{c} ct'\right), \qquad ct=\gamma\left(ct'+\frac{v_\text{f}}{c} z'\right).
\end{equation}

From these Lorentz transformations, we may superimpose laboratory and moving frames onto a same spacetime plot, as done in Fig.~\ref{fig:interface}. The laboratory frame coordinates are chosen to be orthogonal, and so the moving frame coordinates must be skewed, with the axes slopes found as follows. The $z'$ axis corresponds to $ct'=0$. Enforcing this in~\eqref{eq:Lorentz_ct_z} yields $ct=(v_\text{f}/c) z$, which provides $v_\text{f}/c$ for the slope of the $z'$ axis. Similarly, the $ct'$ axis corresponds to $z'=0$, and enforcing this in~\eqref{eq:Lorentz_ct_z} yields the $z=(v_\text{f}/c) ct$ which provides $c/v_\text{f}$ for the $ct'$ axis slope.

\subsubsection{Velocity Addition Formula}

Velocities measured in different frames do not simply differ by the frame velocity, i.e. the velocity measured in the laboratory frame, $v$, is not the sum of the moving frame velocity, $v_\text{f}$, and the velocity measured in the moving frame, $v'=\text{d} z'/\text{d} t'$. The correct relationship between the velocities measured in the different frames is found by applying the chain rule to the definition of $v$, $v=\text{d} z/\text{d} t$, and replacing the resulting terms by the derivative of the first relation of~\eqref{eq:Lorentz_ct_z} with respect to $t'$ and by the derivative of the second relation of~\eqref{eq:Lorentz_ct_z_p} with respect to $t$, i.e.,
\begin{equation}
v=\frac{\text{d} z}{\text{d} t'}\frac{\text{d} t'}{\text{d} t}=
\gamma^2\left(v'+v_\text{f}\right)\left(1-\frac{v_\text{f}v}{c^2}\right),
\end{equation}
and solving for $v$. The result is
\begin{subequations}\label{eq:add_vel}
\begin{equation}\label{eq:addition_velocities}
 v=\frac{v'+v_\text{f}}{1+\dfrac{v'v_\text{f}}{c^2}}.
\end{equation}
and the inverse is
\begin{equation}\label{eq:addition_velocities_inv}
 v'=\frac{v-v_\text{f}}{1-\dfrac{vv_\text{f}}{c^2}}.
\end{equation}
\end{subequations}
\subsubsection{Spectral Lorentz Transformations}\label{sec:spectral_lorentz}

The phase of a wave is observer-independent~\cite{pauli1981theory}. For instance, different observers agree whether a wave is at a maximum or at a minimum, although they do not agree on the distance between two maxima (wavelength) and the time between two maxima (period). Thus
\begin{equation}\label{eq:phase_invariance}
{\phi^\pm}'=\phi^\pm,
\end{equation}
with the phases defined as
\begin{equation}
\phi^\pm=k^\pm z\mp\omega^\pm t, \qquad {\phi^\pm}'={k^\pm}' z'\mp{\omega^\pm}' t'.
\end{equation}
Using the expressions for $z$ and $t$ in~\eqref{eq:Lorentz_ct_z}, and resolving in~\eqref{eq:phase_invariance} provides the frequency and wavenumber transformations
\begin{equation}\label{eq:Lorentz_w_k}
{k^\pm}'=\gamma\left(k^\pm\mp\frac{v_\text{f}}{c} \frac{\omega^\pm}{c}\right), \qquad \frac{{\omega^\pm}'}{c}=\gamma\left(\frac{\omega^\pm}{c}\mp\frac{v_\text{f}}{c} k^\pm\right).
\end{equation}
Comparing these transformation with~\eqref{eq:Lorentz_ct_z_p} reveals that $k^+$ and $\omega^+/c$ transform as $z$ and $ct$.

\subsubsection{Field Transformations}

Consider harmonic plane waves propagating along $z$ with the electric field phasor $\mathbf{E}=E_x \hat{x}$. The field propagating in medium $p$ in the $\pm z$ direction may be written as
\begin{subequations}\label{eq:field_def}
\begin{equation}\label{eq:E_field_def}
E_{xp}^\pm=A_p^\pm e^{\pm i\phi_p^{\pm}},
\enspace\text{with}\enspace
\phi_p^{\pm}={k_p^{\pm}}z\mp{\omega_p^{\pm}}t.
\end{equation}
For later use, we relate the other fields to $E_{xp}^\pm$ through the wave impedance $\eta_p=\sqrt{\mu_p/\epsilon_p}$ and velocity \mbox{$v_p=1/\sqrt{\epsilon_p \mu_p}$} as
\begin{equation}\label{eq:constitutive_def}
H_{yp}^\pm=\mp \frac{1}{\eta_p}E_{xp}^\pm,
\end{equation}
\begin{equation}
B_{yp}^\pm=\mp \frac{1}{v_p}E_{xp}^\pm,
\end{equation}
\begin{equation}\label{eq:constitutive_def}
D_{xp}^\pm=\epsilon_pE_{xp}^\pm=\frac{1}{v_p\eta_p}E_{xp}^\pm,
\end{equation}
\end{subequations}
which follow from Maxwell equations, with the assumption of the media being purely dielectric in the second equality in~\eqref{eq:constitutive_def}.

The fields measured in the laboratory and moving frames are related through~\cite{einstein1905elektrodynamik}
 \begin{subequations}\label{eq:Lorentz_fields}
 \begin{equation}\label{eq:E_H_Lorentz_fields}
E'_x=\gamma\left(E_y-v_\text{f}B_y\right), \quad H'_y=\gamma\left(H_x-v_\text{f}D_x\right),
 \end{equation}
 \begin{equation}\label{eq:D_B_Lorentz_fields}
D'_x=\gamma\left(D_x-\frac{v_\text{f}}{c^2}H_y\right), \quad B'_y=\gamma\left(B_y-\frac{v_\text{f}}{c^2}E_x\right),
 \end{equation}
 \end{subequations}
with the inverse relations again obtained by changing the sign of $v_\text{f}$ and interchanging the primed and unprimed quantities, where the primed counterpart of~\eqref{eq:E_field_def} reads
\begin{equation}\label{eq:E_field_def_p}
E_{xp}^{\pm '}=A_p^{\pm '} e^{\pm i\phi_p^{\pm '}},
\enspace\text{with}\enspace
\phi_p^{\pm '}={k_p^{\pm '}}z'\mp{\omega_p^{\pm '}}t',
\end{equation}

and the primed counterparts of ~\eqref{eq:E_field_def}(b)-(d) are presented in Supp. Mat.~\ref{sec:fizeau}. For later use, we note that the phase velocity $v_p^{\pm'}$ transforms as~\eqref{eq:add_vel}~\cite{laue1907,pauli1981theory}, but with a negative sign for $v_\text{f}$ since media appears to be moving in the $-z$ direction if the frame is moving in the $+z$ direction. Thus, the phase velocity in the moving frame is
\begin{equation}\label{eq:vpresult}
v_p^{\pm'}=\frac{ v_p^\pm+v_\text{f}}{1+ \frac{ v_p^\pm v_\text{f}}{c^2}}.
\end{equation}

\subsection{Resolution Strategy}\label{sec:resolution_strategy}

Moving media problems are conventionally solved in the moving frame where the spacetime system is stationary. The solutions are then converted to the laboratory frame using Lorentz transformations. The situation for modulated media is more subtle. A modulated interface appears purely spatial in the moving frame, but the media on both sides of it appear to be in motion -- since they are motionless in the laboratory frame -- and are therefore bianisotropic~\cite{fizeau1851hypotheses,pauli1981theory,kong2008theory}(Supp. Mat.~\ref{sec:fizeau}). The complexity associated with this bianisotropy defeats the original simplification purpose of the Lorentz transformation. We resolve here this issue by resorting to the following hybrid strategy: we first apply the field continuity conditions in the moving frame, next transform the results to the laboratory frame, and finally apply the constitutive relations or the dispersion relations in the laboratory frame.

We stated in Sec.~\ref{sec:interface_st_diag} that the moving frames for the subluminal and superluminal problems were chosen such that the problems appeared purely spatial or purely temporal, respectively, in the moving frames. We now revisit this statement, at the light of the space and time Lorentz transformations recalled in Sec.~\ref{sec:Lorentz_st}.

In the subluminal regime [Fig.~\ref{fig:interface}(a)], we select the moving frame so that it is co-moving with the interface, at a velocity equal to that of the interface, $v_\text{f}=v_\text{m}$, as conventionally done. The interface is then purely spatial in that frame, positioned here at $z'=0$, which corresponds to the $ct'$ axis being parallel to the interface.

In the superluminal regime [Fig.~\ref{fig:interface}(b)], as announced in Sec.~\ref{sec:interface_st_diag}, the selection of a co-moving ($v_\text{f}=v_\text{m}$) frame is not possible, since Lorentz transformations apply only to frames moving at subluminal velocities. Indeed, the corresponding superluminal Lorentz transformations would result into nonphysical imaginary space and time quantities, as can be seen by setting $v_\text{f}>c$ in~\eqref{eq:gamma_def}. Instead, we choose a frame where the interface appears to be purely temporal, occurring here at $t'=0$, which corresponds to the $z'$ axis being parallel to the interface~\cite{deck2017superluminal}. We then graphically find that $c/v_\text{m}=v_\text{f}/c$. We may also find this result mathematically by inspecting~\eqref{eq:addition_velocities} and finding the condition for $v_\text{m}'=\infty$, which was shown to correspond to an instantaneous interface in Sec.~\ref{sec:interface_st_diag}. Since $v_\text{m}>c$, we have $v_\text{f}<c$, and the Lorentz transformations are now applicable.

\section{Spacetime Interface}\label{sec:interface}

\subsection{Scattering Coefficients}\label{sec:scat_coef}

In this section, we compute the scattering coefficients denoted by lowercase Greek symbols in Fig.~\ref{fig:interface}. Let us start with the subluminal regime following the strategy outlined in Sec.~\ref{sec:resolution_strategy}. First, we derive the continuity conditions at the interface in the moving frame. Since the interface appears purely spatial in the moving frame, the usual boundary conditions apply in it, i.e.,
\begin{equation}\label{eq:continuity_fields_sub_p}
\left.{E_{xi}}'={E_{xj}}'\right|_{z'=0}, \qquad \left.{H_{yi}}'={H_{yj}}'\right|_{z'=0},
\end{equation}
which involve the total fields in media $i$ and $j$, corresponding to $\psi_p=\psi_p^++\psi_p^-$ ($p=i,j$) in Fig.~\ref{fig:interface}(a). We next express these fields in terms of their unprimed counterparts using the Lorentz transformations~\eqref{eq:E_H_Lorentz_fields}, which yields the continuity conditions
\begin{subequations}\label{eq:continuity_fields_sub}
\begin{equation}
\left.E_{xi}-v_\text{f}B_{yi}=E_{xj}-v_\text{f}B_{yj}\right|_{z-v_\text{f}t=0},
\end{equation}
\begin{equation}
\left.H_{yi}-v_\text{f}D_{xi}=H_{yj}-v_\text{f}D_{xj}\right|_{z-v_\text{f}t=0}.
\end{equation}
\end{subequations}
Finally, the frame velocity is set equal to the modulation velocity, i.e. $v_\text{f}=v_\text{m}$, as was suggested in Sec.~\ref{sec:resolution_strategy}.

To study the forward problem, corresponding to incidence $\psi_i^+\neq0$ and $\psi_j^-=0$ (See Fig.~\ref{fig:interface}), we enforce the continuity conditions~\eqref{eq:continuity_fields_sub} by inserting~\eqref{eq:field_def} into~\eqref{eq:continuity_fields_sub} with $\psi_j^-=0$, which yields
\begin{subequations}\label{eq:coeff_sub}
\begin{equation}\label{eq:coeff_refl_sub}
\gamma_{iji}=\left.\frac{A_i^-}{A_j^+}\right|_{A_j^-=0}
=\frac{\eta_j-\eta_i}{\eta_j+\eta_i}\left(\frac{1-v_\text{m}/v_i}{1+v_\text{m}/v_i }\right),
\end{equation}
\begin{equation}\label{eq:coeff_trans_sub}
\tau_{ji}=\left.\frac{A_j^+}{A_i^+}\right|_{A_j^-=0}
=\frac{2\eta_j}{\eta_i+n_j}\left(\frac{1-v_\text{m}/v_i }{1-v_\text{m}/v_j}\right).
\end{equation}
\end{subequations}
The subscripts in the reflection coefficient indicate that the wave travels from medium $i$, reaches medium $j$, and then reflects back to medium $i$. This notation will be useful later on to distinguish forward and backward waves in multiple-interface problems.

The results of~\eqref{eq:coeff_sub} are valid across the subluminal range, which was shown to extend from $-v_j$ to $v_j$ in Sec.~\ref{sec:velocity_regimes}. According to~\eqref{eq:coeff_sub} the transmission coefficient increases monotonically while the reflection coefficients decreases monotonically across this velocity range. In the contradirectional case, the reflection coefficient is greater than in the purely spatial case, while in the codirectional case, the reflection is less than in the purely spatial case: the moving interface ``cushions'' the wave.

We next study the backward problem, corresponding to the incidence $\psi_i^+=0$ and $\psi_j^-\neq0$. Inserting~\eqref{eq:field_def} into~\eqref{eq:continuity_fields_sub}, or alternatively reversing the sign of $v_\text{m}$ and exchanging $i$ and $j$ in~\eqref{eq:coeff_sub}, results into
\begin{subequations}\label{eq:bar_sub}
\begin{equation}\label{eq:gamma_bar_sub}
\bar{\gamma}_{jij}(v_\text{m})=\left.\frac{A_j^+}{A_j^-}\right|_{A_i^+=0}=\frac{\eta_i-\eta_j}{\eta_i+\eta_j}\left(\frac{1+v_\text{m}/v_j}{1-v_\text{m}/v_j }\right),
\end{equation}
\begin{equation}
\bar{\tau}_{ij}(v_\text{m})=\left.\frac{A_i^-}{A_j^-}\right|_{A_i^+=0}=\frac{2\eta_i}{\eta_j+n_i}\left(\frac{1+v_\text{m}/v_j }{1+v_\text{m}/v_i}\right).
\end{equation}
\end{subequations}

We now turn to the superluminal regime. In the moving frame, where the interface appears purely temporal, the continuous fields are~\cite{morgenthaler1958,felsen1970wave}
\begin{equation}\label{eq:continuity_fields_sup_p}
\left.{D_{xi}}'={D_{xj}}'\right|_{t'=0}, \ \left.{B_{yi}}'={B_{yj}}'\right|_{t'=0}.
\end{equation}
Applying the Lorentz transformation~\eqref{eq:D_B_Lorentz_fields} yields the continuity relations
\begin{subequations}\label{eq:continuity_sup}
\begin{equation}
\left.D_{xi}-\frac{v_\text{f}}{c^2}H_{yi}=D_{xj}-\frac{v_\text{f}}{c^2}H_{yj}\right|_{t-\frac{v_\text{f}}{c^2}z=0}, \end{equation}
\begin{equation}
\left.B_{yi}-\frac{v_\text{f}}{c^2}E_{xi}=B_{yj}-\frac{v_\text{f}}{c^2}E_{xj}\right|_{t-\frac{v_\text{f}}{c^2}z=0}.
\end{equation}
\end{subequations}

We start by studying the forward problem, corresponding to incidence $\psi_i^+\neq0$ and $\psi_i^-=0$. The scattering coefficients are found by inserting~\eqref{eq:constitutive_def} into~\eqref{eq:continuity_sup} or equivalently \eqref{eq:continuity_fields_sub} with $\psi_i^-=0$, which yields the later backward and later forward coefficients
\begin{subequations}\label{eq:coeff_sup}
\begin{equation}\label{eq:coeff_refl_sup}
\zeta_{ji}=\left.\frac{A_j^-}{A_i^+}\right|_{A_i^-=0}=\frac{\eta_i-\eta_j}{2\eta_i}\left(\frac{1-v_\text{m}/v_i}{1+v_\text{m}/v_j}\right),
\end{equation}
\begin{equation}\label{eq:coeff_trans_sup}
\xi_{ji}=\left.\frac{A_j^+}{A_i^+}\right|_{A_i^-=0}=\frac{\eta_i+\eta_j}{2\eta_i}\left(\frac{1-v_\text{m}/v_i}{1-v_\text{m}/v_j}\right).
\end{equation}
\end{subequations}

The scattering coefficients for the backward problem are found by reversing the velocity sign, but not the exchanging the media, since both waves originate in the same medium, contrarily to the subluminal case, and read
\begin{subequations}\label{eq:bar_sup}
\begin{equation}
\bar{\zeta}_{ji}=\left.\frac{A_j^+}{A_i^-}\right|_{A_i^+=0}=\zeta_{ji}(-v_\text{m}).
\end{equation}
\begin{equation}
\bar{\xi}_{ji}=\left.\frac{A_j^-}{A_i^-}\right|_{A_i^+=0}=\xi_{ji}(-v_\text{m}), \end{equation}
\end{subequations}

Note that~\eqref{eq:continuity_fields_sub} with the substitution of $v_\text{f}=v_\text{m}$ is identical to~\eqref{eq:continuity_sup} with the substitution $v_\text{f}/c^2=1/v_\text{m}$, which is the substitution that was suggested in Sec.~\ref{sec:resolution_strategy}. Therefore the same boundary conditions apply for all regimes. However, we note the results in the subluminal and the superluminal regime are different: this is due to the fact that in the subluminal case, for the forward problem, $\psi_i^++\psi_i^-=\psi_j^+$ while for the superluminal case, the reflection is on the other side of the equality sign: $\psi_i^+=\psi_j^++\psi_j^-$. Supp. Mat.~\ref{sec:cont_conditions} provides an alternative derivation of the continuity conditions for all regimes.

\subsection{Generalization of the Stokes Principle}

To facilitate the later study of multiple scattering (Sec.~\ref{sec:ST_slab}), we now establish relations between the scattering coefficients of forward and backward problems derived in Sec.~\ref{sec:scat_coef}. This leads to a generalization of the conventional Stokes relations~\cite{stokes_2009} from stationary interfaces to spacetime interfaces.

\begin{figure}
\centering
\includegraphics[width=1\columnwidth]{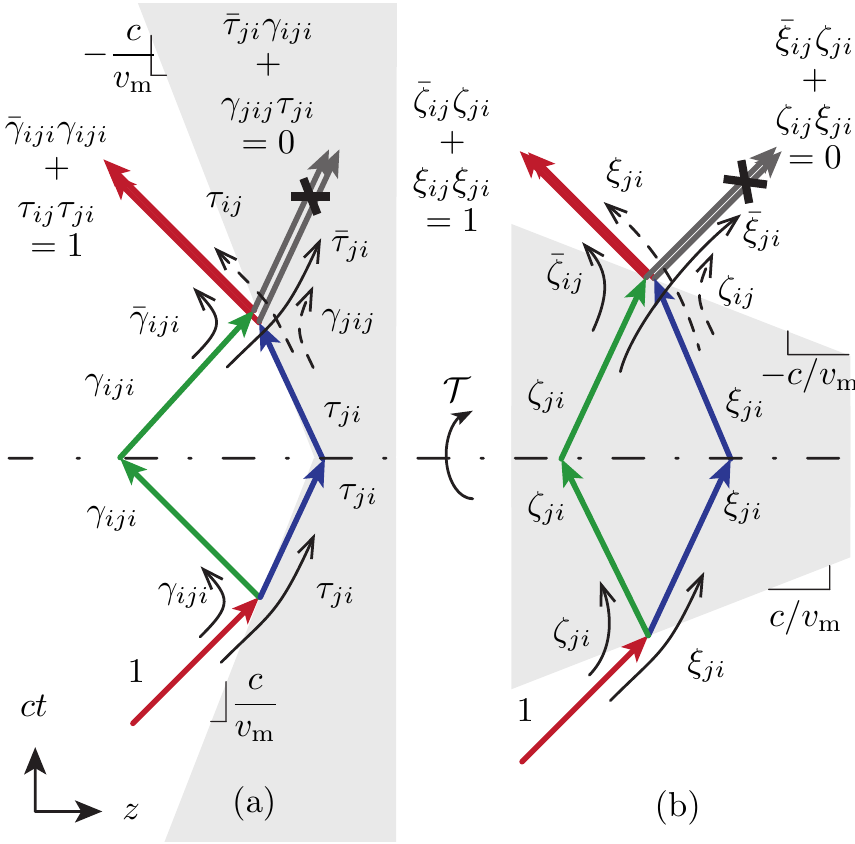}{
\psfrag{a}[c][c]{(a)}
\psfrag{b}[c][c]{(b)}
\psfrag{z}[c][c]{$z$}
\psfrag{t}[c][c]{$ct$}
\psfrag{r}[c][c]{$\mathcal{T}$}
\psfrag{v}[c][c]{$\dfrac{c}{v_\text{m}}$}
\psfrag{V}[c][c]{$-\dfrac{c}{v_\text{m}}$}
\psfrag{c}[c][c]{$c/v_\text{m}$}
\psfrag{C}[c][c]{$-c/v_\text{m}$}
\psfrag{1}[c][c]{$i$}
\psfrag{2}[c][c]{$j$}
\psfrag{i}[c][c]{$n_i$}
\psfrag{j}[c][c]{$n_j$}
\psfrag{A}[c][c]{$1$}
\psfrag{G}[c][c]{$\gamma_{iji}$}
\psfrag{T}[c][c]{$\tau_{ji}$}
\psfrag{g}[r][r]{$\bar{\gamma}_{iji}$}
\psfrag{R}[c][c]{$\gamma_{jij}$}
\psfrag{p}[c][c]{$\tau_{ij}$}
\psfrag{P}[c][c]{$\bar{\tau}_{ji}$}
\psfrag{o}[c][c]{\begin{tabular}{c} $\bar{\gamma}_{iji}\gamma_{iji}$\\
$+$\\ $\tau_{ij}\tau_{ji}$\\ $=1$ \end{tabular}}
\psfrag{n}[c][c]{\begin{tabular}{c} $\bar{\tau}_{ji}\gamma_{iji}$\\
$+$\\ $\gamma_{jij}\tau_{ji}$\\ $=0$ \end{tabular}}
\psfrag{O}[c][c]{\begin{tabular}{c} $\bar{\zeta}_{ij}\zeta_{ji}$\\
$+$\\ $\xi_{ij}\xi_{ji}$\\ $=1$ \end{tabular}}
\psfrag{N}[c][c]{\begin{tabular}{c} $\bar{\xi}_{ij}\zeta_{ji}$\\
$+$\\ $\zeta_{ij}\xi_{ji}$\\ $=0$ \end{tabular}}
\psfrag{B}[c][c]{$\zeta_{ji}$}
\psfrag{F}[c][c]{$\xi_{ji}$}
\psfrag{f}[r][r]{$\xi_{ji}$}
\psfrag{h}[c][c]{$\bar{\xi}_{ji}$}
\psfrag{H}[c][c]{$\bar{\zeta}_{ij}$}
\psfrag{d}[c][c]{$\zeta_{ij}$}
}
\caption{Generalization of the Stokes principle. (a)~Subluminal regime. (b)~Superluminal regime.}
\label{fig:stokes}
\end{figure}

The conventional Stokes relations are derived upon the basis of a time-reversal symmetry argument: time-reversing, or equivalently, reversing the direction of the waves scattered from an interface returns them to their origin. We extend here this argument to the spacetime interface, with the help of Fig.~\ref{fig:stokes}. The subluminal problem is represented in Fig.~\ref{fig:stokes}(a). The bottom part of the figure shows the scattering problem that we have already solved, while the top part shows its time-reversed counterpart, where the velocity signs of the two scattered waves have been reversed, along with the velocity sign of the interface. The scattered waves must return to their origin with the same amplitude, from the time-reversal symmetry argument, and the additional scattering possibility must therefore cancel out. Enforcing the symmetry yields then the subluminal generalized Stokes relations, which may be directly read out from the figure as
\begin{subequations}\label{eq:stokes}
\begin{equation}\label{eq:stokes_a}
\bar{\gamma}_{iji}\gamma_{iji}+\tau_{ij}\tau_{ji}=1,
\end{equation}
\begin{equation}
\bar{\tau}_{ji}\gamma_{iji}+\gamma_{jij}\tau_{ji}=0.
\end{equation}
From these two generalized Stokes conditions we find a third relation, which will be used in Sec.~\ref{sec:ST_slab} to simplify the problem of total scattering from a double interface: barring the unbarred coefficients and unbarring the barred coefficients in~\eqref{eq:stokes_a}, and comparing the resulting equations with the original equation of~\eqref{eq:stokes_a}, we find that
\begin{equation}\label{eq:stokes_c}
\bar{\tau}_{ji}\bar{\tau}_{ji}=\tau_{ji}\tau_{ji}.
\end{equation}
\end{subequations}

The superluminal case is represented in Fig.~\ref{fig:stokes}(b), with the initial problem at the bottom and the time-reversed problem at the top. The generalized superluminal Stokes relations are again read out from the figure, which gives now
\begin{subequations}\label{eq:stokes_sup}
\begin{equation}\label{eq:stokes_sup_a}
\bar{\zeta}_{ij}\zeta_{ji}+\xi_{ij}\xi_{ji}=1,
\end{equation}
\begin{equation}\label{eq:stokes_sup_b}
\bar{\xi}_{ij}\zeta_{ji}+\zeta_{ij}\xi_{ji}=0.
\end{equation}
which are of the same form as their subluminal counterparts. An additional relation, which will be used in Sec.~\ref{sec:ST_slab} to calculate the total scattering from a slab, is found by rewriting~\eqref{eq:stokes_sup_b} as $\bar{\xi}_{ij}=-\xi_{ji}\zeta_{ij}/\zeta_{ji}$, exchanging $i$ and $j$ in this relation to obtain $\bar{\xi}_{ji}=-\xi_{ij}\zeta_{ji}/\zeta_{ij}$, and taking the product of these two, to find
\begin{equation}\label{eq:stokes_sup_c}
\bar{\xi}_{ij}\bar{\xi}_{ji}=\xi_{ji}\xi_{ij}.
\end{equation}
Barring the unbarred coefficients and unbarring the barred coefficients in~\eqref{eq:stokes_sup_a} and using the relation~\eqref{eq:stokes_sup_c} leads then to
\begin{equation}\label{eq:stokes_sup_d} \bar{\zeta}_{ij}\zeta_{ji}=\zeta_{ij}\bar{\zeta}_{ji}.
\end{equation}
\end{subequations}

\subsection{Frequency Transitions}\label{sec:freq_transitions_interface}

The waves scattered from a spacetime interface undergo Doppler-like frequency transitions. We calculate here these frequency transitions, starting again with the subluminal regime. Inserting the moving-frame fields~\eqref{eq:E_field_def_p} into~\eqref{eq:continuity_fields_sub_p} leads to the result that the frequencies are conserved in the moving frame, consistently with the fact that the interface is stationary in that frame. We denote this conserved frequency $\omega_\text{e}'$, with
\begin{subequations}\label{eq:freq_sub}
\begin{equation}\label{eq:freq_cons_sub}
{\omega_\text{e}}'={\omega_i^+}'={\omega_i'}^-={\omega_j'}^+.
\end{equation}
Applying the frequency Lorentz transformation~\eqref{eq:Lorentz_w_k} to~\eqref{eq:freq_cons_sub} with $k_{i,j}^\pm=\omega_{i,j}^\pm /v_{i,j}$ yields
\begin{equation}\label{eq:sub_doppler}
\omega_\text{e}=\omega_i^+\left(1-\frac{v_\text{m}}{v_i}\right)=\omega_i^-\left(1+\frac{v_\text{m}}{v_i}\right)=\omega_j^+\left(1- \frac{v_\text{m}}{v_j}\right),
\end{equation}
\end{subequations}
where the relation between ${\omega_\text{e}}'$ and $\omega_\text{e}$ is found by dividing~\eqref{eq:freq_cons_sub} by~\eqref{eq:sub_doppler},
\begin{equation}\label{eq:gamma}
{\omega_\text{e}}'/\omega_\text{e}=\gamma,
\end{equation}
with $\gamma$ being the Lorentz factor given in~\eqref{eq:gamma_def}. The frequency transitions for a backward problem, with the incident wave on the right of the interface, in medium $j$, would be found by changing the signs of the superscripts, the sign of $v_\text{m}$ and exchanging $i$ and $j$ in~\eqref{eq:sub_doppler}.

\begin{figure}[h]
\centering
\includegraphics[width=1\columnwidth]{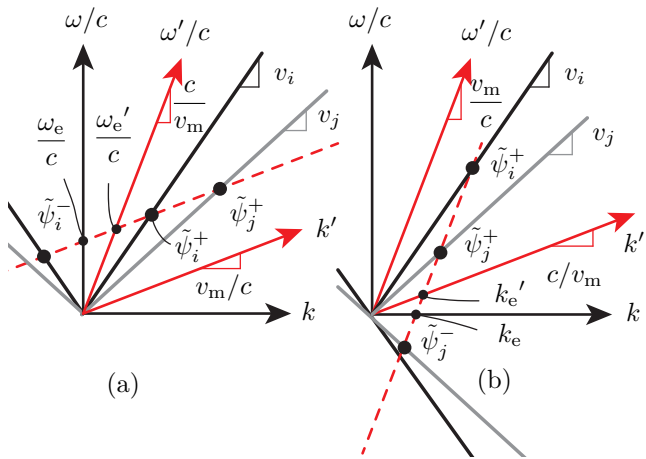}{
\psfrag{w}[c][c]{$\omega/c$}
\psfrag{k}[c][c]{$k$}
\psfrag{W}[c][c]{$\omega'/c$}
\psfrag{K}[c][c]{$k'$}
\psfrag{g}[c][c]{$\dfrac{\omega_\text{e}}{c}$}
\psfrag{h}[c][c]{$\dfrac{{\omega_\text{e}}'}{c}$}
\psfrag{G}[c][c]{${k_\text{e}}$}
\psfrag{H}[c][c]{${k_\text{e}}'$}
\psfrag{e}[c][c]{$\dfrac{c}{v_\text{m}}$}
\psfrag{E}[c][c]{$v_\text{m}/c$}
\psfrag{S}[c][c]{$\dfrac{v_\text{m}}{c}$}
\psfrag{s}[c][c]{$c/v_\text{m}$}
\psfrag{a}[c][c]{(a)}
\psfrag{b}[c][c]{(b)}
\psfrag{1}[c][c]{$v_i$}
\psfrag{2}[c][c]{$v_j$}
\psfrag{A}[c][c][1]{$\tilde{\psi}_i^+$}
\psfrag{B}[c][c][1]{$\tilde{\psi}_i^-$}
\psfrag{C}[c][c][1]{$\tilde{\psi}_j^+$}
\psfrag{D}[c][c][1]{$\tilde{\psi}_j^-$}
}
\caption{Frequency transitions at a spacetime interface (inverse spacetime diagrams) corresponding to Fig.~\ref{fig:interface}. (a)~Subluminal case. (b)~Superluminal case.}
\label{fig:interface_inverse}
\end{figure}

Equations~\eqref{eq:freq_sub} are graphically represented in the inverse spacetime diagram of Fig.~\ref{fig:interface_inverse}(a). The moving-frame axes $k'$ and $\omega'$ are superimposed to the laboratory-frame axes $k$ and $\omega$. The slopes of the moving frame axes, which are $v_\text{f}/c$ and $c/v_\text{f}$, from~\eqref{eq:Lorentz_w_k}, are written in terms of the interface velocity by setting $v_\text{f}=v_\text{m}$. The dispersion relations $\omega_{i,j}^\pm=k_{i,j}^\pm v_{i,j}$ of the two involved media are plotted in the laboratory frame. To find the spectra of the scattered waves, we trace a curve parallel to the $k_z'$ axis with intercept $\omega_\text{e}$, and locate the spectra of the incident and scattered waves at the intersection of this curve and the dispersion diagrams. Reading out the spectra in the laboratory frame reveals that the reflected wave, $\psi_i^-$, is downshifted while the transmitted wave, $\psi_j^+$, is upshifted, consistently with~\eqref{eq:sub_doppler} for $v_i>v_j$.

For the superluminal regime, we insert the moving-frame fields~\eqref{eq:E_field_def_p} into~\eqref{eq:continuity_fields_sup_p}, and find the wavenumbers are conserved in the moving frame
\begin{subequations}\label{eq:freq_sup}
\begin{equation}\label{eq:freq_cons_sup}
{k_\text{e}}'={k_i^+}'={k_j'}^+={k_j'}^-.
\end{equation}
where ${k_\text{e}}'$ is the conserved wavenumber.

Applying the wavenumber Lorentz transformation~\eqref{eq:Lorentz_w_k} to~\eqref{eq:freq_cons_sup}, replacing the frame velocity by the modulation velocity through $v_\text{m}=c^2/v_\text{f}$ with $\omega_{i,j}^\pm=k_{i,j}^\pm v_{i,j}$ provides
\begin{equation}\label{eq:k_doppler_sup}
k_\text{e}=k_i^+\left(1-\frac{v_i}{v_\text{m}}\right)=k_j^+\left(1-\frac{v_j}{v_\text{m}}\right)=k_j^-\left(1+\frac{v_j}{v_\text{m}}\right).
\end{equation}
\end{subequations}
Dividing~\eqref{eq:freq_cons_sup} by \eqref{eq:k_doppler_sup} gives the relation
\begin{equation}\label{eq:ke}
{k_\text{e}}'/k_\text{e}=\gamma.
\end{equation}
The frequency transitions for a backward problem, with the incident wave coming from the right, in medium $i$, would be found by changing the signs of the superscripts and the sign of $v_\text{m}$ in~\eqref{eq:k_doppler_sup}.

Equations~\eqref{eq:freq_sup} are represented graphically in Fig.~\ref{fig:interface_inverse}(b). Once again, the axes of the moving frame and the laboratory frame are superimposed. The slopes of the moving frame, which are again $v_\text{f}/c$ and $c/v_\text{f}$ from~\eqref{eq:Lorentz_w_k}, are written in terms of the interface velocity, $v_\text{m}$, by setting $v_\text{f}/c=c/v_\text{m}$. The dispersion curves of the two media are drawn, and the conservation of $k'$ is enforced by tracing a line parallel to $\omega'$, with intercept $k_\text{e}$. The solutions are located at the intersections of this line and the dispersion diagram of medium $j$, remembering that both scattered waves are in medium $j$. Reading out the solutions in the laboratory frame reveals that both waves are downshifted, with $\psi_j^-$ having a negative frequency, associated with time-reversal~\cite{fink1992time}.

The results of this section are summarized in Tab.~\ref{tab:interface}.

\bgroup
\def\arraystretch{2}
\begin{table}[h]  \centering
\begin{tabular}{cc}
Subluminal & Superluminal \\ \colrule
\toprule
 \multicolumn{2}{c}{Coefficients (Fig.~\ref{fig:interface})}\\ \colrule
 $\tau_{ji}\overset{\eqref{eq:coeff_trans_sub}}{=}\dfrac{2\eta_j}{\eta_i+\eta_j}\left(\dfrac{1-\alpha_i }{1-\alpha_j }\right)$& $\xi_{ji}\overset{\eqref{eq:coeff_trans_sup}}{=}\dfrac{\eta_i+\eta_j}{2\eta_i}\left(\dfrac{1-\alpha_i}{1-\alpha_j}\right)$\\
 $\gamma_{iji}\overset{\eqref{eq:coeff_refl_sub}}{=}\dfrac{\eta_j-\eta_i}{\eta_i+\eta_j}\left(\dfrac{1-\alpha_i }{1+\alpha_i }\right)$  &$\zeta_{ji}\overset{\eqref{eq:coeff_refl_sup}}{=}\dfrac{\eta_i-\eta_j}{2\eta_i}\left(\dfrac{1-\alpha_i}{1+\alpha_j}\right)$\\\colrule
\multicolumn{2}{c}{Frequencies (Fig.~\ref{fig:interface_inverse})}\\ \colrule
$\omega_i^-\overset{\eqref{eq:sub_doppler}}{=}\omega_i^+\dfrac{1-\alpha_i}{1+\alpha_i}$ & $k_j^-\overset{\eqref{eq:k_doppler_sup}}{=}k_i^+\dfrac{1-1/\alpha_j}{1+1/\alpha_i}$\\
$\omega_j^+\overset{\eqref{eq:sub_doppler}}{=}\omega_i^+\dfrac{1-\alpha_i}{1-\alpha_j}$ & $k_j^+\overset{\eqref{eq:k_doppler_sup}}{=}k_i^+\dfrac{1-1/\alpha_i}{1-1/\alpha_i}$\\ \colrule
\multicolumn{2}{c}{$\alpha_{i,j}=v_\text{m}/v_{i,j}$} \\ \botrule
\end{tabular}
\caption{Summary of the scattering formulas, derived in Sec.~\ref{sec:interface}, for a spacetime interface.}
\label{tab:interface}
\end{table}
\egroup

\section{Spacetime Slab}\label{sec:ST_slab}

\subsection{Scattering Phenomenology}

\begin{figure}[h]
\centering
\includegraphics[width=1\columnwidth]{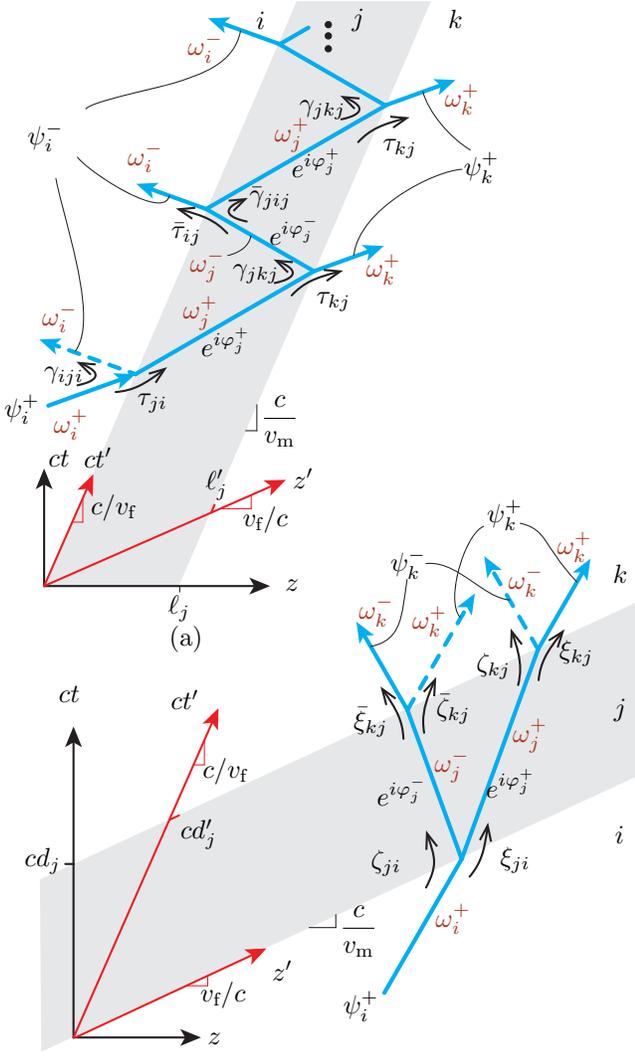}{
\psfrag{a}[c][c]{(a)}
\psfrag{b}[c][c]{(b)}
\psfrag{1}[c][c]{$i$}
\psfrag{2}[c][c]{$j$}
\psfrag{3}[c][c]{$k$}
\psfrag{z}[c][c]{$z$}
\psfrag{t}[c][c]{$ct$}
\psfrag{Z}[c][c]{$z'$}
\psfrag{T}[c][c]{$ct'$}
\psfrag{v}[c][c]{$\dfrac{c}{v_\text{m}}$}
\psfrag{l}[c][c]{$\ell'_j$}
\psfrag{d}[c][c]{$cd'_j$}
\psfrag{L}[c][c]{$cd_j$}
\psfrag{w}[c][c]{$\ell_j$}
\psfrag{e}[c][c]{$c/v_\text{f}$}
\psfrag{E}[c][c]{$v_\text{f}/c$}
\psfrag{p}[c][c]{$\tau_{ji}$}
\psfrag{P}[c][c]{$\tau_{kj}$}
\psfrag{q}[c][c]{$\gamma_{iji}$}
\psfrag{Q}[c][c]{$\gamma_{jkj}$}
\psfrag{M}[c][c]{$\bar{\gamma}_{jij}$}
\psfrag{n}[c][c]{$\bar{\tau}_{ij}$}
\psfrag{x}[c][c]{$\zeta_{kj}$}
\psfrag{X}[c][c]{$\bar{\zeta}_{kj}$}
\psfrag{C}[c][c]{$\xi_{ji}$}
\psfrag{F}[c][c]{$\xi_{kj}$}
\psfrag{S}[c][c]{$\zeta_{ji}$}
\psfrag{m}[c][c]{$\bar{\xi}_{kj}$}
\psfrag{i}[c][c]{$\psi_i^+$}
\psfrag{c}[c][c]{$\psi_i^-$}
\psfrag{R}[c][c]{$\psi_k^+$}
\psfrag{G}[c][c]{$\psi_k^-$}
\psfrag{g}[c][c]{$\psi_k^+$}
\psfrag{I}[c][c]{$\textcolor{Mahogany}{\omega_i^+}$}
\psfrag{r}[c][c]{$\textcolor{Mahogany}{\omega_i^-}$}
\psfrag{J}[c][c]{$\textcolor{Mahogany}{\omega_j^+}$}
\psfrag{j}[c][c]{$\textcolor{Mahogany}{\omega_j^-}$}
\psfrag{K}[c][c]{$\textcolor{Mahogany}{\omega_k^+}$}
\psfrag{k}[c][c]{$\textcolor{Mahogany}{\omega_k^-}$}
\psfrag{f}[c][c]{$e^{i\varphi_j^+}$}
\psfrag{B}[c][c]{$e^{i\varphi_j^-}$}
\psfrag{h}[c][c]{$cd_{\text{e}j}^\text{sub}$}
\psfrag{H}[c][c]{$\ell_{\text{e}j}^\text{sup}$}
}
\caption{Multiple-reflection description of the scattering phenomenology in spacetime slabs. Changes in line type (solid $\leftrightarrow$ dashed) denote phase reversals. (a)~Subluminal slab, with phase change occurring upon reflection to a lower impedance medium ($\eta_j<\eta_i$), according to~\eqref{eq:coeff_sub}. Note that the slope of the trajectories have been altered for representation convenience. (b)~Superluminal slab, with phase change occurring upon reflection to a higher impedance medium ($\eta_k>\eta_j$), according to~\eqref{eq:coeff_sup}.}
\label{fig:3layer_sub}
\end{figure}

Noting that the succession of two interfaces corresponds to a slab, we can now address the slab problem upon the basis of Sec.~\ref{sec:interface}. Figure~\ref{fig:3layer_sub} shows the multiple-reflections that occur in sub- and superluminal spacetime slabs. The slabs consist of a medium $j$ sandwiched between media $i$ and $k$, with media $i$ and $k$ sharing the same parameters. The two interfaces bounding the slab propagate at the same velocity and are thus parallel in the spacetime diagram. The slabs are illuminated by a wave incident from medium $i$.

The multiple reflections in the subluminal case, represented in Fig.~\ref{fig:3layer_sub}(a), and superluminal case, represented in Fig.~\ref{fig:3layer_sub}(b) are strikingly different. In the subluminal case, the incident wave is divided into a reflected and a transmitted wave at the first interface. The transmitted wave propagates to the second interface, where it splits into new reflected and transmitted waves, with the former traveling back to the first interface, where it splits again in two parts, while the latter reaches the other side of the slab. This process repeats indefinitely, with the amplitudes of the exiting waves decreasing at each round trip. The transmitted wave, $\psi_k^+$ and the reflected wave, $\psi_i^-$, are found by summing the infinite contributions.

For the superluminal case, represented in Fig.~\ref{fig:3layer_sub}(b), the wave incident from medium $i$ splits into a later forward and a later backward wave at the first interface. The two waves propagate then to the second interface, where they split again. However, there is no further scattering, so that the total later forward and later backward waves, $\psi_k^+$ and $\psi_k^-$, are the sum of only four contributions.


We next calculate the frequency transitions and the phase shifts in the spacetime slabs, to later calculate the amplitudes of the scattered waves.

\subsection{Frequency Transitions}\label{sec:freq_slab}

Let us start with the subluminal slab. The frequencies, indicated in Fig.~\ref{fig:3layer_sub}(a), are calculated by noting, as in~\eqref{eq:freq_cons_sub}, that the frequencies in the moving frame are conserved, and are all equal to $\omega'_\text{e}$, i.e.,
\begin{equation}\label{eq:freq_cons_sub_slab_p}
{\omega_\text{e}}'={\omega_i^+}'={\omega_i^-}'={\omega_j^+}'={\omega_j^-}'={\omega_k^+}'.
\end{equation}
Applying the frequency Lorentz transformation in~\eqref{eq:Lorentz_w_k}, we obtain the same result as in~\eqref{eq:sub_doppler}, but with the additional frequencies $\omega_j^-$ and $\omega_k^+$ scattered from the second interface
\begin{equation}\label{eq:freq_sub_slab}
\begin{split}
\omega_\text{e}=\omega_i^+&\left(1-\frac{v_\text{m}}{v_i}\right) =\omega_i^-\left(1+\frac{v_\text{m}}{v_i}\right)=\omega_j^+\left(1-\frac{v_\text{m}}{v_j}\right)\\&
=\omega_j^-\left(1+\frac{v_\text{m}}{v_j}\right)=\omega_k^+\left(1-\frac{v_\text{m}}{v_k}\right).
\end{split}
\end{equation}

Equation~\eqref{eq:freq_sub_slab} reveals that when media $i$ and $k$ share the same parameters, the wave transmitted through the slab has the same frequency as the incident wave, $\omega_k^+=\omega_i^+$: it is upshifted to $\omega_j^+$ at the first interface and then downshifted by the same amount at the second interface.

In the superluminal case, the frequencies, indicated in Fig.~\ref{fig:3layer_sub}(b), are calculated by noting, as in~\eqref{eq:freq_cons_sup}, that the wavenumbers in the moving frame are conserved and are all equal to $k_\text{e}$, i.e.,
\begin{equation}\label{eq:k_cons_sup_slab_p}
k_\text{e}'={k_i^+}'={k_j^+}'={k_j^-}'={k_k^+}'={k_k^-}'.
\end{equation}
Applying the wavenumber Lorentz transformation~\eqref{eq:Lorentz_w_k}, these equalities are expressed in the laboratory frame as
\begin{equation}\label{eq:k_cons_sup_slab}
\begin{split}
k_\text{e}=k_i^+&\left(1-\frac{v_i}{v_\text{m}}\right)=k_j^+\left(1-\frac{v_j}{v_\text{m}}\right)=k_j^-\left(1+\frac{v_j}{v_\text{m}}\right)\\&
=k_k^+\left(1-\frac{v_k}{v_\text{m}}\right)=k_k^-\left(1+\frac{v_k}{v_\text{m}}\right).
\end{split}
\end{equation}
Equations \eqref{eq:k_cons_sup_slab} reveal that the later forward wave, $\psi_k^+$, has the same wavenumber as the incident wave, $\psi_i^+$, i.e.  $k_k^+=k_i^+$, if media $i$ and $k$ share the same parameters.

\subsection{Phase Shift}\label{sec:phase_acc}

We start, as usual, with the subluminal case. In the moving frame, where the slab appears stationary, the phase shift of the forward and backward waves is
\begin{equation}\label{eq:phase_p_sub}
{\varphi_j^\pm}'={k_j^\pm}'\ell_j'=\frac{\omega_\text{e}'}{{v_j^\pm}'}\ell_j',
\end{equation}
where $\ell_j'$ is the length of the slab in the moving frame, as shown in Fig.~\ref{fig:3layer_sub}(a), and the second equality was found by using the dispersion relation ${k_j^\pm}'={\omega_j^\pm}'/{v_j^\pm}'=\omega_\text{e}'/{v_j^\pm}'$, from frequency conservation~\eqref{eq:freq_cons_sub_slab_p} and with $v_j'$ the phase velocity in the moving frame. We decompose this phase shift as
\begin{subequations}\label{eq:phase_diff_sum_sub}
\begin{equation}\label{eq:phase_def}
{\varphi_j^\pm}'=\bar{\varphi}'_j\pm\Delta\varphi'_j,
\end{equation}
with the average and difference parts
\begin{equation}\label{eq:phase_diff_sum}
\bar{\varphi}'_j=\frac{{\varphi_j^+}'+{\varphi_j^-}'}{2}\quad\text{and}\quad\Delta\varphi'_j=\frac{{\varphi_j^+}'-{\varphi_j^-}'}{2},
\end{equation}
\end{subequations}
where the bar should not be confused with the bar for the scattering coefficients of the backward problem. Using~\eqref{eq:phase_p_sub}, these average and difference parts may be written  as
\begin{subequations}\label{eq:phase_sum_diff_p}
\begin{equation}\label{eq:phase_sum_p}
\bar{\varphi}_j' = \frac{{\omega_\text{e}}'\ell_j'}{2}\left(\frac{1}{{v_j^+}'}+\frac{1}{{v_j^-}'}\right),
\end{equation}
\begin{equation}\label{eq:phase_diff_p}
\Delta{\varphi}_j'
=\frac{{\omega_\text{e}}'\ell_j'}{2}\left(\frac{1}{{v_j^+}'}-\frac{1}{{v_j^-}'}\right).
\end{equation}
\end{subequations}

Let us now derive the laboratory-frame counterparts of~\eqref{eq:phase_p_sub} and~\eqref{eq:phase_sum_diff_p}. From phase invariance (Eq.~\eqref{eq:phase_invariance}), $\varphi_j=\varphi_j'$, $\bar{\varphi}_j=\bar{\varphi}_j'$ and $\Delta\varphi_j=\Delta\varphi_j'$. The three primed variables $\omega'_\text{e}$, $v_j^{\pm '}$ and $\ell_j'$ in~\eqref{eq:phase_p_sub} and~\eqref{eq:phase_sum_diff_p} may be converted to their unprimed counterparts using respectively ${\omega_\text{e}}'=\gamma{\omega_\text{e}}$ (Eq.~\eqref{eq:gamma}), the equation for ${v_j^\pm}'$ in~\eqref{eq:vpresult}, and the length contraction formula
\begin{equation}\label{eq:lengths}
\ell_j'=\gamma\ell_j,
\end{equation}
where the lengths $\ell_j$ and $\ell_j'$ are indicated in Fig.~\ref{fig:3layer_sub}(a) to corresponds to the spatial separation between the two parallel lines separating the slab. Relation~\eqref{eq:lengths} is obtained by setting $z'=\ell'_j$, $z=\ell_j$ and $ct=0$ in the left expression of~\eqref{eq:Lorentz_ct_z_p}. Applying the three substitutions results in the phase shift counterpart of~\eqref{eq:phase_p_sub}
\begin{subequations}\label{eq:phase_sub_all}
\begin{equation}\label{eq:phase_sub}
{\varphi_j^\pm}=\gamma^2\omega_\text{e}\ell_j\frac{1\mp v_\text{m}v_j/c^2}{v_j\mp v_\text{m}},
\end{equation}
and the average and difference phase counterparts of~\eqref{eq:phase_sum_diff_p}
\begin{equation}\label{eq:phase_sum}
\begin{split}
\bar{\varphi}_j &=\omega_\text{e}\frac{\ell_j v_j}{v_j^2-v_\text{m}^2},
\end{split}
\end{equation}
\begin{equation}\label{eq:phase_diff}
\begin{split}
\Delta{\varphi}_j&=\omega_\text{e}\ell_j\gamma^2v_\text{m}\frac{1-v_j^2/c^2}{v_j^2-v_\text{m}^2}.
\end{split}
\end{equation}
\end{subequations}

For the case of the superluminal slab, the phase shift of the forward and backward waves is found by inserting~\eqref{eq:E_field_def_p} into~\eqref{eq:continuity_fields_sup_p}
\begin{subequations}\label{eq:phase_p_sup}
\begin{equation}\label{eq:phase_def_sup}
{\varphi_j^\pm}'={\omega_j^\pm}'d_j'=k_\text{e}'v_j^{\pm '}d_j'.
\end{equation}
with $d_j'$ is the duration in the moving frame, drawn in Fig.~\ref{fig:3layer_sub}(b). Expressing the phase shift in the form of an average and a difference, as in~\eqref{eq:phase_def}, we find
\begin{equation}\label{eq:phase_sum_p_sup}
\bar{\varphi}_j'=\frac{{k_\text{e}}'d_j'}{2}\left({v_j^+}'+{v_j^-}'\right),
\end{equation}
\begin{equation}\label{eq:phase_diff_p_sup}
\Delta{\varphi}_j'=\frac{{k_\text{e}}'d_j'}{2}\left({v_j^+}'-{v_j^-}'\right).
\end{equation}
\end{subequations}

The superluminal results~\eqref{eq:phase_p_sup} are expressed in terms of laboratory-frame quantities by replacing $k_\text{e}'=\gamma k_\text{e}$ [Eq.~\eqref{eq:ke}], the equation for ${v_j^\pm}'$ in~\eqref{eq:vpresult} in which we substitute $v_\text{f}=c^2/v_\text{m}$, and the duration
\begin{equation}\label{eq:durations}
 d_j'=\gamma d_j,
\end{equation}
where $d_j'$ and $d_j$ are indicated in Fig.~\ref{fig:3layer_sub}(b) to correspond to the time separation between the two parallel trajectories delimiting the slab. Equation~\eqref{eq:durations} is obtained by setting $t'=d'_j$, $t=d_j$, and $z=0$ in the right expression of~\eqref{eq:Lorentz_ct_z_p}. Note that~\eqref{eq:durations} does \emph{not} correspond to the time dilation equation (which would read $ d_j'=\gamma^{-1} d_j$), since time dilation compares the time separation between two fixed events, whereas we are comparing the time separation between two trajectories. Applying the substitutions for $k_\text{e}'$ ${v_j^\pm}'$ and $d_j'$ into~\eqref{eq:phase_p_sup} yield the phase shift
\begin{subequations}\label{eq:phase_sup}
\begin{equation}\label{eq:phase_sup}
{\varphi_j^\pm}=\gamma^2k_\text{e}d_j\frac{1\mp c^2/(v_jv_\text{m})}{1/v_j\mp 1/v_\text{m}},
\end{equation}
and the average and difference quantities
\begin{equation}\label{eq:phase_sum_sup}
\bar{\varphi}_j={k_\text{e}}\frac{d_jv_jv_\text{m}^2}{v_\text{m}^2-v_j^2},
\end{equation}

\begin{equation}\label{eq:phase_diff_sup}
\Delta{\varphi}_j={k_\text{e}}d_j\gamma^2v_\text{m}c^2\frac{1-v_j^2/c^2}{v_j^2-v_\text{m}^2}.
\end{equation}
\end{subequations}

Note that Eqs.~\eqref{eq:phase_sup} could have been alternatively found by applying the substitutions of Tab.~\ref{tab:duality} into~\eqref{eq:phase_sub_all}.

\subsection{Scattering Coefficients}

For the subluminal case, the reflection coefficient for the forward problem is found by summing up the amplitude of the wave reflected at the first interface, the amplitude of the wave having traveled one round trip, the amplitude of the wave having traveled two round trips, and so on, which yields
\begin{equation}\label{eq:reflection_slab_sub}
\begin{split}
\Gamma_{ii}=\frac{|\psi_i^-|}{|\psi_i^+|}=
\gamma_{iji}&
+
\bar{\tau}_{ij}\gamma_{jkj}\tau_{ji}e^{2i\bar{\varphi}}\\
&+\bar{\tau}_{ij}\gamma_{jkj}\bar{\gamma}_{jij}\gamma_{jkj}\tau_{ji}
e^{4i\bar{\varphi}}\\
&+\bar{\tau}_{ij}\gamma_{jkj}\left(\bar{\gamma}_{jij}\gamma_{jkj}\right)^2\tau_{ji}
e^{6i\bar{\varphi}}\\
=\gamma_{iji}&+
\bar{\tau}_{ij}\gamma_{jkj}\tau_{ji}e^{2i\bar{\varphi}}
\sum_{n=0}^\infty{\left(\bar{\gamma}_{jij}\gamma_{jkj}e^{2i\bar{\varphi}}\right)^n},
\end{split}
\end{equation}
which is the same expression as that for a stationary slab~\cite{born1980principles}, but with the local generalized coefficients~\eqref{eq:coeff_sub} and~\eqref{eq:bar_sub}, and the phases $\bar{\varphi}_j$ in~\eqref{eq:phase_sum}. Since $\bar{\gamma}_{jij}<1$ and $\gamma_{jkj}<1$, the geometric series~\eqref{eq:reflection_slab_sub} reduces to
\begin{equation}\label{eq:gamma_slab_sub}
\Gamma_{iki}
=\gamma_{iji}+
\frac{\bar{\tau}_{ij}\gamma_{jkj}\tau_{ji}e^{2i\bar{\varphi}_j}}
{{1-\bar{\gamma}_{jij}\gamma_{jkj}e^{2i\bar{\varphi}_j}}}=
\gamma_{iji}\frac{1-e^{2i\bar{\varphi}_j}}
{1-\bar{\gamma}_{jij}\gamma_{jkj}e^{2i\bar{\varphi}_j}},
\end{equation}
where the second equality was obtained by using the generalized Stokes condition~\eqref{eq:stokes}.

The transmission coefficient for the forward problem wave is found by summing the wave traveling through the slab, the wave reflected at the second interface and traveling one round trip, the wave traveling two round trips, and so on, i.e,
\begin{equation}\label{eq:transmission_slab_sub}
\begin{split}
T_{ki}=\frac{|\psi_k^+|}{|\psi_i^+|}=&
\tau_{kj}\tau_{ji}e^{i\varphi_j^+}\\
&+\tau_{kj}\bar{\gamma}_{jij}\gamma_{jkj}\tau_{ji}e^{i\varphi_j^+}e^{2i\bar{\varphi_j}}\\
&+\tau_{kj}\left(\bar{\gamma}_{jij}\gamma_{jkj}\right)^2\tau_{ji}e^{i\varphi_j^+}e^{2i\bar{\varphi_j}}\\
=&\tau_{kj}\tau_{ji}e^{i\varphi_j^+}\sum_{n=0}^\infty{\left(\bar{\gamma}_{jij}\gamma_{jkj}e^{2i\bar{\varphi_j}}\right)^n}\\
=&\frac{\tau_{kj}\tau_{ji}e^{i\varphi_j^+}}{1-\bar{\gamma}_{jij}\gamma_{jkj}e^{2i\bar{\varphi_j}}}.
\end{split}
\end{equation}

The scattering coefficient extrema are found from~\eqref{eq:gamma_slab_sub} and~\eqref{eq:transmission_slab_sub} to correspond to $\bar{\varphi}_j=n\pi/2$, where $n$ is an integer. For $n$ odd, reflection is maximal and transmission is minimal, and conversely for $n$ even.

The reflection and transmission coefficients for the backward problem are found by barring the unbarred coefficients, unbarring the barred coefficients and exchanging $i$ and $k$. The phases $\bar{\varphi}_j$ are unchanged, since they correspond to the round trip phase, but the phases $\varphi^+_j$ are replaced by $\varphi^-_j$. Thus,
\begin{equation}\label{eq:gamma_contra_slab_sub}
\bar{\Gamma}_{kik}=\frac{|\psi_k^+|}{|\psi_k^-|}=
\bar{\gamma}_{kjk}\frac{1-e^{2i\bar{\varphi}_j}}
{1-\bar{\gamma}_{jij}\gamma_{jkj}e^{2i\bar{\varphi}_j}}=\Gamma_{iki}\frac{\bar{\gamma}_{kjk}}{\gamma_{iji}}.
\end{equation}
\begin{equation}\label{T_Tbar_NR}
\bar{T}_{ik}=\frac{|\psi_i^-|}{|\psi_k^-|}=\frac{\bar{\tau}_{ij}\bar{\tau}_{jk}e^{i\varphi_j^-}}{1-\bar{\gamma}_{jj}\gamma_{jj}e^{2i\bar{\varphi_j}}}=T_{ki}e^{-i\Delta\varphi_j}.
\end{equation}
where the third equality in~\eqref{T_Tbar_NR} was found from a consequence of the Stokes relations~\eqref{eq:stokes_c}.

The fact that transmission coefficients for forward and backward problems are different in~\eqref{T_Tbar_NR} indicates that the structure is nonreciprocal. We remember from Sec.~\ref{sec:freq_slab} that a wave transmitted through a slab undergoes no frequency shift (Eq.~\eqref{eq:freq_sub_slab}). As this is equally true for the forward and the backward problems, and therefore for incidence from both sides of the slab, we conclude that spacetime slabs are frequency-wise reciprocal, and therefore that the transmitted waves of opposite directions differ only by the phase factor $e^{-i\Delta\varphi_j}$ in~\eqref{T_Tbar_NR}. This observation leads us to the rather counterintuitive result  that a \emph{moving} slab is reciprocal in transmission, in contrast to a modulated slab. To verify this, we insert $v_j^{+'}=v_j^{-'}$ into~\eqref{eq:phase_diff_p}, leading to $\Delta\varphi'=0$, and thus $\Delta\varphi=0$ from phase invariance.

In the superluminal case, the situation, although less usual, is simpler, since there are now only three scattering events. Reading out Fig.~\ref{fig:3layer_sub}(b), we find the later backward and later forward coefficients
\begin{subequations}\label{eq:scatt_coeff_slab_sup}
\begin{equation}\label{eq:Z_slab_sup}
Z_{ki}=\frac{|\psi_k^-|}{|\psi_i^+|}=\zeta_{kj}\xi_{ji}e^{i\varphi_j^+}+\bar{\xi}_{kj}\zeta_{ji}e^{-i\varphi_j^-},
\end{equation}
\begin{equation}\label{eq:Xi_slab_sup}
\Xi_{ki}=\frac{|\psi_k^+|}{|\psi_i^+|}=\xi_{kj}\xi_{ji}e^{i\varphi_j^+}+\bar{\zeta}_{kj}\zeta_{ji}e^{-i\varphi_j^-}.
\end{equation}
\end{subequations}
Applying the generalized Stokes relations for superluminal coefficients~\eqref{eq:stokes_sup}, \eqref{eq:scatt_coeff_slab_sup} reduce to
\begin{subequations}\label{eq:scatt_coeff_slab_sup_simpl}
\begin{equation}
Z_{ki}=\zeta_{kj}\xi_{ji}e^{-i\varphi_j^-}\left(e^{i\bar{\varphi}_j}-1\right),
\end{equation}
\begin{equation}
\Xi_{ki}=e^{-i\varphi_j^-}\left(1+\xi_{kj}\xi_{ji}\left(e^{i\bar{\varphi}_j}-1\right)\right).
\end{equation}
\end{subequations}
The scattering coefficient extrema are found from from~\eqref{eq:scatt_coeff_slab_sup_simpl} to correspond, again, to $\bar{\varphi}_j=n\pi/2 $, with $n$ an integer. For $n$ even, both later forward and later backward coefficients are maximal, while for $n$ odd, both coefficients are minimal.

The coefficients for the opposite incidence direction $\bar{Z}_{ki}$ and $\bar{\Xi}_{ki}$ are found by barring and unbarring unbarred and barred coefficients, respectively, and replacing the phase term $\varphi^+_j$ by $\varphi^-_j$:
\begin{subequations}\label{eq:scatt_coeff_slab_sup_neg}
\begin{equation}
\bar{Z}_{ki}=\bar{\zeta}_{kj}\bar{\xi}_{ji}e^{-i\varphi_j^+}\left(e^{i\bar{\varphi}_j}-1\right)=Z_{ki}\frac{\bar{\zeta}_{kj}\bar{\xi}_{ji}}{\zeta_{kj}\xi_{ji}}e^{i\Delta{\varphi}_j},
\end{equation}
\begin{equation}\label{eq:slab_sup_trans}
\bar{\Xi}_{ki}=e^{-i\varphi_j^+}\left(1+\bar{\xi}_{kj}\bar{\xi}_{ji}\left(e^{i\bar{\varphi}_j}-1\right)\right)=\Xi_{ki}e^{-i\Delta{\varphi}_j}.
\end{equation}
\end{subequations}
We observe from~\eqref{eq:slab_sup_trans} that nonreciprocity in transmission is again only due to the phase difference term $\Delta\varphi_j$, since transmission is again frequency-wise reciprocal.

\subsection{Interference Condition}\label{sec:interference_condition}

\begin{figure}[h]
\centering
\includegraphics[width=1\columnwidth]{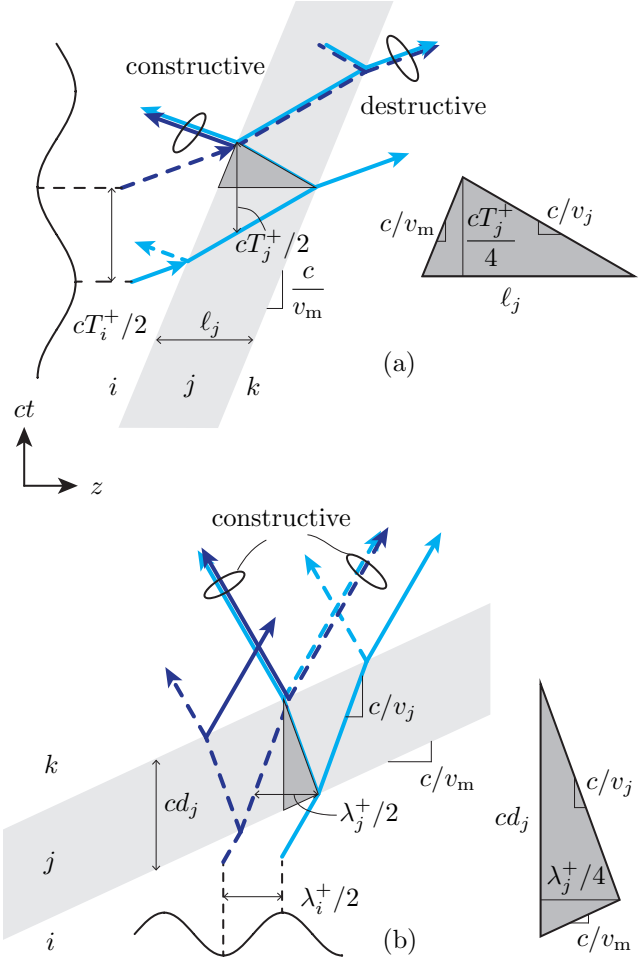}{
\psfrag{a}[c][c]{(a)}
\psfrag{b}[c][c]{(b)}
\psfrag{c}[c][c]{constructive}
\psfrag{i}[c][c]{destructive}
\psfrag{d}[c][c]{$cT_i^+/2$}
\psfrag{A}[c][c]{$cT_j^+/2$}
\psfrag{e}[c][c]{$\ell_j$}
\psfrag{l}[c][c]{$\lambda_i^+/2$}
\psfrag{L}[c][c]{$\lambda_j^+/2$}
\psfrag{D}[c][c]{$\dfrac{cT_j^+}{4}$}
\psfrag{G}[c][c]{$\lambda_j^+/4$}
\psfrag{B}[c][c]{$cd_j$}
\psfrag{1}[c][c][1]{$c/v_i$}
\psfrag{2}[c][c][1]{$c/v_j$}
\psfrag{v}[c][c][1]{$c/v_\text{m}$}
\psfrag{V}[c][c][1]{$\dfrac{c}{v_\text{m}}$}
\psfrag{z}[c][c][1]{$z$}
\psfrag{t}[c][c][1]{$ct$}
\psfrag{I}[c][c]{$i$}
\psfrag{J}[c][c]{$j$}
\psfrag{K}[c][c]{$k$}
}
\caption{Graphical Bragg-like interference argument. The light and dark blue trajectories correspond to the maxima and minima of the incident wave, and changes in line type (solid $\leftrightarrow$ dashed) denote phase reversals. (a)~Subluminal case, with constructive and destructive interference in reflection and transmission, respectively. Note that the slope of the trajectories have been altered for representation convenience. (b)~Superluminal case, with constructive interference in both the later forward and later backward waves.}
\label{fig:bragg}
\end{figure}

The extrema of the scattering coefficients, which were just previously derived, are now analysed in terms of Bragg-like interference conditions, with the help of Fig.~\ref{fig:bragg}.

In the subluminal case, constructive interference occurs in reflection and destructive interference occurs in transmission. The interference condition may be found geometrically by recognizing the triangle with slopes $c/v_\text{m}$ and $c/v_j$ has a total length $\ell_j$ related to the period of the wave $T_j^+$ through
\begin{equation}\label{eq:interference_sub}
\ell_j=\frac{cT_j^+}{4} \left(\frac{v_\text{m}}{c}+\frac{v_j}{c}\right)=\frac{\lambda_j^+}{4} (1+v_\text{m}/v_j),
\end{equation}
where the second equality is found by using $T_j^+=\lambda_j^+/v_j$. This can be seen as a generalization of the quarter-wave condition. Inserting~\eqref{eq:interference_sub} and the definition of $\omega_\text{e}$~\eqref{eq:freq_sub_slab} into~\eqref{eq:phase_sum} retrieves $\bar{\varphi}_j=\pi/2$.

In the superluminal case, shown in Fig.~\ref{fig:bragg}(b), constructive interference occurs for both the later forward and later backward components. The interference condition may be found geometrically by studying the shaded triangle and noticing the total duration $cd_j$ is related to the slopes of the triangle, $c/v_\text{m}$ and $c/v_j$ and to the wavelength $\lambda_j^+/4$ through
\begin{equation}\label{eq:interference_sup}
cd_j=\frac{\lambda_j^+}{4} \left(\frac{c}{v_j}+\frac{c}{v_\text{m}}\right)=\frac{cT_j^+}{4} (1+v_j/v_\text{m}),
\end{equation}
where $\lambda_j^+=v_jT_j^+$ was used. Inserting~\eqref{eq:interference_sup} and the definition of $k_\text{e}$,~\eqref{eq:k_cons_sup_slab}, into~\eqref{eq:phase_sum_p_sup} retrieves $\bar{\varphi}_j=\pi/2$.

The spacetime slab results are summarized in Tab.~\ref{tab:slab}.

\bgroup
\def\arraystretch{2}
\begin{table}[h]  \centering
\begin{tabular}{cc}
 Subluminal regime &Superluminal regime\\ \toprule
\multicolumn{2}{c}{Phase (Fig.~\ref{fig:3layer_sub})}\\ \colrule
\multicolumn{2}{c}{${\varphi_j^\pm}'\overset{\eqref{eq:phase_def}}{=}\bar{\varphi}'_j\pm\Delta\varphi'_j$}\\
$\bar{\varphi}_j\overset{\eqref{eq:phase_sum}}{=}\omega_\text{e}\ell_j\dfrac{ v_j}{v_j^2-v_\text{m}^2}$&$\bar{\varphi}_j\overset{\eqref{eq:phase_sum_sup}}{=}{k_\text{e}d_j}\dfrac{v_jv_\text{m}^2}{v_j^2-v_\text{m}^2}$ \\
$\Delta{\varphi}_j\overset{\eqref{eq:phase_diff}}{=}\dfrac{\bar\varphi_j\gamma^2(1-v_j^2/c^2)v_\text{m}}{v_j}$ & $\Delta{\varphi}_j\overset{\eqref{eq:phase_diff_sup}}{=}\dfrac{\bar\varphi_j\gamma^2(1-v_j^2/c^2)c^2}{v_jv_\text{m}}$\\[3pt]\colrule
\multicolumn{2}{c}{Coefficients (Fig.~\ref{fig:3layer_sub})}\\[3pt] \colrule
$\Gamma_{iki}\overset{\eqref{eq:reflection_slab_sub}}{=}\gamma_{iji}\dfrac{1-e^{2i\bar{\varphi}_j}}
{1-\bar{\gamma}_{jij}\gamma_{jkj}e^{2i\bar{\varphi}_j}}$&$Z_{ki}\overset{\eqref{eq:scatt_coeff_slab_sup_simpl}}{=}\zeta_{kj}\xi_{ji}e^{-i\varphi_j^-}\left(e^{i\bar{\varphi}_j}-1\right)$\\
$T_{ki}\overset{\eqref{eq:transmission_slab_sub}}{=}\dfrac{\tau_{kj}\tau_{ji}e^{i\varphi_j^+}}{1-\bar{\gamma}_{jij}\gamma_{jkj}e^{2i\bar{\varphi_j}}}$&$\Xi_{ki}\overset{\eqref{eq:scatt_coeff_slab_sup_simpl}}{=}\left(1+\xi_{kj}\xi_{ji}\left(e^{i\bar{\varphi}_j}-1\right)\right)e^{-i\varphi_j^-}$\\[3pt]\colrule
\multicolumn{2}{c}{Bragg Interference Condition(Fig.~\ref{fig:bragg})}\\ \colrule
$\ell_j\overset{\eqref{eq:interference_sub}}{=}\dfrac{\lambda_j^+}{4}(1+\dfrac{v_\text{m}}{v_j})$ & $d_j\overset{\eqref{eq:interference_sup}}{=}\dfrac{T_j^+}{4}(1+\dfrac{v_j}{v_\text{m}})$\\\botrule
\end{tabular}
\caption{Summary of spacetime slab results.}
\label{tab:slab}
\end{table}
\egroup

\section{Unbounded Bilayer Crystal}\label{sec:unb_cryst}

\begin{figure}
\centering
\includegraphics[width=0.9\columnwidth]{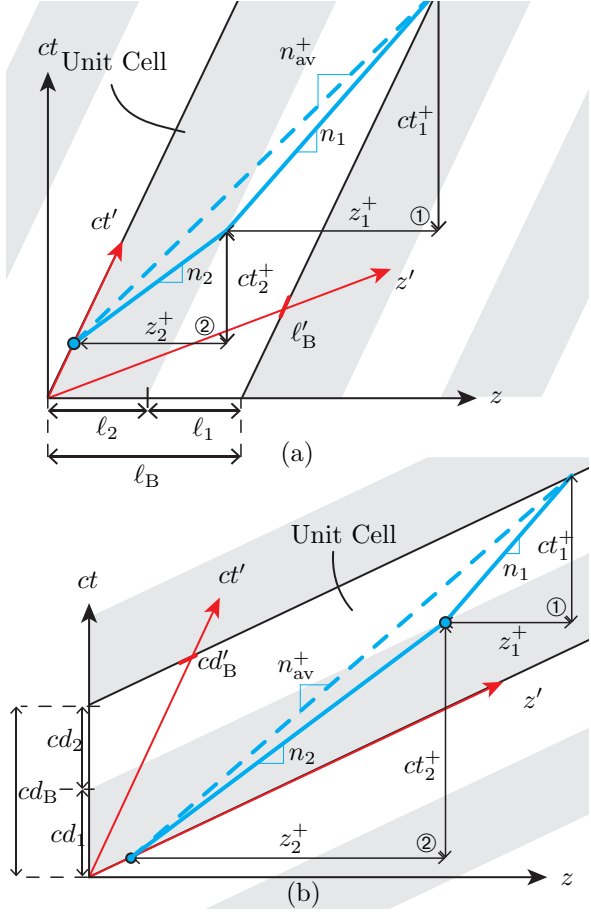}{
\psfrag{z}[c][c]{$z$}
\psfrag{t}[c][c]{$ct$}
\psfrag{A}[c][c]{(a)}
\psfrag{B}[c][c]{(b)}
\psfrag{Z}[c][c]{$z'$}
\psfrag{T}[c][c]{$ct'$}
\psfrag{i}[c][c]{$i$}
\psfrag{j}[c][c]{$j$}
\psfrag{k}[c][c]{$k$}
\psfrag{I}[c][c]{$I_{i,j}$}
\psfrag{J}[c][c]{$I_{j,k}$}
\psfrag{K}[c][c]{$I_{k,l}$}
\psfrag{1}[c][c]{$n_1$}
\psfrag{2}[c][c]{$n_2$}
\psfrag{a}[c][r]{$n_\text{av}^+$}
\psfrag{P}[c][c]{$P_{k}$}
\psfrag{p}[c][c]{$P_{j}$}
\psfrag{m}[c][c]{$T_{kj}$}
\psfrag{M}[c][c]{$T_{ji}$}
\psfrag{F}[c][r]{$\ell_1$}
\psfrag{f}[c][r]{$\ell_2$}
\psfrag{l}[c][c]{$\ell'_\text{B}$}
\psfrag{L}[c][c]{$\ell_\text{B}$}
\psfrag{u}[c][c]{$cd_\text{e}$}
\psfrag{U}[c][r]{$ct_1^+$}
\psfrag{V}[c][r]{$ct_2^+$}
\psfrag{d}[c][r]{\ding{193}}
\psfrag{D}[c][r]{\ding{192}}
\psfrag{e}[c][r]{$z_2^+$}
\psfrag{E}[c][r]{$z_1^+$}
\psfrag{G}[c][r]{$cd_\text{B}$}
\psfrag{g}[c][c]{$cd'_\text{B}$}
\psfrag{H}[c][r]{$cd_2$}
\psfrag{h}[c][r]{$cd_1$}
\psfrag{x}[c][r]{$z_2^+$}
\psfrag{y}[c][r]{$z_1^+$}
\psfrag{X}[c][r]{$ct_2^+$}
\psfrag{Y}[c][r]{$ct_1^+$}
\psfrag{o}[c][r]{$\ding{194}$}
\psfrag{Y}[c][r]{$ct_1^+$}
\psfrag{c}[c][r]{Unit Cell}
}
\caption{Bilayer spacetime crystal with spacetime unit cell and out-of-gap wave trajectories. (a)~Subluminal equal-length crystal, with $\ell_1=\ell_2$. The slopes of the triangles \ding{192} and \ding{193} are $n_1=ct_1^+/z_1^+$ and $n_2=ct_2^+/z_2^+$, so that $z_1^+=ct_1^+/n_1$ and $z_2=ct_2^+/n_2$. Substituting these lengths into the expression for the slope  $n_\text{av}=(ct_1^++ct_2^+)/(z_1^++z_2^+)$, yields~\eqref{eq:length_trajectory}. (b)~Superminal equal-duration crystal, with $d_1=d_2$. From the slopes of the triangles \ding{192} and \ding{193}, given in (a), we have $ct_1^+=z_1^+ n_1$ and $ct_2=z_2^+ n_2$. Substituting these durations into the expression for the average slope, also given in (a), yields \eqref{eq:nav_sup_intermediate}.}
\label{fig:crystal}
\end{figure}

We now move on to the study of bilayer crystals, whose periodic unit cell is composed of a pair of slabs, or layers, as presented in Fig.~\ref{fig:crystal}.
\subsection{Linear Approximation of Dispersion Diagram}\label{sec:nav}
Crystals are largely described by their dispersion diagrams, which consist in periodic alternances of passbands and stopbands, or bandgaps. The bandgaps are produced by the constructive or destructive interference of the wave scattered by the crystal, as described in Sec.~\ref{sec:interference_condition} for the particular case of a slab. Away from these bandgaps, the waves are transmitted through the crystal without attenuation or amplification, and they undergo little dispersion.

Before presenting the exact construction of the dispersion diagrams, whose complexity occults some of the physics of the problem, we shall derive a simple and useful linear approximation of the exact solution. This approximation consists of a diamond-like grid that coincides with the exact diagram away from the gaps and whose nodes correspond to the exact bandgap centers, as will be shown in Sec.~\ref{sec:dd_description}.
\begin{figure}
\centering
\includegraphics[width=1\columnwidth]{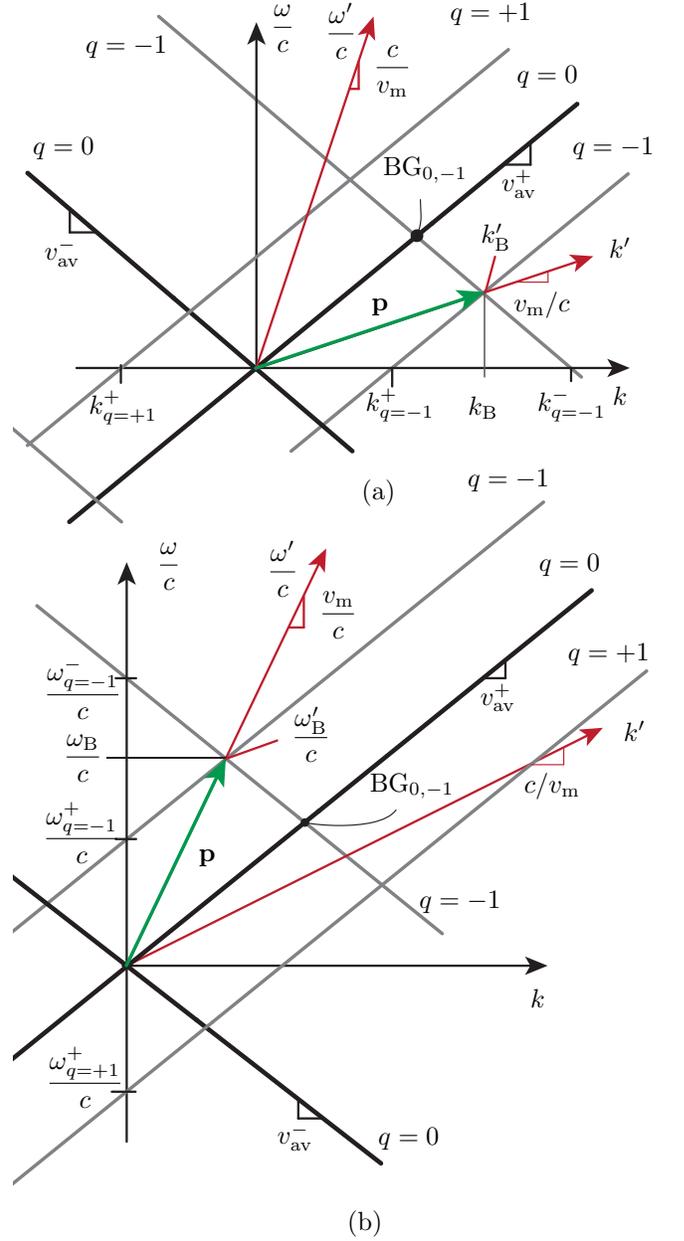}{
\psfrag{A}[c][c][1]{(a)}
\psfrag{B}[c][c][1]{(b)}
\psfrag{w}[c][c][1]{$\dfrac{\omega }{c}$}
\psfrag{k}[c][c][1]{$k$}
\psfrag{p}[c][c][1]{$\dfrac{\omega'}{c}$}
\psfrag{P}[c][c][1]{$k'$}
\psfrag{S}[c][c][1]{$v_\text{m}/c$}
\psfrag{v}[c][c][1]{$\dfrac{c}{v_\text{m}}$}
\psfrag{V}[c][c][1]{$\dfrac{v_\text{m}}{c}$}
\psfrag{s}[c][c][1]{$c/v_\text{m}$}
\psfrag{z}[c][c][1]{$v_\text{BZ}/c$}
\psfrag{Z}[c][c][1]{$\dfrac{v_\text{BZ}}{c}$}
\psfrag{F}[c][c][1]{$v_\text{av}^+$}
\psfrag{N}[c][c][1]{$v_\text{av}^-$}
\psfrag{0}[c][c][1]{$q=0$}
\psfrag{1}[c][c][1]{$q=-1$}
\psfrag{2}[c][c][1]{$q=+1$}
\psfrag{r}[c][c][1]{$\mathbf{p}$}
\psfrag{R}[c][c][1]{$\mathbf{p}$}
\psfrag{W}[c][c][1]{$\text{BG}_{0,-1}$}
\psfrag{g}[c][c][1]{$\text{BG}_{+1,0}$}
\psfrag{y}[c][c][1]{$\text{BG}_{0,-1}$}
\psfrag{Y}[c][c][1]{$\text{BG}_{+1,0}$}
\psfrag{O}[c][c][1]{$k_{{q=-1}}^-$}
\psfrag{o}[c][c][1]{$k_{{q=+1}}^+$}
\psfrag{X}[c][c][1]{$k_{{q=-1}}^+$}
\psfrag{x}[c][c][1]{$\dfrac{\omega_{{q=-1}}^+}{c}$}
\psfrag{Q}[c][c][1]{$\dfrac{\omega_{{q=+1}}^+}{c}$}
\psfrag{q}[c][c][1]{$\dfrac{\omega_{{q=-1}}^-}{c}$}
\psfrag{u}[c][c][1]{$k_\text{B}$}
\psfrag{U}[c][c][1]{$\dfrac{\omega_\text{B}}{c}$}
\psfrag{m}[c][c][1]{$k_\text{B}'$}
\psfrag{M}[c][c][1]{$\dfrac{\omega_\text{B}'}{c}$}
}
\caption{Linear approximation of the dispersion diagram of bilayer crystals with $n_2/n_1=1.5$ and equal electrical lengths. (a)~Subluminal case, with $v=(1/3)c$. (b)~Superluminal case, with $v=2c$.}
\label{fig:disp_diag_approx}
\end{figure}

\subsubsection{Average Velocity}

The slopes of the curves in the dispersion diagram of Fig.~\ref{fig:disp_diag_approx} correspond to the average group velocities, which are inversely proportional to the average refractive indices, $v_\text{av}^\pm=c/n_\text{av}^\pm$. These indices may be found geometrically with the help of Fig.~\ref{fig:crystal} as the slopes of the straight segments connecting the entering and exiting points of the trajectory through the crystal unit cell. This leads to expressions for the average velocities in terms of the refractive indices with weights corresponding to the time spent or distance traveled by the wave in each medium, as shown in Fig.~\ref{fig:crystal}.

In the subluminal case, represented in Fig.~\ref{fig:crystal}(a), the average index is found as
\begin{equation}\label{eq:length_trajectory}
n_\text{av}^\pm=\frac{{t_1^\pm}+{t_2^\pm}}{{t_1^\pm/n_1}+{t_2^\pm /n_2}}
\end{equation}
where the superscript refers to the $\pm z$ wave direction, and where the terms ${d_{1,2}^\pm}$ correspond to the duration of the wave travels in layers $1,2$ in the $\pm$ direction. This average index may be more conveniently written in terms the layer lengths, $\ell_{1,2}$. For this purpose, we express the durations in terms of these lengths, derived in Supp. Mat.~\ref{sec:lengths} as
\begin{equation}\label{eq:duration_pm}
{t_{1,2}^\pm}=\frac{\ell_{1,2}n_{1,2}/c}{1\mp v_\text{m}n_{1,2}/c},
\end{equation}
and insert this relation into~\eqref{eq:length_trajectory}, which yields
\begin{equation}\label{eq:nav}
n_\text{av}^\pm=\frac{n_1\ell_1+n_2\ell_2\mp v_\text{m}n_1n_2/c(\ell_1+\ell_2)}{\ell_1+\ell_2\mp v_\text{m}/c\left(n_2\ell_1+n_1\ell_2\right)}.
\end{equation}
It is also useful, for later use, to write the expression of the average velocity $v_\text{av}^\pm=c/n_\text{av}$ in terms of the velocities in the two media, i.e.,
\begin{equation}\label{eq:vav}
v_\text{av}^\pm=\frac{v_1v_2\left(\ell_1+\ell_2\right)\mp v_\text{m}\left(\ell_1v_1+\ell_2v_2\right)}{\left(\ell_1v_2+\ell_2v_1\right)\mp v_\text{m}\left(\ell_1+\ell_2\right)}.
\end{equation}

In the superluminal case, represented in Fig.~\ref{fig:crystal}(b), the average index is found as
\begin{equation}\label{eq:nav_sup_intermediate}
n_\text{av}^\pm=\frac{z_1^\pm n_1+z_2^\pm n_2}{z_1^\pm+z_2^\pm},
\end{equation}
where the quantities $z_{1,2}^\pm$ correspond to the length traveled by the wave in the $\pm$ direction in layers $1,2$. The average index may be more conveniently written in terms the layer durations, $d_{1,2}$. For this purpose, we express the lengths in terms of these durations, derived in Supp. Mat.~\ref{sec:lengths}, as
\begin{equation}\label{eq:length_pm}
z_{1,2}^\pm=\frac{cd_{1,2}}{n_{1,2}\mp c/v_\text{m}},
\end{equation}
and insert this relation into~\eqref{eq:nav_sup_intermediate}, which yields
\begin{equation}\label{eq:nav_sup}
n_\text{av}^\pm=\frac{d_1+d_2\mp c/v_\text{m}(d_1/n_2+d_2/n_1)}
{d_1/n_1+d_2/n_2\mp c/(v_\text{m}n_1n_2)(d_1+d_2)}.
\end{equation}
It is also useful, for later use, to write the expression of the average velocity in terms of the velocities in the two media, i.e.,
\begin{equation}\label{eq:vav_sup}
v_\text{av}^\pm=\frac{v_\text{m}(d_1v_1+d_2v_2)\mp v_1v_2(d_1+d_2)}
{v_\text{m}(d_1+d_2)\mp (d_1v_2+d_2v_1)}.
\end{equation}
Note that~\eqref{eq:nav_sup} and~\eqref{eq:vav_sup} could have been alternatively found by applying the substitutions of Tab.~\ref{tab:duality} to~\eqref{eq:nav} and~\eqref{eq:vav}.

Table~\ref{tab:nav} lists the forward and backward average refractive indices at special velocity points for unit cells of equal layer length or equal layer duration. The first row gives the average indices for the purely spatial regime, found by setting $v_\text{m}=0$ in~\eqref{eq:nav}. These indices are found to be the same for forward or backward waves, and are the arithmetic average of the two indices.

The second row provides the average indices for the upper limit of the subluminal regime (or lower limit of the interluminal regime, see Sec.~\ref{sec:velocity_regimes}), and are found by setting $v_\text{m}=c/n_2$ in~\eqref{eq:nav}. The forward average index reduces to $n_2$, since the wave has the same velocity as the modulation, and therefore never exits the layer of refractive index $n_2$. The backward wave travels across both media, so that the average index is a function of the two refractive indices.

The third row corresponds to the superluminal lower limit (or upper limit of the interluminal regime, see Sec.~\ref{sec:velocity_regimes}) and is found by setting $v_\text{m}=c/n_1$ in~\eqref{eq:nav_sup}. Now, the forward average index reduces to $n_1$ since the wave remains in medium $n_1$ while the backward index is the same function as in the previous case with exchanged indices.

The fourth row provides the indices for the purely temporal limit, found by setting $v_\text{m}=\infty$ in~\eqref{eq:nav_sup}. The average indices for forward and backward waves are again equal. Now the inverse of the average index average index is the arithmetic average of the inverses of two indices

The series and parallel circuit forms of the spatial and temporal average indices are reminiscent of the polarization-dependent form of the average indices in hyperbolic media~\cite{smith2003electromagnetic}.
\bgroup
\def\arraystretch{2.5}
\begin{table}[h]  \centering
\begin{tabular}{lccc}
 \toprule
  \begin{minipage}[c]{1.8cm}\textbf{Modulation\\regime} \end{minipage}
   &\begin{minipage}[c]{1.6cm}\textbf{Velocity \\($v_\text{m}$)} \end{minipage} & \begin{minipage}[c]{2cm}\textbf{Average index ($n_\text{av}^+$)}\end{minipage} &  \begin{minipage}[c]{2cm}\textbf{Average index ($n_\text{av}^-$)}\end{minipage}\\ \colrule
spatial& $0$ & $\dfrac{1}{2}\left(n_1+n_2\right)$&$\dfrac{1}{2}\left(n_1+n_2\right)$\\
  \begin{minipage}[l]{1.6cm} subluminal\\upper limit\end{minipage}&   $v_2=\dfrac{c}{n_2}$ & $n_2$ & $n_2\dfrac{3n_1+n_2}{n_1+3n_2}$ \\[3pt]
 \begin{minipage}[l]{1.6cm} superluminal\\lower limit\end{minipage}& $v_1=\dfrac{c}{n_1}$ & $n_1$ &$n_1\dfrac{3n_2+n_1}{n_2+3n_1}$\\[3pt]
temporal&  $\infty$ & $\dfrac{1}{2}\left(\dfrac{1}{n_1}+\dfrac{1}{n_2}\right)^{-1}$&$\dfrac{1}{2}\left(\dfrac{1}{n_1}+\dfrac{1}{n_2}\right)^{-1}$ \\[3pt]
 \botrule
\end{tabular}
\caption{Average refractive index for different modulation velocities, from~\eqref{eq:nav}, with $\ell_1=\ell_2$ (upper half) and~\eqref{eq:nav_sup}, with $d_1=d_2$ (lower half).}
\label{tab:nav}
\end{table}
\egroup
\subsubsection{Dispersion Diagram Period}

The first pair of dispersion curves in Fig.~\ref{fig:disp_diag_approx} (solid lines), $\omega=\pm v_\text{av}^\pm k$, has its intercept at the origin, while the other pairs of curves are (obliquely) translated from this first pair by $q\mathbf{p}=q(k_\text{B},\omega_\text{B})$, where $q$ is the integer representing the pair of curves, $\mathbf{p}$ is the spacetime period and $k_\text{B}$ and $\omega_\text{B}$ are the wavenumber and frequency translation quantities.

In the subluminal case, the spacetime period vector is found by recalling that the crystal is stationary in the moving frame, so that the period is simply
\begin{equation}
\mathbf{p}'=\left(k'_\text{B},0\right)=\left(\frac{2\pi}{\ell_\text{B}'},0\right).
\end{equation}
This period is expressed in terms of laboratory-frame quantities by first applying the wavenumber Lorentz transformations $k_\text{B}'=k_\text{B}/\gamma$ and $k_\text{B}'=\omega_\text{B}/(v_\text{m}\gamma)$, which are found by setting $k'=k'_\text{B}$, $\omega'=0$, $k=k_\text{B}$ and $\omega=\omega_\text{B}$ in~\eqref{eq:Lorentz_w_k} with $v_\text{f}=v_\text{m}$ (Sec.~\ref{sec:resolution_strategy}), and next applying the length contraction $\ell_\text{B}'=\gamma \ell_\text{B}$ (as in Eq.~\eqref{eq:lengths}, with replacing $\ell_j, \ell_j'$ by $\ell_\text{B},\ell_\text{B}'$), so that
\begin{equation}\label{eq:kb_wb_sub}
k_\text{B}=\frac{2\pi}{\ell_\text{B}}, \qquad \omega_\text{B}=\frac{2\pi v_\text{m}}{\ell_\text{B}},
\end{equation}
which may be alternatively obtained from Bloch-Floquet theory.

In the superluminal case, the spacetime period vector is
\begin{equation}
\mathbf{p}'=\left(0,\omega'_\text{B}\right)=\left(0,\frac{2\pi}{d'_\text{B}}\right).
\end{equation}
As for the subluminal case, we express this period into laboratory frame quantities by first applying the frequency transformations $\omega_\text{B}'=\omega_\text{B}/\gamma$ and $\omega'_\text{B}=k_\text{B}/(\gamma v_\text{f}/c^2)=k_\text{B}v_\text{m}/\gamma$,
which are found by setting $\omega'=\omega_\text{B}'$, $k'=0$, $k=k_\text{B}$, and $\omega=\omega_\text{B}$ in~\eqref{eq:Lorentz_w_k} with $v_\text{f}=c^2/v_\text{m}$ (Sec.~\ref{sec:resolution_strategy}). We then apply the time contraction $d_\text{B}'=d_\text{B}/\gamma $, as in~\eqref{eq:durations}, replacing $d_j,d_j'$ by $d_\text{B},d_\text{B}'$, and this gives
\begin{equation}\label{eq:kb_wb_sup}
\omega_\text{B}=\frac{2\pi}{d_\text{B}}, \qquad k_\text{B}=\frac{2\pi }{v_\text{m}d_\text{B}}.
\end{equation}

For both the subluminal and the superluminal cases, we may write the equations for the pairs of curves forming the linear approximation grid as
\begin{equation}\label{eq:st_period}
\omega + q \omega_\text{B}=\pm v_\text{av}^\pm (k +qk_\text{B}),
\end{equation}
so that Fig.~\ref{fig:disp_diag_approx} may be simply plotted from $v_\text{av}^\pm$ derived in Sec.~\ref{sec:nav} along with the just derived translation quantities.

\subsection{Bandgap Center Position}

The bandgap centers are located at the nodes of the linear approximation grid we have just derived. The bandgap center $\text{BG}_{0,-1}$ is indicated in Fig.~\ref{fig:disp_diag_approx} at the intersection of curves $q=0$ and $q=-1$.

Let us start with the subluminal case. We first derive the $k$-intercepts of the curves $\omega=v_\text{av}^\pm (k-k_q^\pm)$, with $k_{q}^\pm$ being the $k$-intercepts of the forward-wave and backward-wave solutions, found by setting $\omega=0$ in \eqref{eq:st_period} as
\begin{equation}\label{eq:k_intercept}
k_{q}^\pm=\frac{q\omega_\text{B}}{ \pm v_\text{av}^\pm}- qk_\text{B}=qk_\text{B}\left(\frac{v_\text{m}}{\pm v_\text{av}^\pm}-1\right),
\end{equation}
where the second equation in~\eqref{eq:k_intercept} was found using~\eqref{eq:kb_wb_sub}

The position of bandgap $\text{BG}_{0,-1}$ is then found by intersecting the curves representing the two relevant modes, i.e. setting $\omega=v_\text{av}^+k=-v_\text{av}^-(k-k_{q=-1}^-)$. Inserting~\eqref{eq:k_intercept} into this intersection condition results in the bandgap center position

\begin{equation}\label{eq:bandgap_pos}
k_{0,-1}^\text{c}=\frac{v_\text{m}+v_\text{av}^-}{v_\text{av}^++v_\text{av}^-}k_\text{B},
\end{equation}
where the superscript c refers to the center of the bandgap. The other bandgap centers are found by a similar procedure. This relation reduces to the coupled-mode result in~\cite{seshadri1977asymptotic1,elnaggar2018controlling} under the approximation $v_\text{av}^+=v_\text{av}^-$.

For the superluminal regime, we first calculate the $\omega$-intercepts of the forward and backward curve of each mode. These intercepts are found by setting $k=0$ in~\eqref{eq:st_period}, which yields

\begin{equation}
\omega_{q}^\pm=\pm qk_\text{B}v_\text{av}^\pm-q\omega_\text{B}=q\omega_\text{B}\left(\frac{\pm v_\text{av}^+}{v_\text{m}}-1\right),
\end{equation}
with the second equality again obtained using~\eqref{eq:kb_wb_sup}. From these intercepts, we find the intersection position of the bandgap $\text{BG}_{0,-1}$ by intersecting the two curves as $k=\omega/v_\text{av}^+=-(\omega-\omega_{q=-1}^\text{b})/v_\text{av}^-$, which yields
\begin{equation}\label{eq:bandgap_pos_sup}
\omega_{0,-1}^\text{c}
=\frac{v_\text{av}^+}{v_\text{m}}
\frac{v_\text{av}^-+v_\text{m}}{v_\text{av}^-+v_\text{av}^+}\omega_\text{B},
\end{equation}
which reduce to the results in~\cite{seshadri1977asymptotic2} under the approximation $v_\text{av}^+=v_\text{av}^-$. The other bandgap centers are again found by a similar procedure.

\subsection{Unit-Cell Tranfer Matrix}\label{sec:unitcell}

\begin{figure}
\centering
\includegraphics[width=1\columnwidth]{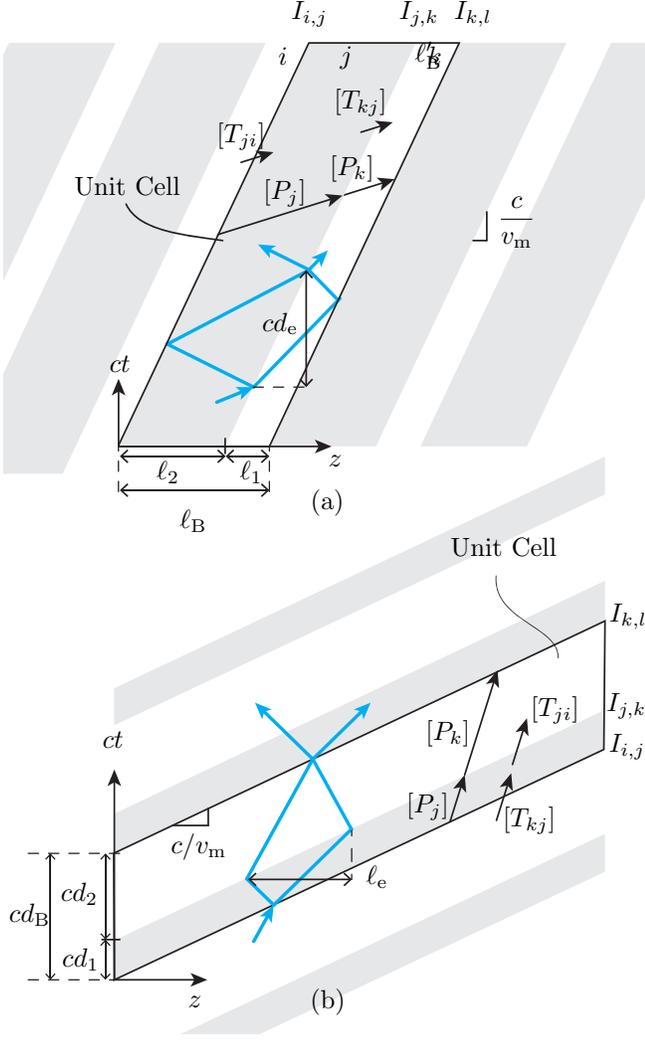}{
\psfrag{z}[c][c]{$z$}
\psfrag{t}[c][c]{$ct$}
\psfrag{A}[c][c]{(a)}
\psfrag{B}[c][c]{(b)}
\psfrag{Z}[c][c]{$z'$}
\psfrag{T}[c][c]{$ct'$}
\psfrag{i}[c][c]{$i$}
\psfrag{j}[c][c]{$j$}
\psfrag{k}[c][c]{$k$}
\psfrag{I}[c][c]{$I_{i,j}$}
\psfrag{J}[c][c]{$I_{j,k}$}
\psfrag{K}[c][c]{$I_{k,l}$}
\psfrag{1}[c][c]{$n_1$}
\psfrag{2}[c][c]{$n_2$}
\psfrag{a}[c][r]{$n_\text{av}^+$}
\psfrag{P}[c][c]{$[P_{k}]$}
\psfrag{p}[c][c]{$[P_{j}]$}
\psfrag{m}[c][c]{$[T_{kj}]$}
\psfrag{M}[c][c]{$[T_{ji}]$}
\psfrag{F}[c][r]{$\ell_1$}
\psfrag{f}[c][r]{$\ell_2$}
\psfrag{l}[c][c]{$\ell'_\text{B}$}
\psfrag{L}[c][c]{$\ell_\text{B}$}
\psfrag{u}[c][c]{$cd_\text{e}$}
\psfrag{U}[c][r]{$ct_1^+$}
\psfrag{V}[c][r]{$ct_2^+$}
\psfrag{d}[c][c]{$\ell_\text{e}$}
\psfrag{e}[c][r]{$z_2^+$}
\psfrag{E}[c][r]{$z_1^+$}
\psfrag{G}[c][r]{$cd_\text{B}$}
\psfrag{g}[c][c]{$cd'_\text{B}$}
\psfrag{D}[c][r]{$cd_1$}
\psfrag{H}[c][r]{$cd_2$}
\psfrag{x}[c][r]{$z_2^+$}
\psfrag{y}[c][r]{$z_1^+$}
\psfrag{X}[c][r]{$ct_2^+$}
\psfrag{Y}[c][r]{$ct_1^+$}
\psfrag{o}[c][r]{$\ding{194}$}
\psfrag{Y}[c][r]{$ct_1^+$}
\psfrag{v}[c][r]{$\dfrac{c}{v_\text{m}}$}
\psfrag{V}[c][r]{$c/v_\text{m}$}
\psfrag{c}[c][r]{Unit Cell}
}

\caption{Bilayer spacetime crystal with spacetime unit cell (equal-phase design). (a)~Subluminal regime. b) Superminal regime.}
\label{fig:equal-phase}
\end{figure}

We now calculate the transfer matrix of a unit cell, extending the classical transfer matrix of stationary structures~\cite{abeles1950theorie}, with the help of Fig.~\ref{fig:equal-phase}. Note that this graph corresponds, for later illustration convenience, to the particular design of an equal-phase crystal, where the forward and backward round-trip trajectories emerge at the same spacetime point. By definition, the transfer matrix relates the fields in media $i$ and $k$ at the interfaces $I_{i,j}$ and $I_{k,l}$ as
\begin{equation}\label{eq:uc_matrix}
\begin{bmatrix}
\psi_k^+\\
\psi_k^-
\end{bmatrix}_{I_{k,l}}=
\left[M_\text{B}\right]\begin{bmatrix}
\psi_i^+\\
\psi_i^-
\end{bmatrix}_{I_{i,j}},
\end{equation}
where the two interfaces separate layers $i,j$ and $k,l$, respectively, as shown in Fig.~\ref{fig:equal-phase}.

The unit-cell matrix is found by multiplying interface matrices, $[T_{ji}]$ and propagation matrices, $[P_{j}]$, according to Fig.~\ref{fig:equal-phase}, i.e.,
\begin{equation}
[M_\text{B}]=
\left[P_{k}\right]\left[T_{kj}\right]\left[P_{j}\right]\left[T_{ji}\right].
\end{equation}
We first calculate the interface matrix, $[T_{m,m-1}]$, which relates the fields at both sides of the interface $I_{m,m-1}$. At the first interface of the unit cell, we have
\begin{equation}\label{eq:interface_matrix}
\begin{bmatrix}
\psi_j^+\\
\psi_j^-
\end{bmatrix}_{I_{i,j}}=[T_{ji}]\begin{bmatrix}
\psi_i^+\\
\psi_i^-
\end{bmatrix}_{I_{i,j}}.
\end{equation}
The interface matrix may be computed using the results of Sec.~\ref{sec:interface} for interface scattering.

In the subluminal case, reading out Fig.~\ref{fig:interface}(a), we write the scattered fields in terms of the incident fields
\begin{subequations}\label{eq:sub_field_relations}
\begin{equation}
\psi_j^+=\tau_{ji}\psi_i^++\bar{\gamma}_{jij}\psi_j^-,
\end{equation}
\begin{equation}
\psi_i^-=\gamma_{iji}\psi_i^++\bar{\tau}_{ij}\psi_j^-,
\end{equation}
\end{subequations}
and rearrange these equations to cast them into the form of~\eqref{eq:interface_matrix}, i.e.,
\begin{equation}\label{eq:sub_TM_coeff}
[T_{ji}^\text{Sb}]=\frac{1}{\bar{\tau}_{ij}}
\begin{bmatrix}
\tau_{ji}\bar{\tau}_{ij}-\bar{\gamma}_{jij}\gamma_{iji}&\bar{\gamma}_{jij}\\
-\gamma_{iji}&1
\end{bmatrix},
\end{equation}
where the superscript Sb stands for subluminal. Substituting the interface scattering coefficients~\eqref{eq:coeff_sub} into~\eqref{eq:sub_TM_coeff} yields
\begin{equation}\label{eq:sub_TM}
\left[T_{ji}^\text{Sb}\right]=
\frac{1}{2\eta_i}\begin{bmatrix}
(\eta_i+\eta_j)\dfrac{1-v_\text{m}/v_i}{1-v_\text{m}/v_j}&(\eta_i-\eta_j)\dfrac{1+v_\text{m}/v_i}{1-v_\text{m}/v_j}\\[2ex]
(\eta_i-\eta_j)\dfrac{1-v_\text{m}/v_i}{1+v_\text{m}/v_j}&(\eta_i+\eta_j)\dfrac{1+v_\text{m}/v_i}{1+v_\text{m}/v_j}\\[2ex]
\end{bmatrix}.
\end{equation}

In the superluminal case, reading out Fig.~\ref{fig:interface}(b), we write the scattered fields in terms of the incident fields as
\begin{subequations}\label{eq:sup_field_relations}
\begin{equation}
\psi_j^+=\xi_{ji}\psi_i^++\bar{\zeta}_{ji}\psi_i^-,
\end{equation}
\begin{equation}
\psi_j^-=\zeta_{ji}\psi_i^++\bar{\xi}_{ji}\psi_i^-.
\end{equation}
\end{subequations}
These equations can then be cast into the matrix form of~\eqref{eq:interface_matrix}
\begin{equation}\label{eq:sup_TM_coeff}
[T_{ji}^\text{Sp}]=
\begin{bmatrix}
\xi_{ji}& \bar{\zeta}_{ji}\\
\zeta_{ji}&\bar{\xi}_{ji}
\end{bmatrix},
\end{equation}
with the superscript Sp standing for superluminal. We note that~\eqref{eq:sup_TM_coeff} is much simpler than~\eqref{eq:sub_TM_coeff}, since in the superluminal case both incident waves are in the first medium and both scattered waves are in the later medium. In other words, the scattering matrix is equal to the interface matrix for the superluminal case. Although~\eqref{eq:sup_TM_coeff} and~\eqref{eq:sub_TM_coeff} have a very different appearance, inserting~\eqref{eq:coeff_sup} into \eqref{eq:sup_TM_coeff}, reveals that they are in fact exactly identical, i.e. $[T_{ji}^\text{Sb}]=[T_{ji}^\text{Sp}]$, consistently with~\cite{biancalana2007dynamics}.

We next derive the propagation matrix, which relates the fields at the first and second interface delimiting a given layer. For instance, the propagation matrix of layer $j$ is given by
\begin{equation}
\left[\begin{matrix}
\psi_j^+\\
\psi_j^-
\end{matrix}\right]_{I_{j,k}}=
[P_j]
\left[\begin{matrix}
\psi_j^+\\
\psi_j^-
\end{matrix}\right]_{I_{i,j}}.
\end{equation}
Since there is no scattering and only a phase shift between the interfaces, the matrix simply reads
\begin{equation}\label{eq:prop_matrix}
[P_j]=
\begin{bmatrix}
e^{i\varphi_j^+}&0\\[2ex]
0&e^{-i\varphi_j^-}\\
\end{bmatrix}=e^{i\Delta\varphi_j}\begin{bmatrix}
e^{i\bar{\varphi}_j}&0\\[2ex]
0&e^{-i\bar{\varphi}_j}\\
\end{bmatrix},
\end{equation}
with the phase terms defined in Sec.~\ref{sec:phase_acc}. Inserting the interface matrix \eqref{eq:sub_TM} and the propagation matrix \eqref{eq:prop_matrix} into~\eqref{eq:uc_matrix} yields the unit-cell transfer matrix
\begin{equation}\label{eq:crystal_M}
[M_\text{B}]=e^{i\Delta\varphi}\begin{bmatrix}
a & ib \\
ic & a^*\\
\end{bmatrix}=e^{i\Delta\varphi}[M_{\text{B}0}],
\end{equation}
where
\begin{subequations}\label{eq:crystal_M_values}
\begin{equation}
a=e^{i\bar{\varphi}_k}\left(\cos\bar{\varphi}_j+\frac{i}{2}\left(\frac{\eta_1}{\eta_2}+\frac{\eta_2}{\eta_1}\right)\sin\bar{\varphi}_j\right),
\end{equation}
\begin{equation}
b=-e^{-i\bar{\varphi}_k}\frac{1}{2}\left(\frac{\eta_1}{\eta_2}-\frac{\eta_2}{\eta_1}\right)\frac{1+v/v_1}{1-v/v_1}\sin\bar{\varphi}_j,
\end{equation}
\begin{equation}
c=e^{i\bar{\varphi}_k}\frac{1}{2}\left(\frac{\eta_1}{\eta_2}-\frac{\eta_2}{\eta_1}\right)\frac{1-v/v_1}{1+v/v_1}\sin\bar{\varphi}_j.
\end{equation}
\end{subequations}
Equations~\eqref{eq:crystal_M_values} are valid for both the subluminal regime and the superluminal regime, with the phase expressions $\bar{\varphi}_{j,k}$ and $\Delta\varphi_{j,k}$ provided in Tab.~\ref{tab:slab}.
Matrix $[M_{B0}]$ is unimodular, and so the determinant of $M_\text{B}$ in~\eqref{eq:crystal_M} is $\det[M_\text{B}]=e^{i\Delta\varphi}$. Note that this matrix satisfies the conditions $be^{i\Delta\phi}=-b^*$ and $ce^{i\Delta\phi}=-c^*$, which is consistent with~\cite{longhi2010pt}, and stems from space-time reversal symmetry, as shown in Supp.~Mat.~\ref{sec:PT}. Identical results were found in~\cite{biancalana2007dynamics} using an alternative approach based on a symmetric form of Maxwell equations.

\subsection{Dispersion Relation Derivation}\label{sec:dd_derivation}
The exact dispersion diagram may now be computed using the unit-cell transfer matrix~\eqref{eq:crystal_M}, which we have just derived. From Bloch-Floquet theory, the fields before and after the unit-cell of a crystal are related through $\psi_k^\pm=e^{i\Phi_\text{B}}\psi_i^\pm$, where $\Phi_\text{B}$ is the Bloch-Floquet phase. This leads to the eignenvalue problem
\begin{equation}\label{eq:bf_matrix}
\begin{bmatrix}
\psi_\text{k}^+\\
\psi_\text{k}^-
\end{bmatrix}_{I_{k,l}}=
\left[M_\text{B}\right]
\begin{bmatrix}
\psi_\text{i}^+\\
\psi_\text{i}^-
\end{bmatrix}_{I_{i,j}}=
e^{i\Phi_\text{B}}
\begin{bmatrix}
\psi_\text{i}^+\\
\psi_\text{i}^-
\end{bmatrix}_{I_{i,j}}.
\end{equation}
The eigensolution to this problem is found by substituting~\eqref{eq:crystal_M} into~\eqref{eq:bf_matrix}, grouping the difference phase $e^{i\Delta\varphi}$ with the Bloch-Floquet phase $e^{i\Phi_\text{B}}$, so that $([M_\text{B0}]-e^{i(\Phi_\text{B}-\Delta\varphi)}[I])[\psi_i]=0$, and setting the determinant to zero, i.e.,
\begin{equation}\label{eq:determinant_bloch}
\begin{vmatrix}
a-e^{i(\Phi_\text{B}-\Delta\varphi)} & b\\[2ex]
c & a^*-e^{i(\Phi_\text{B}-\Delta\varphi)} \\
\end{vmatrix}=0.
\end{equation}
This corresponds to a quadratic equation whose solutions are
\begin{equation}\label{eq:exp_eig}
e^{i(\Phi_\text{B}-\Delta\varphi)}=\frac{a+a^*}{2}\pm\sqrt{\left(\frac{a+a^*}{2}\right)^2-1},
\end{equation}
where the identity $aa^*-bc=1$ associated with the unimodularity of $[M_{B0}]$ has been applied. Equation~\eqref{eq:exp_eig} may be rewritten in a more standard, trigonometric form by applying the Euler identity $e^{i(\Phi_\text{B}+\Delta\varphi)}=\cos(\Phi_\text{B}+\Delta\varphi)+i\sin(\Phi_\text{B}+\Delta\varphi)$ to the left hand side of~\eqref{eq:exp_eig} and writing $i\sin(\Phi_\text{B}+\Delta\varphi)=\pm\sqrt{\cos^2(\Phi_\text{B}+\Delta\varphi)-1}$. Comparing the result with the right hand side of~\eqref{eq:exp_eig} implies
\begin{equation}\label{eq:CF_DD}
\begin{split}
\cos&(\Phi_\text{B}-\Delta\varphi)=\frac{a+a^*}{2}\\
&=\cos\bar{\varphi}_j\cos\bar{\varphi}_k-\frac{1}{2}\left(\frac{\eta_1}{\eta_2}+\frac{\eta_2}{\eta_1}\right)\sin\bar{\varphi}_j\sin\bar{\varphi}_k,
\end{split}
\end{equation}
where~\eqref{eq:crystal_M_values} has been used in the second equality.

Equation~\eqref{eq:CF_DD} is the dispersion relation, whose graphical representation corresponds to the dispersion diagram, which we will construct next. We may already note that the phase $\Phi_\text{B}$ in~\eqref{eq:CF_DD} may be either real or complex, since the right hand side may be less than $1$, and the inverse cosine of a quantity less than one is complex. A real $\Phi_\text{B}$ phase will be associated with pure propagation, while an imaginary $\Phi_\text{B}$ phase will be associated with either attenuation or amplification, corresponding to bandgaps in the dispersion diagram.

\subsection{Construction of Dispersion Diagram}\label{sec:dd_construction}

The dispersion relation~\eqref{eq:CF_DD} essentially relates the Bloch-Floquet phase ($\Phi_\text{B}$) and the phases in the layers of the crystal ($\bar{\varphi}$ and $\Delta\varphi$). To plot the dispersion diagram, we must transform this relation into one of that relates the frequencies and wavenumbers ($\omega$ and $k$). This may be conveniently done using a set of parametric equations.

We start with the subluminal case. First, we write the Bloch-Floquet phase in terms of the frequency and wavenumber, i.e.,
\begin{equation}\label{eq:phase_def_bloch}
\Phi_\text{B}=\Phi_\text{B}'=k'\ell'_\text{B}=\gamma^2\left(k -\frac{v_\text{m}}{c^2}\omega\right)\ell_\text{B},
\end{equation}
where the last equation used the Lorentz wavenumber transformation~\eqref{eq:Lorentz_w_k} and length contraction~\eqref{eq:lengths}. The frequency and wavenumber in~\eqref{eq:phase_def_bloch} are related to the conserved frequency parameter $\omega_\text{e}$ through
\begin{equation}\label{eq:equal-omega}
\omega_\text{e}=\omega-v_\text{m}k,
\end{equation}
from the fact that $\omega_\text{e}'=\omega'$, similarly to~\eqref{eq:freq_sub}. Combining~\eqref{eq:phase_def_bloch} and~\eqref{eq:equal-omega}, and factoring out $\omega$ and $k$ finally yields the parametric equations
\begin{subequations}\label{eq:disp_final_solutions}
\begin{equation}\label{eq:phase_komega_sup}
\omega\left(\omega_\text{e}\right) =\frac{v_\text{m}\Phi_\text{B}\left(\omega_\text{e}\right)}{\ell_\text{B}}+\gamma^2\omega_\text{e},
\end{equation}
\begin{equation}\label{eq:phase_komega_sup} k\left(\omega_\text{e}\right)=\frac{\Phi_\text{B}\left(\omega_\text{e}\right)}{\ell_\text{B}}+\gamma^2\frac{v_\text{m}^2}{c^2}\omega_\text{e}.
\end{equation}
\end{subequations}

\begin{figure}[t]
	\centering
	\includegraphics[width=1\columnwidth]{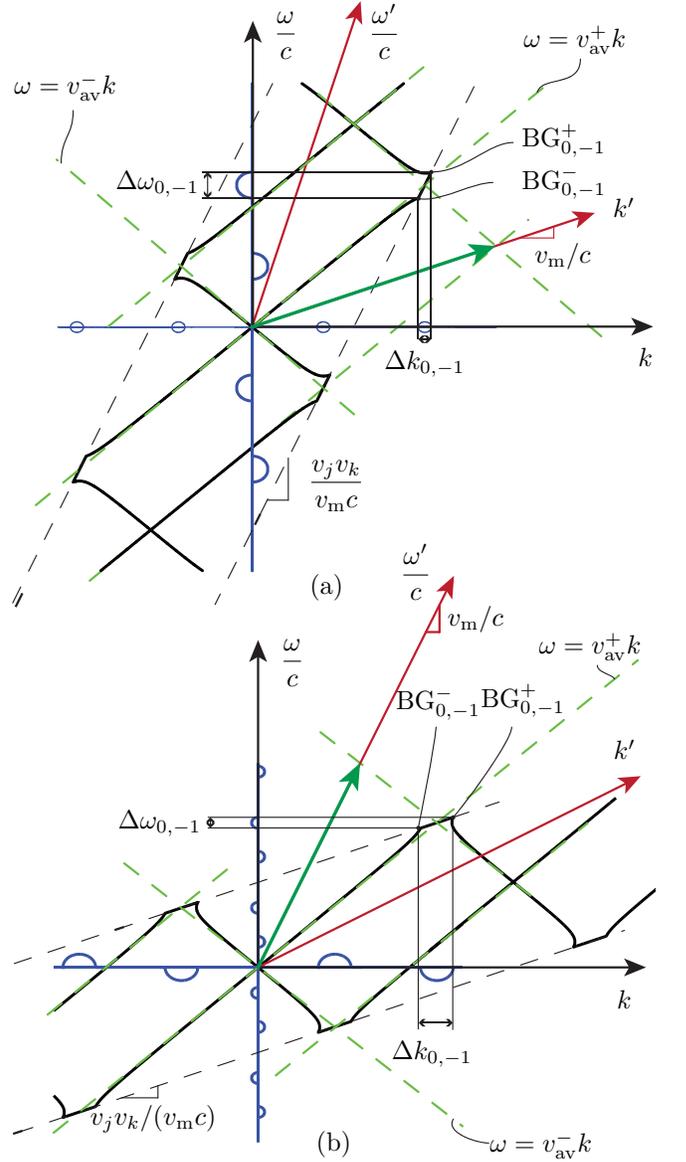}{
		\psfrag{a}[c][c][1]{(a)}
		\psfrag{b}[c][c][1]{(b)}
		\psfrag{w}[c][c][1]{$\dfrac{\omega }{c}$}
		\psfrag{k}[c][c][1]{$k$}
		\psfrag{p}[c][c][1]{$\dfrac{\omega'}{c}$}
		\psfrag{P}[c][c][1]{$k'$}
		\psfrag{v}[c][c][1]{$v_\text{m}/c$}
		\psfrag{V}[c][c][1]{$v_jv_k/(v_\text{m}c)$}
		\psfrag{B}[c][c][1]{$\dfrac{v_jv_k}{v_\text{m}c}$}
		\psfrag{F}[c][c][1]{$\omega=v_\text{av}^+ k$}
		\psfrag{N}[c][c][1]{$\omega=v_\text{av}^- k$}
		\psfrag{u}[c][c][1]{$k_\text{B}$}
		\psfrag{U}[c][c][1]{$\dfrac{\omega_\text{B}}{c}$}
		\psfrag{m}[c][c][1]{$\vec{p}$}
		\psfrag{G}[c][c][1]{$\text{BG}_{0,-1}^-$}
		\psfrag{g}[c][c][1]{$\text{BG}_{0,-1}^+$}
		\psfrag{R}[c][c][1]{$\text{BG}_{0,-1}^+$}
		\psfrag{r}[c][c][1]{$\text{BG}_{0,-1}^-$}
		\psfrag{D}[c][c][1]{$\Delta\omega_{0,-1}$}
		\psfrag{d}[c][c][1]{$\Delta k_{0,-1}$}
	}
	\caption{Dispersion diagram of bilayer crystals with $n_2/n_1=1.5$ and equal phases ($\bar{\varphi}_j=\bar{\varphi}_k$). The solid curves corresponds to the exact solution [Eqs.~\eqref{eq:disp_final_solutions} and~\eqref{eq:disp_final_solutions_sup}], with the black and blue parts respectively corresponding to the real and imaginary parts, while the dashed curves to the linear approximation [Eq.~\eqref{eq:st_period}]. (a)~Subluminal case, with $v=(1/3)c$. (b)~Superluminal case, with $v=3c$.}
	\label{fig:disp_diag}
\end{figure}

To plot the dispersion relations, we first calculate the Bloch-Floquet phase $\Phi_\text{B}(\omega_\text{e})$ using~\eqref{eq:CF_DD}, with $\bar{\varphi}_{i,j}(\omega_\text{e})$ and $\Delta\varphi_{i,j}(\omega_\text{e})$ given in Tab.~\ref{tab:slab}. We then compute $\omega(\omega_\text{e})$ and $k(\omega_\text{e})$ in~\eqref{eq:disp_final_solutions} independently, for a range of frequencies $\omega_\text{e}$, and plot them to obtain the dispersion diagram of Fig.~\ref{fig:disp_diag}(a).

In the superluminal case, the Bloch-Floquet phase is
\begin{equation}\label{eq:phase_def_bloch_sup}
\Phi_\text{B}=\Phi_\text{B}'=\omega'd'_\text{B}=\gamma^2\left(\omega -\frac{c^2}{v_\text{m}}k\right)d_\text{B},
\end{equation}
where the second equality was again found from the Lorentz frequency transformation~\eqref{eq:Lorentz_w_k} and the time contraction~\eqref{eq:durations}. The frequency and wavenumber in~\eqref{eq:phase_def_bloch_sup} are related to the conserved spatial frequency $k_\text{e}$ through
\begin{equation}\label{eq:equal-ke}
k_\text{e}=k-\omega/v_\text{m},
\end{equation}
from the fact that $k_\text{e}'=k'$ as in~\eqref{eq:freq_sup}. Combining~\eqref{eq:phase_def_bloch_sup} with~\eqref{eq:equal-ke} leads to the parametric equations
\begin{subequations}\label{eq:disp_final_solutions_sup}
\begin{equation}\label{eq:phase_komega_sup}
\omega\left(k_\text{e}\right) =\frac{\Phi_\text{B}\left(k_\text{e}\right)}{d_\text{B}}+\gamma^2\frac{c^2}{v_\text{m}}k_\text{e},
\end{equation}
and
\begin{equation}\label{eq:phase_komega_sup} k\left(k_\text{e}\right)=\frac{\Phi_\text{B}\left(k_\text{e}\right)}{v_\text{m}d_\text{B}}+\gamma^2k_\text{e}.
\end{equation}
\end{subequations}
The dispersion diagram corresponding to the parametric equations~\eqref{eq:disp_final_solutions_sup} is plotted Fig.~\ref{fig:disp_diag}(b) following the same procedure as for the subluminal case, i.e., calculating $\Phi_\text{B}(k_\text{e})$ from~\eqref{eq:CF_DD}, with $\bar{\varphi}(k_\text{e})$ and $\Delta\varphi(k_\text{e})$ given in Tab.~\ref{tab:slab}, and then computing the parametric equations~\eqref{eq:disp_final_solutions_sup} for $\omega(k_\text{e})$ and $k(k_\text{e})$.

Figure~\ref{fig:disp_diag} plots not only the exact dispersion diagram, but also the linear approximation derived in Sec.~\ref{sec:nav}. It may be seen that latter coincides with the former as expected.

\subsection{Equal-Phase Crystals}\label{sec:equal-phase}

The formulas of the previous sections are completely general and readily amenable to computation. For presentation simplicity, the rest of the paper studies the particular design of an equal-phase crystal. In this section, we find the closed-form expressions of the average and difference phase of equal-phase crystals.

We define a bilayer equal-phase crystal as a crystal whose two layers constituting the unit cell induce the same round-trip phase shift, i.e.
\begin{equation}\label{eq:equal-length_phase}
\bar{\varphi}=\bar{\varphi}_j=\bar{\varphi}_k.
\end{equation}
For stationary crystals, this corresponds to the condition $n_i\ell_i=n_j\ell_j$, which is called the quarter-wave stack condition when $n_i\ell_i=n_j\ell_j=\lambda_0/4$~\cite{born1980principles}.

In the subluminal case, the expression of the average phase is written in terms of the lengths, the modulation velocity and the wave velocities by inserting~\eqref{eq:phase_sum} into~\eqref{eq:equal-length_phase}, yielding
\begin{equation}\label{eq:equal-phase}
\bar{\varphi}= \omega_\text{e}\frac{\ell_jv_j}{v_j^2-v_\text{m}^2}
=\omega_\text{e}\frac{\ell_kv_k}{v_k^2-v_\text{m}^2}
=\omega_\text{e}d_\text{e}/2,
\end{equation}
where the quantity $d_\text{e}$ corresponds to the time a wave takes to accomplish a round trip in a layer, as can be found geometrically from the plot of $d_\text{e}$ in Fig.~\ref{fig:crystal}(a). For later use, we express this duration in terms of the length $\ell_\text{B}$ as
\begin{equation}\label{eq:de_lb}
d_\text{e}=\frac{2v_jv_k}{\left(v_j+v_k\right)
\left(v_jv_k-v_\text{m}^2\right)}\ell_\text{B},
\end{equation}
which may also be found geometrically from Fig.~\ref{fig:crystal}(a).

The total phase difference $\Delta \varphi$ between forward and backward waves in the entire unit-cell is
\begin{equation}\label{eq:delta_phi}
\begin{split}
 \Delta\varphi&=\Delta\varphi_j+ \Delta\varphi_k\\
&=v_\text{m}\gamma^2\frac{(v_j+v_k)(1-v_jv_k/c^2)}{v_jv_k}\omega_\text{e}d_\text{e}\\
&=2v_\text{m}\gamma^2\frac{1-v_jv_k/c^2}{v_jv_k-v_\text{m}^2}\omega_\text{e}\ell_\text{B}.
\end{split}
\end{equation}
where~\eqref{eq:phase_diff} and~\eqref{eq:de_lb} have been used in the second and third equalities.

In the superluminal case, the equal-phase condition is written in terms of the durations of the layers, the modulation velocity and the wave velocities by inserting~\eqref{eq:phase_sum_sup} into~\eqref{eq:equal-length_phase}, yielding
\begin{equation}\label{eq:equal-phase_sup}
\bar{\varphi}=k_\text{e}\frac{d_jv_\text{m}^2v_j}{v_\text{m}^2-v_j^2}
=k_\text{e}\frac{d_kv_\text{m}^2v_k}{v_\text{m}^2-v_k^2}
=k_\text{e}\ell_{\text{e}}/2,
\end{equation}
where $\ell_{\text{e}}$ corresponds to the total travel length of the forward and backward waves in a slab, as can be found from Fig.~\ref{fig:crystal}(b), and is related to the duration $d_\text{B}$ through
\begin{equation}\label{eq:le_db}
\ell_\text{e}=
\frac{2v_\text{m}^2v_jv_k}{\left(v_j+v_k\right)
\left(v_\text{m}^2-v_jv_k\right)}d_\text{B}.
\end{equation}

The total phase difference, $\Delta \varphi$, is
\begin{equation}\label{eq:delta_phi_sup}
\begin{split}
 \Delta\varphi&=\Delta\varphi_j+ \Delta\varphi_k\\
&=\gamma^2\frac{c^2}{v_\text{m}^2}\frac{(v_j+v_k)(1-v_jv_k/c^2)}{v_jv_k}k_\text{e}\ell_\text{e}\\
&=v_\text{m}c^2\gamma^2\frac{1-v_jv_k/c^2}{v_jv_k-v_\text{m}^2}k_\text{e}d_\text{B}
\end{split}
\end{equation}
where~\eqref{eq:phase_diff_sup} and~\eqref{eq:le_db} have been used in the second and third equalities.

\subsection{Description of the Dispersion Diagram}\label{sec:dd_description}

\subsubsection{Brillouin Zone}

The dispersion diagrams corresponding to the exact solutions~\eqref{eq:disp_final_solutions} and~\eqref{eq:disp_final_solutions_sup} are plotted only for first Brillouin zones in Fig.~\ref{fig:disp_diag}, since the rest of the diagrams follows by periodicity. To find the limits of the Brillouin zones, we study equation~\eqref{eq:CF_DD}, and see that this function is periodic in $\Phi_\text{B}-\Delta\varphi=2\pi n$. We therefore define the Brillouin zones as delimited by
\begin{equation}\label{eq:phiB}
\Phi_\text{B}-\Delta\varphi=\pm\pi.
\end{equation}

In the subluminal case, this can be expressed in terms of the frequency and wavenumber by substituting~\eqref{eq:phase_def_bloch} and~\eqref{eq:delta_phi} with~\eqref{eq:equal-omega} into~\eqref{eq:phiB}, yielding
\begin{equation}\label{eq:bz}
k= \frac{v_\text{m}}{v_jv_k}\omega\pm\frac{\pi (v_j v_k-v_\text{m}^2)}{v_jv_k\ell_\text{B}}.
\end{equation}
This is the expression for the two oblique lines delimiting the Brillouin zone. Their slope is $v_jv_k/v_\text{m}$, and the $k$-intercepts are $\pm\pi (v_j v_k-v_\text{m}^2)/(v_jv_k\ell_\text{B})$, reducing, as expected, to $\pm\pi/\ell_\text{B}$ in the stationary case. Note that the limits of the Brillouin zone are not parallel to the $\omega'$ axis, as they would be for the case of a moving medium. In the case of a moving isotropic medium, the crystal would appear stationary in the moving frame and so the Brillouin zone would be parallel to the $\omega'$ axis.

In the superluminal case, substituting~\eqref{eq:phase_def_bloch_sup} and~\eqref{eq:delta_phi_sup} with~\eqref{eq:equal-ke} into~\eqref{eq:phiB} yields
\begin{equation}\label{eq:bz_sup}
\omega= \frac{v_jv_k}{v_\text{m}}k\pm\frac{\pi (v_\text{m}^2-v_j v_k)}{v_\text{m}^2d_\text{B}}.
\end{equation}
with the slope being again $v_jv_k/v_\text{m}$.

\subsubsection{Complex Nature of the Bandgaps}

The bandgaps correspond to the imaginary part of the solution in Fig.~\ref{fig:disp_diag}, with the imaginary frequencies alternating along the $k$ axis and the imaginary wavenumbers alternating along the $\omega$ axis. We see that the frequencies and the wavenumbers are simultaneously complex, for both the subluminal and superluminal regimes. This is a feature that seems to have been previously overlooked in the literature, although it is a straightforward consequence of the Lorentz transformations.

In the subluminal case, the moving-frame solution has the complex wavenumbers $k'=k_\text{r}'+ik_\text{i}'$, with $k_\text{r}'$, $k_\text{i}'$ the real and imaginary parts, and purely real frequencies, so that $\omega_\text{i}'=0$. Applying the frequency Lorentz transformation~\eqref{eq:Lorentz_w_k} to the imaginary frequency yields $\omega_\text{i}-v_\text{m}k_\text{i}=0$, or $\omega_\text{i}=v_\text{m}k_\text{i}$, which shows that the complexity of $k_\text{i}$ indeed implies the complexity of $\omega_\text{i}$, with $\omega_\text{i}$ non negligible for relativistic velocities.

In the superluminal case, the moving-frame solution has the complex wavenumbers $\omega'=\omega_\text{r}'+i\omega_\text{i}'$, with $\omega_\text{r}'$, $\omega_\text{i}'$ the real and imaginary parts, and purely real wavenumbers, so that $k_\text{i}'=0$. Applying the wavenumber Lorentz transformation~\eqref{eq:Lorentz_w_k} to the imaginary wavenumber, with $v_\text{f}=c^2/v_\text{m}$, yields \mbox{$k_i-(1/v_\text{m})\omega_i=0$}, or $k_i=\omega_i/v_\text{m}$, so that the complexity of $\omega_i$ implies the complexity of $k_i$, with $k_i$ decreasing as the modulation velocity increases.


\subsubsection{Bandgap Edges}\label{sec:bandgap_edges}

We now calculate the edges of the bandgaps, indicated as $\text{BG}_{p,q}^\pm$ in Fig.~\ref{fig:disp_diag}, where the subscripts indicate the two modes involved, and the superscript $\pm$ refers to the top and bottom edges. Theses edges are found by analyzing the exact solution~\eqref{eq:CF_DD}, which reduces to
\begin{equation}\label{eq:CF_DD_equal}
\cos(\Phi_\text{B}-\Delta\varphi)=
\cos^2\bar{\varphi}-\frac{1}{2}\left(\frac{\eta_1}{\eta_2}+\frac{\eta_2}{\eta_1}\right)\sin^2\bar{\varphi}
\end{equation}
in the equal-phase case, $\bar{\varphi}_1=\bar{\varphi}_2=\bar{\varphi}$ (Sec.~\ref{sec:equal-phase}).

The bandgaps correspond to complex Bloch-Floquet phases, and therefore to the right-hand side of~\eqref{eq:CF_DD_equal} being less than $-1$. Thus, the bandgap edges occur at the limit
\begin{equation}
\cos^2\bar{\varphi}-\frac{1}{2}\left(\frac{\eta_1}{\eta_2}+\frac{\eta_2}{\eta_1}\right)\sin^2\bar{\varphi}=-1.
\end{equation}
Applying the trigonometric relation $\sin^2(\bar{\varphi})=1-\cos^2(\bar{\varphi})$ and grouping the cosine terms, we find that the bandgap edge condition reduces to
\begin{equation}\label{eq:bandgap_edge_1}
\bar{\varphi}=\arccos\left(\pm\frac{\eta_2-\eta_1}{\eta_1+\eta_2}\right)
=\arccos(\pm\upsilon)\overset{\upsilon\ll1} {\approx} \frac{\pi}{2}\mp\upsilon.
\end{equation}
This condition delimits two curves on which the bandgap edges lie. In addition, the bandgap edges lie on the Brillouin zone limits, as can be seen in Fig.~\ref{fig:disp_diag}. This is found by noting that the inverse cosine of a quantity less than $-1$ has the form $\pi+ix$, and at the limit is $\pm\pi$. Setting the inverse cosine of the right-hand side of~\eqref{eq:CF_DD_equal} equal to $\pm\pi$ retrieves~\eqref{eq:phiB}.

In the subluminal case, the condition for the bandgap edge may be written in terms of frequency and wavenumber by inserting~\eqref{eq:equal-phase} with~\eqref{eq:equal-omega} into~\eqref{eq:bandgap_edge_1} as
\begin{equation}\label{eq:bandgap_edge_1_sol}
\omega=v_\text{m}k+\frac{1}{d_\text{e}}\arccos(\pm\upsilon),
\end{equation}
which represent the equations of the parallel straight curves, with slope $v_\text{m}$ and $\omega$-intercept $1/d_\text{e}\arccos(\pm\upsilon)$, that delimit the bandgap. The Brillouin zone limits are written in terms of frequencies and wavenumbers in~\eqref{eq:bz}. Solving this system of equations provides the bandgap edge positions $\text{BG}_{0,-1}$ as
\begin{equation}\label{eq:bandgap_edge_2}
\omega_{0,-1}^\pm=\frac{1}{\ell_\text{B}}\left(v_\text{m}\pi+(v_j+v_k)\arccos\left(\pm\upsilon\right)\right),
\end{equation}
\begin{equation}\label{eq:bandgap_edge_2}
k_{0,-1}^\pm=\frac{1}{\ell_\text{B}}\left(\pi+\frac{v_\text{m}(v_j+v_k)}{v_jv_k}\arccos\left(\pm\upsilon\right)\right).
\end{equation}

From these results, we may express the bandgap width $\Delta \omega_{0,-1} = \omega_{0,-1}^+-\omega_{0,-1}^-$ as
\begin{equation}
\begin{split}
\Delta \omega_{0,-1} &=\frac{v_j+v_k}{\ell_\text{B}}\left(\arccos\upsilon-\arccos(-\upsilon)\right)\\
&\overset{\upsilon\ll1} {\approx}2\frac{vj+v_k}{\ell_\text{B}}|\upsilon|,
\end{split}
\end{equation}
and the corresponding wavenumber width $\Delta k_{0,-1} = k_{0,-1}^+-k_{0,-1}^-$, simply found as $\Delta k_{0,-1}=\Delta \omega_{0,-1}v_\text{m}/(v_jv_k)$ from the geometrical construction in Fig.~\ref{fig:disp_diag}(a).

Note that the bandgap widths are the same for all the bandgaps in the case of equal-phase crystals, so that the edges of all the bandgaps may be found from the bandgap width expressions and the center positions provided in~\eqref{eq:bandgap_pos}.

In the superluminal case, the bandgap edges lie at the intersection of the straight line given by~\eqref{eq:bandgap_edge_1} and the Brillouin zone limits, the former of which is rewritten in terms of frequency and wavenumber by inserting~\eqref{eq:equal-phase_sup} with~\eqref{eq:equal-ke} into~\eqref{eq:bandgap_edge_1} as
\begin{equation}\label{eq:bandgap_edge_1_sol}
k=\frac{\omega}{v_\text{m}}+\frac{1}{\ell_\text{e}}\arccos(\pm\upsilon),
\end{equation}
and the latter of which corresponds to~\eqref{eq:bz_sup}. Solving for this system of equations, we find the positions of the limits of the gap $\text{BG}_{0,-1}$, denoted by $\text{BG}_{0,-1}^\pm$ in Fig.~\ref{fig:disp_diag}(b), as
\begin{equation}\label{eq:bandgap_edge_2}
k_{0,-1}^\pm=\frac{1}{d_\text{B}}\left(\frac{\pi}{v_\text{m}}+\frac{v_j+v_k}{v_jv_k}\arccos\left(\pm\upsilon\right)\right),
\end{equation}
\begin{equation}\label{eq:bandgap_edge_2}
\omega_{0,-1}^\pm=\frac{1}{d_\text{B}}\left(\pi+\frac{v_j+v_k}{v_\text{m}}\arccos\left(\pm\upsilon\right)\right).
\end{equation}
From these results, we may express the bandgap width $\Delta k_{0,-1} = k_{0,-1}^+-k_{0,-1}^-$ as
\begin{equation}
\begin{split}
\Delta k_{0,-1} &=\frac{v_j+v_k}{v_jv_kd_\text{B}}\left(\arccos\upsilon-\arccos(-\upsilon)\right)\\
&\overset{\upsilon\ll1} {\approx}2\frac{v_j+v_k}{v_jv_kd_\text{B}}|\upsilon| ,
\end{split}
\end{equation}
and the frequency width $\Delta \omega_{0,-1} = \omega_{0,-1}^+-\omega_{0,-1}^-$ is then simply found as $\Delta \omega_{0,-1}=\Delta k_{0,-1}v_jv_k/v_\text{m}$, from the geometrical construction in Fig.~\ref{fig:disp_diag}(b).

\section{Truncated Crystal}\label{sec:truncated}

\subsection{Types of Truncation}

As for conventional crystals, practical spacetime crystals are necessarily finite, and may be truncated along different directions, including now also the time direction.

\begin{figure}[h]
	\centering \includegraphics[width=1\columnwidth]{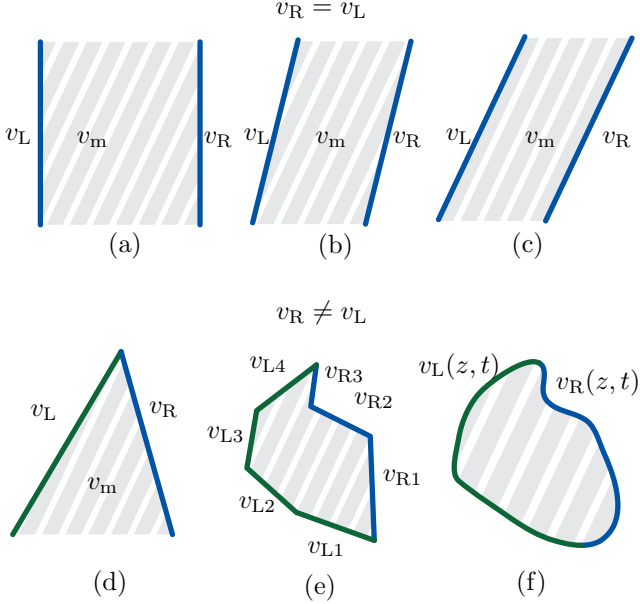}{
        \psfrag{A}[c][c][1]{$v_\text{R}=v_\text{L}$}
		\psfrag{B}[c][c][1]{$v_\text{R}\neq v_\text{L}$}
		\psfrag{a}[c][c][1]{(a)}
		\psfrag{b}[c][c][1]{(b)}
        \psfrag{c}[c][c][1]{(c)}
		\psfrag{d}[c][c][1]{(d)}
        \psfrag{e}[c][c][1]{(e)}
        \psfrag{f}[c][c][1]{(f)}
	   \psfrag{m}[c][c][1]{$v_\text{m}$}
        \psfrag{l}[c][c][1]{$v_\text{L}$}
        \psfrag{r}[c][c][1]{$v_\text{R}$}
		\psfrag{z}[c][c][1]{$z$}
		\psfrag{1}[c][c][1]{$v_{\text{L}1}$}
\psfrag{2}[c][c][1]{$v_{\text{L}2}$}
\psfrag{3}[c][c][1]{$v_{\text{L}3}$}
\psfrag{4}[c][c][1]{$v_{\text{L}4}$}
\psfrag{5}[c][c][1]{$v_{\text{R}1}$}
\psfrag{6}[c][c][1]{$v_{\text{R}2}$}
\psfrag{7}[c][c][1]{$v_{\text{R}3}$}
\psfrag{L}[c][c][1]{$v_\text{L}(z,t)$}
\psfrag{R}[c][c][1]{$v_\text{R}(z,t)$}
	}
	\caption{Examples of spacetime crystal truncation by a pair of spactime interfaces of velocities $v_\text{L}$ and $v_\text{R}$ for the left and right interfaces, respectively. Top row: the two interfaces have the same velocity, $v_\text{L}=v_\text{R}$. Bottom row: the two interfaces have different velocities, $v_\text{L}\neq v_\text{R}$. (a)~Purely spatial truncation, $v_\text{L}=v_\text{R}=0$. (b)~Truncation with velocity different from the modulation velocity, $v_\text{L}=v_\text{R}\neq v_\text{m}$. (c)~Co-moving truncation, $v_\text{L}=v_\text{R}= v_\text{m}$. (d)~Antiparallel truncation, $v_\text{L}\neq v_\text{R}\neq v_\text{m}$. (e)~Spacetime cavity with piecewise constant velocities. (f)~Spacetime cavity with continuously varying velocity.}
	\label{fig:truncation}
\end{figure}

Figure~\ref{fig:truncation} describes examples of spacetime crystal truncation. In the top row (Figs.~\ref{fig:truncation}(a)-(c)), the crystals are truncated on the left and on the right by interfaces sharing the same velocity. In the bottom row (Figs.~\ref{fig:truncation}(d)-(f)) the crystals are truncated on the left and on the right by interfaces of different velocities. The truncations of Fig.~\ref{fig:truncation} represent only a few of an infinite number of possibilities, including for instance multiple truncations separating the crystal in different parts, truncations delimiting holes, or spacetime defects, in the crystal, or periodic truncation leading to a multiscale spacetime crystal.

\begin{figure*}[t]
	\centering \includegraphics[width=1\textwidth]{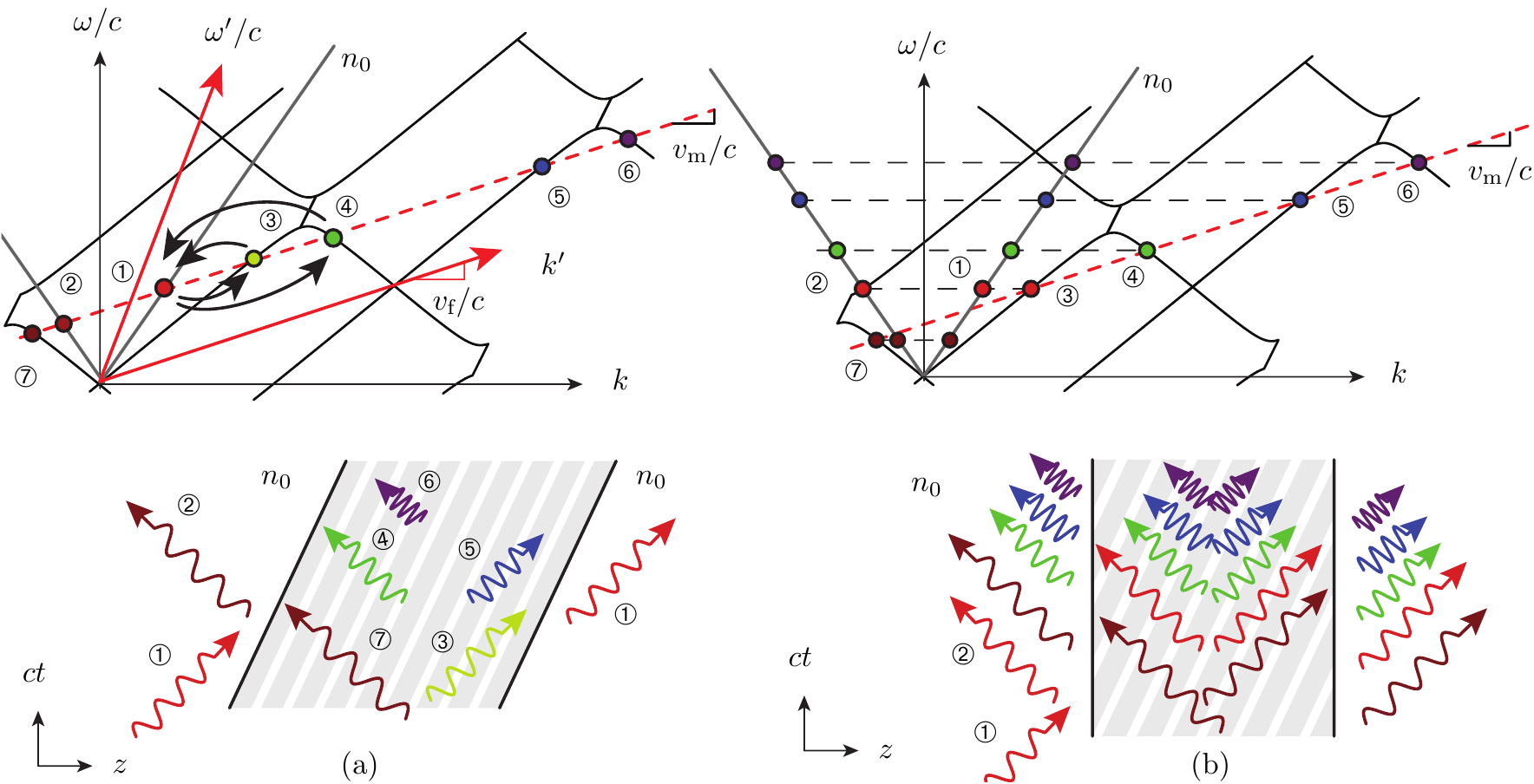}{
		\psfrag{A}[c][c][1]{(a)}
		\psfrag{B}[c][c][1]{(b)}
		\psfrag{a}[c][c][1]{\ding{192}}
        \psfrag{f}[c][c][1]{\ding{193}}
		\psfrag{b}[c][c][1]{\ding{194}}
        \psfrag{c}[c][c][1]{\ding{195}}
        \psfrag{d}[c][c][1]{\ding{196}}
        \psfrag{e}[c][c][1]{\ding{197}}
        \psfrag{g}[c][c][1]{\ding{198}}
		\psfrag{w}[c][c][1]{$\omega/c$}
		\psfrag{k}[c][c][1]{$k$}
		\psfrag{W}[c][c][1]{$\omega'/c$}
		\psfrag{K}[c][c][1]{$k'$}
		\psfrag{z}[c][c][1]{$z$}
        \psfrag{t}[c][c][1]{$ct$}
		\psfrag{1}[c][c][1]{$n_0$}
        \psfrag{m}[c][c][1]{$v_\text{m}/c$}
        \psfrag{E}[c][c][1]{$v_\text{f}/c$}
		\psfrag{G}[c][c][1]{\textcolor{blue}{$\omega_{\text{i}_\text{g}}$}}
	}
	\caption{Scattering from two canonical truncated spacetime crystals. In both cases, the crystal is subluminal, and the medium surrounding it is a simple nondispersive dielectric medium of refractive index $n_0$. Top row: dispersion diagrams with transition frequencies. Bottom row: spacetime diagrams with scattered waves, with labels corresponding the solutions of the top panels. Note that the drawn scattered waves correspond to the waves seen in the laboratory frame, and would completely different in the moving frame. (a)~Co-moving truncation. A moving-frame observer would see a stationary crystal bounded by stationary interfaces, and hence measure a unique frequency everywhere, inside and outside the crystal. The arrows indicate the up and down frequency and wavenumber transitions at the two interfaces. (b)~Purely spatial truncation. A moving-frame observer would see a stationary crystal bounded by moving interfaces, and hence measure an infinity of frequencies.}
	\label{fig:truncation_scat}
\end{figure*}

Figure~\ref{fig:truncation_scat} illustrates the scattering phenomenology for truncated spacetime crystals, with Fig.~\ref{fig:truncation_scat}(a) corresponding to the co-moving structure of Fig.~\ref{fig:truncation}(c) and Fig.~\ref{fig:truncation_scat}(b) corresponding to the purely spatially truncated crystal of Fig.~\ref{fig:truncation}(a). The top panels show the frequency transitions in the dispersion diagrams, while the bottom panels show the different scattered waves in the spacetime diagrams.

Let us start with the co-moving truncated crystal (Fig.~\ref{fig:truncation_scat}(a)). The transition frequencies are obtained trough the following construction. First, we plot the dispersion diagrams of the spacetime crystal (Sec.~\ref{sec:dd_construction}) and of the surrounding medium, which is here a simple pair of straight lines in the present nondispersive case. Second, we place the point corresponding to the incident wave, labeled \ding{192}, on the dispersion diagram of the incident medium. Third, we trace an oblique line of slope $v_\text{m}/c$, corresponding to the moving-frame conserved frequency (see Sec.~\ref{sec:freq_transitions_interface}). The intersections of this line with the two dispersion diagrams provide the scattered frequencies. The corresponding scattered waves are represented in the spacetime diagram below the dispersion diagram.

We notice that, although infinitely many frequencies are excited inside the crystal (solutions labeled \ding{194}-\ding{198}), the transmitted and reflected waves are monochromatic. Indeed, in the moving frame, the interfaces appear stationary, so that a single frequency exists throughout the structure. Therefore the reflected and transmitted solutions in the laboratory frame, found by applying the frequency Lorentz transformation~\eqref{eq:Lorentz_w_k}, are also monochromatic. The incident and transmitted waves both transform in the same way, since they both propagate in the same direction, whereas the reflected wave, propagating in the opposite direction, acquires a different frequency. In the case of Fig.~\ref{fig:truncation_scat}(a) it is downshifted, as expected from the Doppler effect for a receding structure.

We now proceed to the analysis of the purely spatially truncated crystal (Fig.~\ref{fig:truncation_scat}(b)). The transition frequencies are obtained through the following construction. First, we plot again the dispersion diagrams of the spacetime crystal and the surrounding media, along with the incident wave, labeled \ding{192}. Second, we enforce frequency conservation at the first interface by tracing horizontal lines, which leads to the reflected wave labeled \ding{193} and the transmitted wave labeled \ding{194}. The frequency conservation in the moving frame involves then all the harmonic waves along the oblique line of slope $v_\text{m}/c$ (labeled \ding{195}-\ding{198}). Due to frequency conservation at the left and right interfaces, each of these harmonics produces a corresponding wave in the surrounding medium, so that the scattered waves contain multiple harmonics.

\subsection{Scattering Coefficients}

We now apply the transfer matrix technique to solve the problem of the co-moving truncation spacetime crystal (Fig.~\ref{fig:truncation_scat}(a)). The spatially truncated spacetime crystal may be studied using conventional phase matching and Bloch-Floquet techniques~\cite{fante1971transmission,chamanara2017optical}.

The transmission and reflection coefficients of a spacetime crystal truncated by co-moving interfaces of $N$ periods may be obtained by simply multiplying unit-cell transfer matrix~\eqref{eq:crystal_M} $N$ times, and then extracting the coefficients. An alternative approach to the matrix multiplication is the application of the Chebyshev identity~\cite{born1980principles}
\begin{equation}
\begin{split}
[&M_{\text{B}}]^N = e^{iN\Delta\phi}[M_{\text{B}0}] \\
     & =e^{iN\Delta\phi}\begin{bmatrix}
a\mathcal{U}_{N-1}(x)-\mathcal{U}_{N-2}(x) & ib\mathcal{U}_{N-1}(x)\\
ic\mathcal{U}_{N-1}(x) & a^*\mathcal{U}_{N-1}(x)-\mathcal{U}_{N-2}(x)\\
\end{bmatrix},
\end{split}
\end{equation}
where $\mathcal{U}_{N}$ are the Chebyshev polynomials of the second kind, defined as
\begin{equation}
\mathcal{U}_{N}(x)=\frac{\sin[(N+1)\cos^{-1}x]}{\sqrt{1-x^2}}
\end{equation}
with the argument
\begin{equation}
x=\frac{a+a^*}{2},
\end{equation}
where we recall that $a$, $a^*$, $b$, and $c$ are the elements of the unit-cell matrix $M_\text{B}$, given in~\eqref{eq:crystal_M_values}.

To extract the total transmission and reflection coefficients, the scattering matrix is written in terms of transfer-matrix parameters.
\begin{equation}\label{eq:S}
\begin{split}
\begin{bmatrix}
\psi_1^-\\
\psi_N^-
\end{bmatrix} =&
\begin{bmatrix}
\Gamma_{11}&\bar{T}_{1N}\\
T_{N1}&\bar{\Gamma}_{NN}
\end{bmatrix}
\begin{bmatrix}
\psi_1^+\\
\psi_N^+
\end{bmatrix}\\=&
\dfrac{1}{A^*}\begin{bmatrix}
-A&1\\
AA^*-BC&A
\end{bmatrix}
\begin{bmatrix}
\psi_1^+\\
\psi_N^+
\end{bmatrix},
\end{split}
\end{equation}

with $A,A^*,B,C$ corresponding to the $a,a^*,b,c$ elements of the $[M_{\text{B}}]^N$ matrix.

\begin{figure}[]
\includegraphics[width=1\columnwidth]{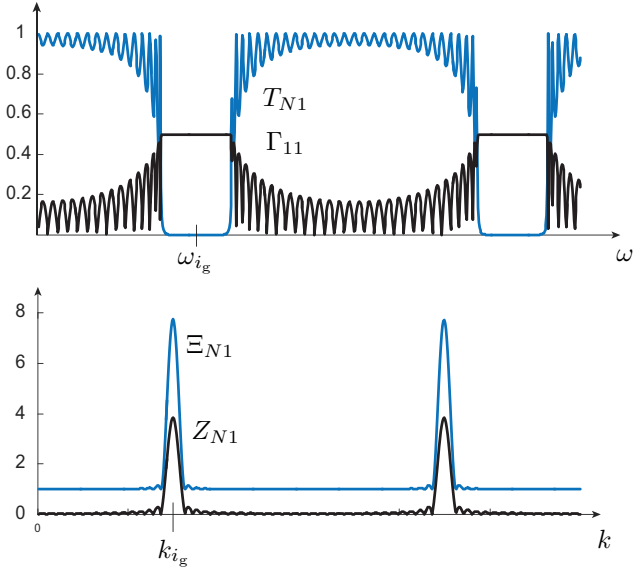}{
\psfrag{t}[c][c]{$T_{N1}$}
\psfrag{g}[c][c]{$\Gamma_{11}$}
\psfrag{G}[c][c]{$Z_{N1}$}
\psfrag{T}[c][c]{$\Xi_{N1}$}
\psfrag{w}[c][c]{$\omega$}
\psfrag{k}[c][c]{$k $}
\psfrag{W}[c][c]{$\omega_{i_\text{g}}$}
\psfrag{K}[c][c]{$k_{i_\text{g}}$}
\psfrag{a}[c][c]{(a)}
\psfrag{b}[c][c]{(b)}
}
\centering
\caption{Transmission and reflection coefficients for an \mbox{$N=15$}-layer crystal (corresponding to Fig.~\ref{fig:truncation}(c) and Fig.~\ref{fig:truncation_scat}(a)) with $\bar{\phi}_{1,2}=\pi/2$. The gap centers $\omega_\text{i}$, $k_\text{i}$ are provided in Supp. Mat~\ref{sec:centers}. (a)~Subluminal case, with $v=1/3c$ and $n_2/n_1=2$, with attenuation in the bandgaps. (b)~Superluminal case, with $v=3c$ and $n_2/n_1=1.2$ with amplification in the bandgaps.}
\label{fig:crystal_amplitude}
\end{figure}

The results are plotted in Fig.~\ref{fig:crystal_amplitude}(a). We notice that in the gap, the reflected wave is partly absorbed by the structure.

For the superluminal crystal, the scattering matrix is the same as the transfer matrix, since waves in the last medium are also the scattered waves. The coefficients are thus
\begin{subequations}\label{eq:tr_sup}
\begin{equation}
{Z_{N1}}=A,
\end{equation}
\begin{equation}
{\Xi_{N1}}=C,
\end{equation}
\end{subequations}
and are plotted in Fig.~\ref{fig:crystal_amplitude}. We see that in the gap, both the transmitted and the reflected wave are amplified. This is consistent with the analysis of the interference for a superluminal slab in Sec.\ref{sec:ST_slab}: both scattered waves are in constructive interference in the bandgap.

We note that the bandgap centers in Fig.~\ref{fig:crystal_amplitude} do \emph{not} correspond to the bandgap centers calculated in~\eqref{eq:bandgap_pos} and~\eqref{eq:bandgap_pos_sup}. This is due to the effect illustrated in Fig.~\ref{fig:truncation_scat}(a): the frequencies horizontally aligned with the bandgaps will not see the bandgap. Instead, the frequencies aligned on an oblique line of slope $v_\text{m}$ and passing through the bandgap will see the bandgap.

\section{Conclusion}\label{sec:concl}

We have performed a comprehensive analysis of uniform-velocity bilayer spacetime crystals. We started with the problem of spacetime interfaces, then addressed the problem of double-interfaces, or spacetime slabs, next solved the problem of unbounded crystals, and finally studied the problem of truncated crystals.

Throughout the paper, we have made an extensive use of special relativity concepts, including the graphical tool of spacetime diagrams, which offered remarkable resolution simplicity, provided deep insight into the physics, and led us to uncover a number of new results. The insights include a vivid Bragg-type description of the bandgap interference phenomenology, a quick linear approximation of dispersion diagrams, and the explanation of the effect of the truncation. The results include the generalization of the Stokes principle to spacetime interfaces and the description of the simultaneous frequency and wavenumber complexity in the bandgaps.

This work may be extended to multi-layer-cell spacetime crystals, spacetime crystals made of dispersive media, and higher dimension crystals.

\clearpage
\appendix

\section{Bianisotropy , Wave Impedance and Wave Velocity in the Moving Frame}\label{sec:fizeau}

\begin{figure}[h]
\centering
\includegraphics[width=1\columnwidth]{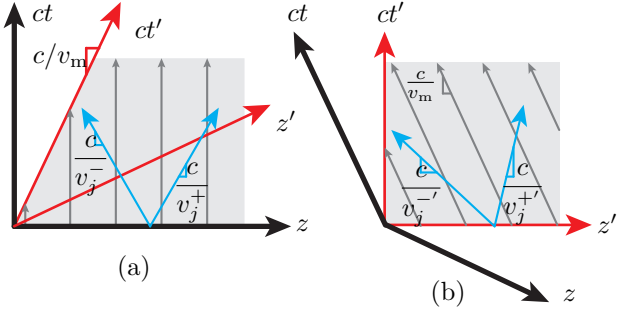}{
\psfrag{a}[c][c]{(a)}
\psfrag{b}[c][c]{(b)}
\psfrag{z}[c][c]{$z$}
\psfrag{t}[c][c]{$ct$}
\psfrag{Z}[c][c]{$z'$}
\psfrag{T}[c][c]{$ct'$}
\psfrag{m}[c][c]{$\frac{c}{v_\text{m}}$}
\psfrag{M}[c][c]{$c/v_\text{m}$}
\psfrag{v}[c][c][1]{$\dfrac{c}{v_j^+}$}
\psfrag{V}[c][c][1]{$\dfrac{c}{v_j^-}$}
\psfrag{p}[c][c][1]{$\dfrac{c}{v_j^{-'}}$}
\psfrag{P}[c][c][1]{$\dfrac{c}{v_j^{+'}}$}
}
\caption{Spacetime interface represented in two inertial frames. The arrows represent the trajectories of the media particles.  In both (a) and (b), the interfaces of the spacetime variation are parallel to the $ct'$ axis and the media trajectories are parallel to the $ct$ axis. (a) Laboratory frame viewpoint. Media appears at rest. wave velocity is independent of direction: $v_j^+=v_j^-$  (b) Moving frame viewpoint. Media appears to be moving in $-z$ direction. Wave velocities are direction dependent: $v_j^+\neq v_j^-$, with $|v_j^-|>|v_j^+|$.}
\label{fig:drag_frames}
\end{figure}

We derive here the constitutive relations for homogeneous isotropic media as seen in a moving frame. Figure~\ref{fig:drag_frames} represents a modulated spacetime interface, from the viewpoint of the laboratory frame in Fig.~\ref{fig:drag_frames}(a) and from the viewpoint of the moving frame in Fig.~\ref{fig:drag_frames}(b). In Fig.~\ref{fig:drag_frames}(a), the interface of the spacetime modulation moves to the right but its constitutive particles have vertical trajectories, i.e. appear stationary. In Fig.~\ref{fig:drag_frames}(b), the interface appears stationary, but the particles appear in motion, moving to the left.

We first write the constitutive relations of the bulk stationary medium in the laboratory frame
\begin{subequations}
\begin{equation}\label{eq:constitutive_D}
D_{xj}=\epsilon_j E_{xj},
\end{equation}
\begin{equation}\label{eq:constitutive_B}
B_{yj}=\mu_j H_{yj}.
\end{equation}
\end{subequations}
Applying the Lorentz transformations~\eqref{eq:Lorentz_fields} to these relations yields
\begin{subequations}
\begin{equation}\label{eq:constitutive_Dp}
D_{xj}'-\frac{v_\text{f}}{c^2}H'_{yj}=\epsilon_j\left(E'_{xj} -v_\text{f}B'_{yj}\right)
\end{equation}
and
\begin{equation}\label{eq:constitutive_Bp}
B'_{yj}-\frac{v_\text{f}}{c^2}E'_{xj}=\mu_j\left(H'_{yj} -v_\text{f}D'_{xj}\right).
\end{equation}
\end{subequations}
Solving~\eqref{eq:constitutive_Dp} and~\eqref{eq:constitutive_Bp} for
$D_y'$ and $B_x'$, we obtain the constitutive relations
\begin{subequations}\label{eq:constitutive}
\begin{equation}\label{eq:constitutive_Dp_simplified}
\begin{split}
D_{xj}'&=\epsilon_j\frac{1-v_\text{f}^2/c^2}{1- v_\text{f}^2/v_j^2}E'_{xj}+\frac{v_\text{f}}{c^2}\frac{1-c^2/v_j^2}{1-v_\text{f}^2/v_j^2}H_{yj}'\\
&=\epsilon_j'E_{xj}'+\chi_j'H_{yj}',
\end{split}
\end{equation}
and
\begin{equation}\label{eq:constitutive_Bp_simplified}
\begin{split}
B_{yj}'&=\mu_j\frac{1-v_\text{f}^2/c^2}{1-v_\text{f}^2/v_j^2}H'_{yj}+\frac{v_\text{f}}{c^2}\frac{1-c^2/v_j^2}{1-v_\text{f}^2/v_j^2}E_{xj}'\\
&=\mu_j'H'_{yj}+\chi_j' E_{xj}'.
\end{split}
\end{equation}
\end{subequations}
where $v_j$ is the velocity of light in the medium in the laboratory frame, $v_j=1/\sqrt{\epsilon_j\mu_j}$. These equations reveal that the medium is biisotropic in the moving frame, since the magneto-electric coupling terms are scalar, and, more specifically, a Tellegen medium, since the there is a unique variable $\chi_j$ for both the electro-magnetic and the magneto-electric coupling terms. More generally, moving volumes appear bianisotropic in the moving frame. Indeed, the media does not transform the same way in the directions parallel and perpendicular to the motion of the medium.

We next calculate the wave impedance and the wave velocity of moving isotropic media. We first write the Maxwell curl equations for a plane monochromatic wave,
\begin{equation}\label{eq:Max_1D}
 \omega_j' D_{xj}' = k_j'H_{yj}', \qquad \omega_j'B_{yj}'= k_j'E_{xj}'.
\end{equation}
Dividing the left equation of~\eqref{eq:Max_1D} by the right equation of~\eqref{eq:Max_1D}, we find
\begin{equation}\label{eq:int_max1D}
\frac{D_{xj}'}{H_{yj}'}=\frac{B_{yj}'}{E_{xj}'}.
\end{equation}

Inserting the constitutive relations~\eqref{eq:constitutive_Bp_simplified} into~\eqref{eq:int_max1D}, we find the impedance $\eta_j'$, defined as the ratio of the $E$ and $H$ fields:
\begin{equation}\label{eq:impedancep}
\eta_j'=\frac{E_{xj}'}{H_{yj}'}=\sqrt\frac{\mu_j'}{\epsilon_j'}=\sqrt\frac{\mu_j}{\epsilon_j}=\eta_j,
\end{equation}
where the third equality comes from the definitions of $\mu'$, $\epsilon'$ in~\eqref{eq:constitutive_Dp_simplified} and~\eqref{eq:constitutive_Bp_simplified}. We thus conclude from~\eqref{eq:impedancep} that the impedance is invariant under a change of frames.

The wave velocity, defined as $v_j^{\pm'}=\omega_j'/k_j'$, is found from~\eqref{eq:Max_1D} as
\begin{equation}\label{eq:vp}
v_j^{\pm'}=\frac{\omega_j'}{k_j'}=\frac{H_{yj}'}{D_{xj}'}=\frac{E_{xj}'}{B_{yj}'}.
\end{equation}
inserting~\eqref{eq:impedancep} and~\eqref{eq:constitutive} into~\eqref{eq:vp} yields
\begin{equation}\label{eq:vpresult_supp}
v_j^{\pm'}=\frac{H_{yj}'}{D_{xj}'}
=\frac{1}{\epsilon_j'\eta_j'+\chi_j'}
=\frac{v_j^\pm + v_\text{f}}{1+v_j^\pm v_\text{f}/c^2}.
\end{equation}
Comparing this result with the addition law of velocities provided in~\eqref{eq:addition_velocities_inv}, we see a sign difference of $v_\text{f}$, due to the fact that for a positive modulation velocity, $v_\text{m}>0$, the medium appears to be moving in the $-z'$ direction in the moving frame, as drawn in Fig.~\ref{fig:drag_frames}(b).
%

\section{Alternative Derivation of the Continuity Conditions}\label{sec:cont_conditions}

We start with one-dimensional Maxwell equations,
\begin{subequations}\label{eq:we}
\begin{equation}\label{eq:we_Ey}
\frac{\D E_x}{\D z}=-\frac{\D B_y}{\D t}=-\frac{\D \mu(z-v_\text{m}t) H_y}{\D t},
\end{equation}

\begin{equation}\label{eq:we_Bx}
\frac{\D H_y}{\D z}=-\frac{\D D_x}{\D t}=-\frac{\D \epsilon(z-v_\text{m}t)E_x}{\D t},
\end{equation}
\end{subequations}
and apply the following coordinate transformation:
\begin{equation}\label{eq:zp}
z'=z-v_\text{m}t, \qquad t'=t.
\end{equation}

This does \emph{not} correspond to a Lorentz transformation, and therefore does not lead to invariance. The partial derivatives of~\eqref{eq:we} are found as
\begin{subequations}\label{eq:dzdt}
\begin{equation}
\frac{\D }{\D z}=\frac{\D}{\D z'}\frac{\D z' }{\D z}+\frac{\D}{\D t'}\frac{\D t' }{\D z}=\frac{\D}{\D z'},
\end{equation}
\begin{equation}
\frac{\D }{\D t}=\frac{\D}{\D t'}\frac{\D t' }{\D t}+\frac{\D}{\D z'}\frac{\D z'}{\D t}=\frac{\D}{\D t'}-v_\text{m}\frac{\D}{\D z'}.
\end{equation}
\end{subequations}

Substituting these relations into \eqref{eq:we} yields
\begin{subequations}\label{eq:inter}
\begin{equation}
\frac{\D E_x}{\D z'} = -\left(\frac{\D}{\D t'}-v_\text{m}\frac{\D}{\D z'} \right) \mu(z')H_y,
\end{equation}
\begin{equation}
\frac{\D H_y}{\D z'}=-\left(\frac{\D }{\D t'}-v_\text{m}\frac{\D}{\D z'}\right)\epsilon(z')E_x.
\end{equation}
\end{subequations}
Rearranging~\eqref{eq:inter}, we find
\begin{subequations}\label{eq:fin}
\begin{equation}\label{eq:Eyp_fin}
\frac{\D }{\D z'}\left(E_x-v_\text{m}\mu(z')H_y\right) = -\frac{\D}{\D t'} \mu(z')H_y,
\end{equation}
\begin{equation}\label{eq:Hxp_fin}
\frac{\D }{\D z'}\left(H_y-v_\text{m}\epsilon(z')E_x\right)=-\frac{\D }{\D t'}\epsilon(z')E_x.
\end{equation}
\end{subequations}

Inspecting these equations, we can deduce the continuity conditions by reasoning ad absurbdum: if the arguments of the $\D z'$ derivative on the left-hand side in~\eqref{eq:fin}, $E_x+v_\text{m}\mu(z')H_y$ or $H_y+v_\text{m}\epsilon(z')E_x$, vary as a step function, then the right-hand side would be an impulse function, which is unphysical. Therefore, quantities $E_x+v_\text{m}\mu(z')H_y$ and $H_y+v_\text{m}\epsilon(z')E_x$, or  $E_x+v_\text{m}B_y$  and $H_y+v_\text{m}D_x$, must be continuous at the interface. We can see that in the limiting case when $v_\text{m}=0$, the quantities $E_x, H_y$ are continuous, as expected. In the limiting case $v_\text{m}=\infty$, the quantities $D_x, B_y$ are continuous, as reported in~\cite{morgenthaler1958}.

\section{Unit-cell Matrix Symmetry}\label{sec:PT}

\begin{figure}
\includegraphics[width=1\columnwidth]{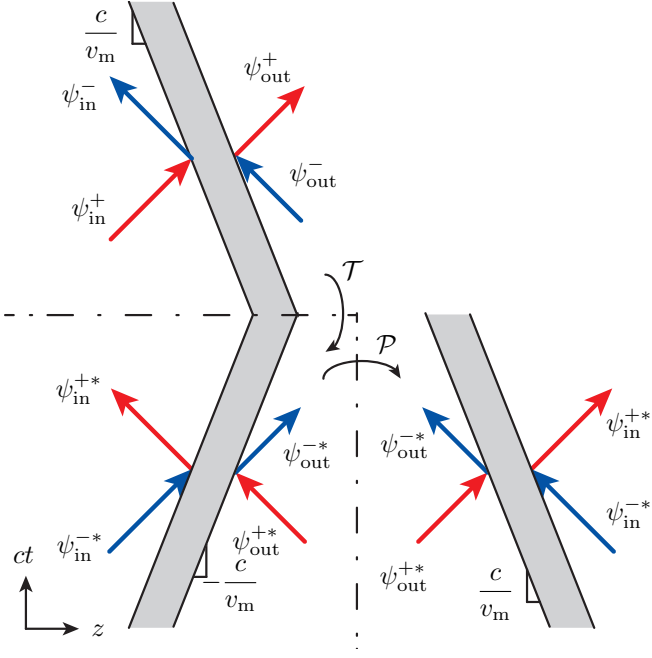}{
\psfrag{y}[c][c][1]{$\dfrac{c}{v_\text{m}}$}
\psfrag{Y}[c][c][1]{$-\dfrac{c}{v_\text{m}}$}
\psfrag{A}[c][c][1]{$\psi_\text{in}^+$}
\psfrag{B}[c][c][1]{$\psi_\text{in}^-$}
\psfrag{C}[c][c][1]{$\psi_\text{out}^+$}
\psfrag{D}[c][c][1]{$\psi_\text{out}^-$}
\psfrag{a}[c][c][1]{$\psi_\text{in}^{+*}$}
\psfrag{b}[c][c][1]{$\psi_\text{in}^{-*}$}
\psfrag{c}[c][c][1]{$\psi_\text{out}^{+*}$}
\psfrag{d}[c][c][1]{$\psi_\text{out}^{-*}$}
\psfrag{1}[c][c][1]{$\psi_\text{out}^{+*}$}
\psfrag{2}[c][c][1]{$\psi_\text{out}^{-*}$}
\psfrag{3}[c][c][1]{$\psi_\text{in}^{+*}$}
\psfrag{4}[c][c][1]{$\psi_\text{in}^{-*}$}
\psfrag{z}[c][c][1]{$z$}
\psfrag{t}[c][c][1]{$ct$}
\psfrag{r}[c][c][1]{$\mathcal{T}$}
\psfrag{p}[c][c][1]{$\mathcal{P}$}}
\centering
\caption{Successive application of time reversal ($\mathcal{T}$) and space reversal ($\mathcal{P}$).}
\label{fig:PT}
\end{figure}

We deduce the properties of the unit-cell transfer matrix,
\begin{equation}\label{eq:conj_matrix_app}
\left[\begin{matrix}
\psi_\text{out}^{+}\\
\psi_\text{out}^{-}
\end{matrix}\right]=[M_\text{B}]\left[\begin{matrix}
\psi_\text{in}^{+}\\
\psi_\text{in}^{-}
\end{matrix}\right],
\end{equation}
by applying a spacetime symmetry argument, with the help of Fig.~\ref{fig:PT}. The initial problem of a spacetime slab is represented in the top left panel of the figure. A time reversal and a space reversal are successively applied to this problem, resulting in the bottom right panel, which retrieves the initial problem. We therefore expect the matrix relating the fields on the left to the fields on the right of the slab to be the same for the space-time reversed problem and for the initial problem. According to the bottom right panel of Fig.~\ref{fig:PT},
\begin{equation}\label{eq:TR_matrix_app}
\left[\begin{matrix}
\psi_\text{in}^{+*}\\
\psi_\text{in}^{-*}
\end{matrix}\right]=[M_\text{B}]\left[\begin{matrix}
\psi_\text{out}^{+*}\\
\psi_\text{out}^{-*}
\end{matrix}\right].
\end{equation}
Taking the complex conjugate of this relation, and multiplying the resulting equation by the inverse of $M_\text{B}^*$ reveals the following property:
\begin{equation}
[M_\text{B}^*]^{-1}=M_\text{B},
\end{equation}
or, in terms of the matrix elements,
\begin{equation}
\begin{bmatrix}
a&b\\
c&d
\end{bmatrix}=
\frac{1}{\det [M]}\begin{bmatrix}
d^*&-b^*\\
-c^*&a^*
\end{bmatrix}.
\end{equation}
Since for spacetime slab, $\det [M_\text{B}]=e^{i\Delta\phi}$ (see Sec.~\ref{sec:unitcell},
\begin{equation}
ae^{i\Delta\phi}=d^*,
\end{equation}
\begin{equation}
be^{i\Delta\phi}=-b^*, \qquad ce^{i\Delta\phi}=-c^*
\end{equation}
which is consistent with~\cite{longhi2010pt} and with \eqref{eq:crystal_M_values} of Sec.~\ref{sec:unitcell}.

\section{Travel Length or Duration across the Unit Cell of the Crystal}\label{sec:lengths}

\begin{figure}[h]
	\centering
	\includegraphics[width=1\columnwidth]{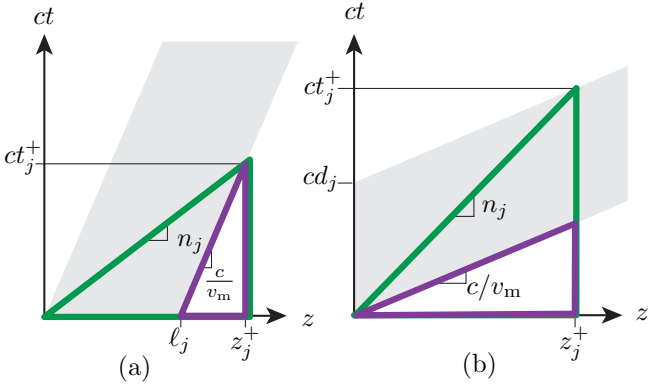}{
		\psfrag{a}[c][c]{(a)}
		\psfrag{b}[c][c]{(b)}
		\psfrag{d}[c][c]{$cd_j$}
		\psfrag{w}[c][c]{$\ell_j$}
		\psfrag{n}[c][c]{$n_j$}
		\psfrag{l}[c][c]{$z_j^+$}
		\psfrag{L}[c][c]{$\lambda_j^+/2$}
		\psfrag{D}[c][c]{$ct_j^+$}
		\psfrag{V}[c][c][1]{$\frac{c}{v_\text{m}}$}
		\psfrag{v}[c][c][1]{$c/v_\text{m}$}
		\psfrag{z}[c][c][1]{$z$}
		\psfrag{t}[c][c][1]{$ct$}
		\psfrag{I}[c][c]{$i$}
		\psfrag{J}[c][c]{$j$}
		\psfrag{K}[c][c]{$k$}
	}
	\caption{Graphical derivation of the travel length or duration across the unit cell of the crystal. (a)~Subluminal regime (length). (b)~Superluminal regime (duration).}
	\label{fig:lengths}
\end{figure}

Consider Fig.~\ref{fig:lengths}. Each graph involves two triangles, one with slope $n_j$ and the other with slope $c/v_\text{m}$.

In the subluminal case, represented in Fig.~\ref{fig:lengths}(a), we have
\begin{equation}
n_j=\frac{ct_j^+}{z_j^+}, \qquad \frac{c}{v_\text{m}}=\frac{ct_j^+}{z_j^+-\ell_j}.
\end{equation}
Isolating $z_j^+$ in both relations,
\begin{equation}
z_j^+=\frac{ct_j^+}{n_j}, \qquad z_j^+=\ell_j+ct_j^+\frac{v_\text{m}}{c},
\end{equation}
and equating these results yields
\begin{equation}
\frac{ct_j^+}{n_j}=\ell_j+ct_j^+\frac{v_\text{m}}{c}, \qquad t_j^+=\frac{\ell_j n_j/c}{1-v_\text{m}n_j/c}.
\end{equation}

Similarly, in superluminal case, represented in Fig.~\ref{fig:lengths}(b), we have
\begin{equation}
n_j=\frac{ct_j^+}{z_j^+}, \qquad \frac{c}{v_\text{m}}=\frac{ct_j^+-cd_j}{z_j^+}.
\end{equation}
Isolating $ct_j^+$ in both relations,
\begin{equation}
ct_j^+=z_j^+ n_j, \qquad ct_j^+=\frac{c}{v_\text{m}}z_j^++cd_j.
\end{equation}
and equating these results yields
\begin{equation}
z_j^+ n_j=\frac{c}{v_\text{m}}z_j^++cd_j, \qquad z_j^+=\frac{cd_j}{n_j- c/v_\text{m}}.
\end{equation}

\section{Frequency Centers for Parallel Truncation}\label{sec:centers}

\begin{figure}[h]
\centering
\includegraphics[width=1\columnwidth]{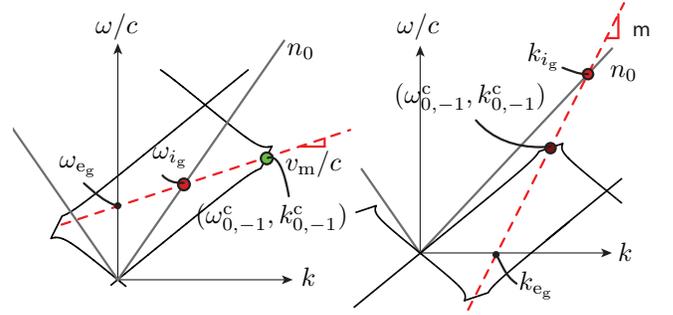}{
\psfrag{a}[c][c]{(a)}
\psfrag{b}[c][c]{(b)}
\psfrag{1}[c][c]{$n_0$}
\psfrag{w}[c][c]{$\omega/c$}
\psfrag{k}[c][c]{$k$}
\psfrag{i}[c][c]{$\omega_{i_\text{g}}$}
\psfrag{I}[c][c]{$k_{i_\text{g}}$}
\psfrag{e}[c][c]{$\omega_{\text{e}_\text{g}}$}
\psfrag{E}[c][c]{$k_{\text{e}_\text{g}}$}
\psfrag{M}[c][c]{$v_\text{m}/c$}
\psfrag{c}[c][c]{$(\omega^\text{c}_{0,-1},k^\text{c}_{0,-1})$}
}
\caption{Construction to find the frequencies aligned with the bandgaps. (a)Subluminal regime. (b)Superluminal regime}
\label{fig:centers}
\end{figure}

Here, we calculate the incident frequencies and wavenumbers aligned with the bandgaps in the case of a co-moving truncation. Figure~\ref{fig:centers} provides the graphical construction, with the dispersion of the incident medium of refractive index $n_0$ and the dispersion diagram of the spacetime crystals.

We start with the subluminal case. We first deduce $\omega_{\text{e}_\text{g}}$, which is found from Fig.~\ref{fig:centers}(a) to satisfy
\begin{subequations}\label{eq:w_bg}
\begin{equation}
\omega_{\text{e}_\text{g}}=\omega^\text{c}_{0,-1}-v_\text{m}k^\text{c}_{0,-1}=(v_\text{av}^+-v_\text{m})\frac{v_\text{m}+v_\text{av}^-}{v_\text{av}^++v_\text{av}^-}k_\text{B},
\end{equation}
where the expressions for the bandgap centers of~\eqref{eq:bandgap_pos} in the paper were used. The incident frequencies aligned with the center of the bandgap, $\omega_{i_\text{g}}$, are expressed as a function of $\omega_{\text{e}_\text{g}}$ as
\begin{equation}\
\omega_{i_\text{g}}
=\frac{v_\text{i}}{v_\text{i}-v_\text{m}}\omega_{\text{e}_\text{g}}.
\end{equation}
\end{subequations}

Similarly, in the superluminal case, $k_{\text{e}_\text{g}}$ is found from Fig.~\ref{fig:centers}(b) as
\begin{subequations}\label{eq:k_bg}
\begin{equation}
k_{\text{e}_\text{g}}
=k^\text{c}_{0,-1}-\frac{\omega^\text{c}_{0,-1}}{v_\text{m}}
=\left(\frac{1}{v_\text{av}^+}-\frac{1}{v_\text{m}}\right)\frac{v_\text{av}^+}{v_\text{m}}
\frac{v_\text{av}^-+v_\text{m}}{v_\text{av}^-+v_\text{av}^+}\omega_\text{B},
\end{equation}
where the bandgap center expressions of~\eqref{eq:bandgap_pos_sup} in the paper were used for the second equality. The incident wavenumbers aligned with the center of the bandgap, $k_{i_\text{g}}$, are expressed as a function of $k_{\text{e}_\text{g}}$ as
\begin{equation}
k_{i_\text{g}}=\frac{v_\text{m}}{v_\text{m}-v_\text{i}}k_{\text{e}_\text{g}}.
\end{equation}
\end{subequations}

\bibliography{ST_crystals}

\begin{thebibliography}{62}%
\makeatletter
\providecommand \@ifxundefined [1]{%
 \@ifx{#1\undefined}
}%
\providecommand \@ifnum [1]{%
 \ifnum #1\expandafter \@firstoftwo
 \else \expandafter \@secondoftwo
 \fi
}%
\providecommand \@ifx [1]{%
 \ifx #1\expandafter \@firstoftwo
 \else \expandafter \@secondoftwo
 \fi
}%
\providecommand \natexlab [1]{#1}%
\providecommand \enquote  [1]{``#1''}%
\providecommand \bibnamefont  [1]{#1}%
\providecommand \bibfnamefont [1]{#1}%
\providecommand \citenamefont [1]{#1}%
\providecommand \href@noop [0]{\@secondoftwo}%
\providecommand \href [0]{\begingroup \@sanitize@url \@href}%
\providecommand \@href[1]{\@@startlink{#1}\@@href}%
\providecommand \@@href[1]{\endgroup#1\@@endlink}%
\providecommand \@sanitize@url [0]{\catcode `\\12\catcode `\$12\catcode
  `\&12\catcode `\#12\catcode `\^12\catcode `\_12\catcode `\%12\relax}%
\providecommand \@@startlink[1]{}%
\providecommand \@@endlink[0]{}%
\providecommand \url  [0]{\begingroup\@sanitize@url \@url }%
\providecommand \@url [1]{\endgroup\@href {#1}{\urlprefix }}%
\providecommand \urlprefix  [0]{URL }%
\providecommand \Eprint [0]{\href }%
\providecommand \doibase [0]{http://dx.doi.org/}%
\providecommand \selectlanguage [0]{\@gobble}%
\providecommand \bibinfo  [0]{\@secondoftwo}%
\providecommand \bibfield  [0]{\@secondoftwo}%
\providecommand \translation [1]{[#1]}%
\providecommand \BibitemOpen [0]{}%
\providecommand \bibitemStop [0]{}%
\providecommand \bibitemNoStop [0]{.\EOS\space}%
\providecommand \EOS [0]{\spacefactor3000\relax}%
\providecommand \BibitemShut  [1]{\csname bibitem#1\endcsname}%
\let\auto@bib@innerbib\@empty
\bibitem [{\citenamefont {Ashcroft}\ and\ \citenamefont
  {Mermin}(1976)}]{ashcroft1976introduction}%
  \BibitemOpen
  \bibfield  {author} {\bibinfo {author} {\bibfnamefont {N.}~\bibnamefont
  {Ashcroft}}\ and\ \bibinfo {author} {\bibfnamefont {N.}~\bibnamefont
  {Mermin}},\ }\href@noop {} {\emph {\bibinfo {title} {Introduction to Solid
  State Physics}}}\ (\bibinfo  {publisher} {Saunders, Philadelphia},\ \bibinfo
  {year} {1976})\BibitemShut {NoStop}%
\bibitem [{\citenamefont {Joannopoulos}\ \emph {et~al.}(2011)\citenamefont
  {Joannopoulos}, \citenamefont {Johnson}, \citenamefont {Winn},\ and\
  \citenamefont {Meade}}]{joannopoulos2011photonic}%
  \BibitemOpen
  \bibfield  {author} {\bibinfo {author} {\bibfnamefont {J.~D.}\ \bibnamefont
  {Joannopoulos}}, \bibinfo {author} {\bibfnamefont {S.~G.}\ \bibnamefont
  {Johnson}}, \bibinfo {author} {\bibfnamefont {J.~N.}\ \bibnamefont {Winn}}, \
  and\ \bibinfo {author} {\bibfnamefont {R.~D.}\ \bibnamefont {Meade}},\
  }\href@noop {} {\emph {\bibinfo {title} {Photonic crystals: {M}olding the
  flow of light}}}\ (\bibinfo  {publisher} {Princeton {U}niversity {P}ress},\
  \bibinfo {year} {2011})\BibitemShut {NoStop}%
\bibitem [{\citenamefont {Saleh}\ and\ \citenamefont
  {Teich}(2007)}]{saleh2007}%
  \BibitemOpen
  \bibfield  {author} {\bibinfo {author} {\bibfnamefont {B.~E.~A.}\
  \bibnamefont {Saleh}}\ and\ \bibinfo {author} {\bibfnamefont {M.~C.}\
  \bibnamefont {Teich}},\ }\href@noop {} {\emph {\bibinfo {title} {Fundamentals
  of Photonics}}}\ (\bibinfo  {publisher} {Wiley},\ \bibinfo {year} {2007})\
  \bibinfo {note} {, {C}hap.~7}\BibitemShut {NoStop}%
\bibitem [{\citenamefont {Bragg}\ and\ \citenamefont
  {Bragg}(1913)}]{bragg1913reflection}%
  \BibitemOpen
  \bibfield  {author} {\bibinfo {author} {\bibfnamefont {W.~H.}\ \bibnamefont
  {Bragg}}\ and\ \bibinfo {author} {\bibfnamefont {W.~L.}\ \bibnamefont
  {Bragg}},\ }\bibfield  {title} {\enquote {\bibinfo {title} {The reflection of
  {X}-rays by crystals},}\ }\href@noop {} {\bibfield  {journal} {\bibinfo
  {journal} {Proc. R. Soc. Lond. A.}\ }\textbf {\bibinfo {volume} {88}},\
  \bibinfo {pages} {428--438} (\bibinfo {year} {1913})}\BibitemShut {NoStop}%
\bibitem [{\citenamefont {Holberg}\ and\ \citenamefont
  {Kunz}(1966)}]{holberg1966parametric}%
  \BibitemOpen
  \bibfield  {author} {\bibinfo {author} {\bibfnamefont {D.}~\bibnamefont
  {Holberg}}\ and\ \bibinfo {author} {\bibfnamefont {K.}~\bibnamefont {Kunz}},\
  }\bibfield  {title} {\enquote {\bibinfo {title} {Parametric properties of
  fields in a slab of time-varying permittivity},}\ }\href@noop {} {\bibfield
  {journal} {\bibinfo  {journal} {IEEE Trans. Antennas. Propag.}\ }\textbf
  {\bibinfo {volume} {14}},\ \bibinfo {pages} {183--194} (\bibinfo {year}
  {1966})}\BibitemShut {NoStop}%
\bibitem [{\citenamefont {Fante}(1971)}]{fante1971transmission}%
  \BibitemOpen
  \bibfield  {author} {\bibinfo {author} {\bibfnamefont {R.}~\bibnamefont
  {Fante}},\ }\bibfield  {title} {\enquote {\bibinfo {title} {Transmission of
  electromagnetic waves into time-varying media},}\ }\href@noop {} {\bibfield
  {journal} {\bibinfo  {journal} {IEEE Trans. Antennas. Propag.}\ }\textbf
  {\bibinfo {volume} {19}},\ \bibinfo {pages} {417--424} (\bibinfo {year}
  {1971})}\BibitemShut {NoStop}%
\bibitem [{\citenamefont {Kamal}(1960)}]{kamal1960parametric}%
  \BibitemOpen
  \bibfield  {author} {\bibinfo {author} {\bibfnamefont {A.}~\bibnamefont
  {Kamal}},\ }\bibfield  {title} {\enquote {\bibinfo {title} {A parametric
  device as a nonreciprocal element},}\ }\href@noop {} {\bibfield  {journal}
  {\bibinfo  {journal} {Proc. IRE}\ }\textbf {\bibinfo {volume} {48}},\
  \bibinfo {pages} {1424--1430} (\bibinfo {year} {1960})}\BibitemShut {NoStop}%
\bibitem [{\citenamefont {Baldwin}(1961)}]{baldwin1961nonreciprocal}%
  \BibitemOpen
  \bibfield  {author} {\bibinfo {author} {\bibfnamefont {L.}~\bibnamefont
  {Baldwin}},\ }\bibfield  {title} {\enquote {\bibinfo {title} {Nonreciprocal
  parametric amplifier circuits},}\ }\href@noop {} {\bibfield  {journal}
  {\bibinfo  {journal} {Proc. IRE}\ }\textbf {\bibinfo {volume} {49}},\
  \bibinfo {pages} {1075} (\bibinfo {year} {1961})}\BibitemShut {NoStop}%
\bibitem [{\citenamefont {Zurita-S\'anchez}\ \emph {et~al.}(2009)\citenamefont
  {Zurita-S\'anchez}, \citenamefont {Halevi},\ and\ \citenamefont
  {Cervantes-Gonz\'alez}}]{halevi2009}%
  \BibitemOpen
  \bibfield  {author} {\bibinfo {author} {\bibfnamefont {J.~R.}\ \bibnamefont
  {Zurita-S\'anchez}}, \bibinfo {author} {\bibfnamefont {P.}~\bibnamefont
  {Halevi}}, \ and\ \bibinfo {author} {\bibfnamefont {J.~C.}\ \bibnamefont
  {Cervantes-Gonz\'alez}},\ }\bibfield  {title} {\enquote {\bibinfo {title}
  {Reflection and transmission of a wave incident on a slab with a
  time-periodic dielectric function $\epsilon(t)$},}\ }\href@noop {} {\bibfield
   {journal} {\bibinfo  {journal} {Phys. Rev. A}\ }\textbf {\bibinfo {volume}
  {79}},\ \bibinfo {pages} {053821} (\bibinfo {year} {2009})}\BibitemShut
  {NoStop}%
\bibitem [{\citenamefont {Vila}\ \emph {et~al.}(2017)\citenamefont {Vila},
  \citenamefont {Pal}, \citenamefont {Ruzzene},\ and\ \citenamefont
  {Trainiti}}]{vila2017bloch}%
  \BibitemOpen
  \bibfield  {author} {\bibinfo {author} {\bibfnamefont {J.}~\bibnamefont
  {Vila}}, \bibinfo {author} {\bibfnamefont {R.~K.}\ \bibnamefont {Pal}},
  \bibinfo {author} {\bibfnamefont {M.}~\bibnamefont {Ruzzene}}, \ and\
  \bibinfo {author} {\bibfnamefont {G.}~\bibnamefont {Trainiti}},\ }\bibfield
  {title} {\enquote {\bibinfo {title} {A {B}loch-based procedure for dispersion
  analysis of lattices with periodic time-varying properties},}\ }\href@noop {}
  {\bibfield  {journal} {\bibinfo  {journal} {J. Sound Vib.}\ }\textbf
  {\bibinfo {volume} {406}},\ \bibinfo {pages} {363--377} (\bibinfo {year}
  {2017})}\BibitemShut {NoStop}%
\bibitem [{\citenamefont {Koutserimpas}\ and\ \citenamefont
  {Fleury}(2018)}]{koutserimpas2018electromagnetic}%
  \BibitemOpen
  \bibfield  {author} {\bibinfo {author} {\bibfnamefont {T.}~\bibnamefont
  {Koutserimpas}}\ and\ \bibinfo {author} {\bibfnamefont {R.}~\bibnamefont
  {Fleury}},\ }\bibfield  {title} {\enquote {\bibinfo {title} {Electromagnetic
  waves in a time periodic medium with step-varying refractive index},}\
  }\href@noop {} {\bibfield  {journal} {\bibinfo  {journal} {IEEE Trans.
  Antennas. Propag.}\ }\textbf {\bibinfo {volume} {66}},\ \bibinfo {pages}
  {5300--5307} (\bibinfo {year} {2018})}\BibitemShut {NoStop}%
\bibitem [{\citenamefont {Kalluri}(2010)}]{kalluri2010electromagnetics}%
  \BibitemOpen
  \bibfield  {author} {\bibinfo {author} {\bibfnamefont {D.~K.}\ \bibnamefont
  {Kalluri}},\ }\href@noop {} {\emph {\bibinfo {title} {Electromagnetics of
  Time Varying Complex Media: Frequency and Polarization Transformer}}}\
  (\bibinfo  {publisher} {CRC Press},\ \bibinfo {year} {2010})\BibitemShut
  {NoStop}%
\bibitem [{\citenamefont {Shlivinski}\ and\ \citenamefont
  {Hadad}(2018)}]{shlivinski2018beyond}%
  \BibitemOpen
  \bibfield  {author} {\bibinfo {author} {\bibfnamefont {Amir}\ \bibnamefont
  {Shlivinski}}\ and\ \bibinfo {author} {\bibfnamefont {Yakir}\ \bibnamefont
  {Hadad}},\ }\bibfield  {title} {\enquote {\bibinfo {title} {Beyond the
  bode-fano bound: Wideband impedance matching for short pulses using temporal
  switching of transmission-line parameters},}\ }\href@noop {} {\bibfield
  {journal} {\bibinfo  {journal} {Phys. Rev. Lett.}\ }\textbf {\bibinfo
  {volume} {121}},\ \bibinfo {pages} {204301} (\bibinfo {year}
  {2018})}\BibitemShut {NoStop}%
\bibitem [{\citenamefont {Mirmoosa}\ \emph {et~al.}(2019)\citenamefont
  {Mirmoosa}, \citenamefont {Ptitcyn}, \citenamefont {Asadchy},\ and\
  \citenamefont {Tretyakov}}]{mirmoosa2019time}%
  \BibitemOpen
  \bibfield  {author} {\bibinfo {author} {\bibfnamefont {M.}~\bibnamefont
  {Mirmoosa}}, \bibinfo {author} {\bibfnamefont {G.}~\bibnamefont {Ptitcyn}},
  \bibinfo {author} {\bibfnamefont {V.}~\bibnamefont {Asadchy}}, \ and\
  \bibinfo {author} {\bibfnamefont {S.}~\bibnamefont {Tretyakov}},\ }\bibfield
  {title} {\enquote {\bibinfo {title} {Time-varying reactive elements for
  extreme accumulation of electromagnetic energy},}\ }\href@noop {} {\bibfield
  {journal} {\bibinfo  {journal} {Phys. Rev. Applied}\ }\textbf {\bibinfo
  {volume} {11}},\ \bibinfo {pages} {014024} (\bibinfo {year}
  {2019})}\BibitemShut {NoStop}%
\bibitem [{\citenamefont {Shapere}\ and\ \citenamefont
  {Wilczek}(2012)}]{wilczek2012classical}%
  \BibitemOpen
  \bibfield  {author} {\bibinfo {author} {\bibfnamefont {A.}~\bibnamefont
  {Shapere}}\ and\ \bibinfo {author} {\bibfnamefont {F.}~\bibnamefont
  {Wilczek}},\ }\bibfield  {title} {\enquote {\bibinfo {title} {Classical time
  crystals},}\ }\href@noop {} {\bibfield  {journal} {\bibinfo  {journal} {Phys.
  Rev. Let.}\ }\textbf {\bibinfo {volume} {109}},\ \bibinfo {pages} {160402}
  (\bibinfo {year} {2012})}\BibitemShut {NoStop}%
\bibitem [{\citenamefont {Skorobogatiy}\ and\ \citenamefont
  {Joannopoulos}(2000)}]{skorobogatiy2000rigid}%
  \BibitemOpen
  \bibfield  {author} {\bibinfo {author} {\bibfnamefont {M.}~\bibnamefont
  {Skorobogatiy}}\ and\ \bibinfo {author} {\bibfnamefont {J.~D.}\ \bibnamefont
  {Joannopoulos}},\ }\bibfield  {title} {\enquote {\bibinfo {title} {Rigid
  vibrations of a photonic crystal and induced interband transitions},}\
  }\href@noop {} {\bibfield  {journal} {\bibinfo  {journal} {Phys. Rev. B}\
  }\textbf {\bibinfo {volume} {61}},\ \bibinfo {pages} {5293} (\bibinfo {year}
  {2000})}\BibitemShut {NoStop}%
\bibitem [{\citenamefont {Wang}\ \emph {et~al.}(2013)\citenamefont {Wang},
  \citenamefont {Zhou}, \citenamefont {Guo}, \citenamefont {Zhang},
  \citenamefont {Evers},\ and\ \citenamefont {Zhu}}]{wang2013optical}%
  \BibitemOpen
  \bibfield  {author} {\bibinfo {author} {\bibfnamefont {Da-Wei}\ \bibnamefont
  {Wang}}, \bibinfo {author} {\bibfnamefont {Hai-Tao}\ \bibnamefont {Zhou}},
  \bibinfo {author} {\bibfnamefont {Miao-Jun}\ \bibnamefont {Guo}}, \bibinfo
  {author} {\bibfnamefont {Jun-Xiang}\ \bibnamefont {Zhang}}, \bibinfo {author}
  {\bibfnamefont {J{\"o}rg}\ \bibnamefont {Evers}}, \ and\ \bibinfo {author}
  {\bibfnamefont {Shi-Yao}\ \bibnamefont {Zhu}},\ }\bibfield  {title} {\enquote
  {\bibinfo {title} {Optical diode made from a moving photonic crystal},}\
  }\href@noop {} {\bibfield  {journal} {\bibinfo  {journal} {Phys. Rev. Lett.}\
  }\textbf {\bibinfo {volume} {110}},\ \bibinfo {pages} {093901} (\bibinfo
  {year} {2013})}\BibitemShut {NoStop}%
\bibitem [{\citenamefont {Doppler}(1842)}]{doppler1842}%
  \BibitemOpen
  \bibfield  {author} {\bibinfo {author} {\bibfnamefont {C.}~\bibnamefont
  {Doppler}},\ }\href@noop {} {\emph {\bibinfo {title} {{\"U}ber das farbige
  {L}icht der {D}oppelsterne und einiger anderer {G}estirne des {H}immels}}}\
  (\bibinfo  {publisher} {Calve},\ \bibinfo {year} {1842})\BibitemShut
  {NoStop}%
\bibitem [{\citenamefont {Ramasastry}(1967)}]{ramasastry1967wave}%
  \BibitemOpen
  \bibfield  {author} {\bibinfo {author} {\bibfnamefont {J}~\bibnamefont
  {Ramasastry}},\ }\bibfield  {title} {\enquote {\bibinfo {title} {Wave
  interaction with moving boundaries},}\ }\href@noop {} {\bibfield  {journal}
  {\bibinfo  {journal} {Electron. Lett}\ }\textbf {\bibinfo {volume} {11}},\
  \bibinfo {pages} {479--481} (\bibinfo {year} {1967})}\BibitemShut {NoStop}%
\bibitem [{\citenamefont {Tsai}\ and\ \citenamefont
  {Auld}(1967)}]{tsai1967wave}%
  \BibitemOpen
  \bibfield  {author} {\bibinfo {author} {\bibfnamefont {C.}~\bibnamefont
  {Tsai}}\ and\ \bibinfo {author} {\bibfnamefont {B.}~\bibnamefont {Auld}},\
  }\bibfield  {title} {\enquote {\bibinfo {title} {Wave interactions with
  moving boundaries},}\ }\href@noop {} {\bibfield  {journal} {\bibinfo
  {journal} {J. Appl. Phys.}\ }\textbf {\bibinfo {volume} {38}},\ \bibinfo
  {pages} {2106--2115} (\bibinfo {year} {1967})}\BibitemShut {NoStop}%
\bibitem [{\citenamefont {Bradley}(1729)}]{Bradley_1729}%
  \BibitemOpen
  \bibfield  {author} {\bibinfo {author} {\bibfnamefont {J.}~\bibnamefont
  {Bradley}},\ }\bibfield  {title} {\enquote {\bibinfo {title} {{IV. A letter
  from the Reverend Mr. James Bradley Savilian Professor of Astronomy at
  Oxford, and FRS to Dr. Edmond Halley Astronom. Reg. \&c. giving an account of
  a new discovered motion of the fix'd stars}},}\ }\href@noop {} {\bibfield
  {journal} {\bibinfo  {journal} {Philos. Trans. Royal Soc.}\ }\textbf
  {\bibinfo {volume} {35}},\ \bibinfo {pages} {637--661} (\bibinfo {year}
  {1729})}\BibitemShut {NoStop}%
\bibitem [{\citenamefont {Caloz}\ and\ \citenamefont
  {Deck-L{\'e}ger}(2019)}]{caloz2019spacetime}%
  \BibitemOpen
  \bibfield  {author} {\bibinfo {author} {\bibfnamefont {C.}~\bibnamefont
  {Caloz}}\ and\ \bibinfo {author} {\bibfnamefont {Z.-L.}\ \bibnamefont
  {Deck-L{\'e}ger}},\ }\bibfield  {title} {\enquote {\bibinfo {title}
  {Spacetime metamaterials},}\ }\href@noop {} {\bibfield  {journal} {\bibinfo
  {journal} {arXiv:1905.00560}\ } (\bibinfo {year} {2019})}\BibitemShut
  {NoStop}%
\bibitem [{\citenamefont {Lurie}\ and\ \citenamefont
  {Weekes}(2006)}]{lurie2006wave}%
  \BibitemOpen
  \bibfield  {author} {\bibinfo {author} {\bibfnamefont {K.}~\bibnamefont
  {Lurie}}\ and\ \bibinfo {author} {\bibfnamefont {S.}~\bibnamefont {Weekes}},\
  }\bibfield  {title} {\enquote {\bibinfo {title} {Wave propagation and energy
  exchange in a spatio-temporal material composite with rectangular
  microstructure},}\ }\href@noop {} {\bibfield  {journal} {\bibinfo  {journal}
  {J. Math. Anal. Appl.}\ }\textbf {\bibinfo {volume} {314}},\ \bibinfo {pages}
  {286--310} (\bibinfo {year} {2006})}\BibitemShut {NoStop}%
\bibitem [{\citenamefont {Lurie}\ \emph {et~al.}(2009)\citenamefont {Lurie},
  \citenamefont {Onofrei},\ and\ \citenamefont
  {Weekes}}]{lurie2009mathematicali}%
  \BibitemOpen
  \bibfield  {author} {\bibinfo {author} {\bibfnamefont {K.}~\bibnamefont
  {Lurie}}, \bibinfo {author} {\bibfnamefont {D.}~\bibnamefont {Onofrei}}, \
  and\ \bibinfo {author} {\bibfnamefont {S.}~\bibnamefont {Weekes}},\
  }\bibfield  {title} {\enquote {\bibinfo {title} {Mathematical analysis of the
  waves propagation through a rectangular material structure in space--time},}\
  }\href@noop {} {\bibfield  {journal} {\bibinfo  {journal} {J. Math. Anal.
  Appl.}\ }\textbf {\bibinfo {volume} {355}},\ \bibinfo {pages} {180--194}
  (\bibinfo {year} {2009})}\BibitemShut {NoStop}%
\bibitem [{\citenamefont {Milton}\ and\ \citenamefont
  {Mattei}(2017)}]{milton2017field}%
  \BibitemOpen
  \bibfield  {author} {\bibinfo {author} {\bibfnamefont {G.}~\bibnamefont
  {Milton}}\ and\ \bibinfo {author} {\bibfnamefont {O.}~\bibnamefont
  {Mattei}},\ }\bibfield  {title} {\enquote {\bibinfo {title} {Field patterns:
  a new mathematical object},}\ }\href@noop {} {\bibfield  {journal} {\bibinfo
  {journal} {Proc. R. Soc. A}\ }\textbf {\bibinfo {volume} {473}},\ \bibinfo
  {pages} {20160819} (\bibinfo {year} {2017})}\BibitemShut {NoStop}%
\bibitem [{\citenamefont {Mattei}\ and\ \citenamefont
  {Milton}(2017)}]{mattei2017field}%
  \BibitemOpen
  \bibfield  {author} {\bibinfo {author} {\bibfnamefont {O.}~\bibnamefont
  {Mattei}}\ and\ \bibinfo {author} {\bibfnamefont {G.}~\bibnamefont
  {Milton}},\ }\bibfield  {title} {\enquote {\bibinfo {title} {Field patterns
  without blow up},}\ }\href@noop {} {\bibfield  {journal} {\bibinfo  {journal}
  {New Journal of Physics}\ }\textbf {\bibinfo {volume} {19}},\ \bibinfo
  {pages} {093022} (\bibinfo {year} {2017})}\BibitemShut {NoStop}%
\bibitem [{\citenamefont {Mattei}\ and\ \citenamefont
  {Milton}(2018)}]{mattei2018field2}%
  \BibitemOpen
  \bibfield  {author} {\bibinfo {author} {\bibfnamefont {O.}~\bibnamefont
  {Mattei}}\ and\ \bibinfo {author} {\bibfnamefont {G.}~\bibnamefont
  {Milton}},\ }\bibfield  {title} {\enquote {\bibinfo {title} {Field patterns:
  A new type of wave with infinitely degenerate band structure},}\ }\href@noop
  {} {\bibfield  {journal} {\bibinfo  {journal} {EPL}\ }\textbf {\bibinfo
  {volume} {120}},\ \bibinfo {pages} {54003} (\bibinfo {year}
  {2018})}\BibitemShut {NoStop}%
\bibitem [{\citenamefont {Pierce}(1959)}]{pierce1959use}%
  \BibitemOpen
  \bibfield  {author} {\bibinfo {author} {\bibfnamefont {J.~R.}\ \bibnamefont
  {Pierce}},\ }\bibfield  {title} {\enquote {\bibinfo {title} {Use of the
  principles of conservation of energy and momentum in connection with the
  operation of wave-type parametric amplifiers},}\ }\href@noop {} {\bibfield
  {journal} {\bibinfo  {journal} {J. Appl. Phys.}\ }\textbf {\bibinfo {volume}
  {30}},\ \bibinfo {pages} {1341--1346} (\bibinfo {year} {1959})}\BibitemShut
  {NoStop}%
\bibitem [{\citenamefont {Tien}(1958)}]{tien1958parametric}%
  \BibitemOpen
  \bibfield  {author} {\bibinfo {author} {\bibfnamefont {P.~K.}\ \bibnamefont
  {Tien}},\ }\bibfield  {title} {\enquote {\bibinfo {title} {Parametric
  amplification and frequency mixing in propagating circuits},}\ }\href@noop {}
  {\bibfield  {journal} {\bibinfo  {journal} {J. Appl. Phys.}\ }\textbf
  {\bibinfo {volume} {29}},\ \bibinfo {pages} {1347--1357} (\bibinfo {year}
  {1958})}\BibitemShut {NoStop}%
\bibitem [{\citenamefont {Cullen}(1958)}]{cullen1958travelling}%
  \BibitemOpen
  \bibfield  {author} {\bibinfo {author} {\bibfnamefont {A.}~\bibnamefont
  {Cullen}},\ }\bibfield  {title} {\enquote {\bibinfo {title} {A
  travelling-wave parametric amplifier},}\ }\href@noop {} {\bibfield  {journal}
  {\bibinfo  {journal} {Nature}\ }\textbf {\bibinfo {volume} {181}},\ \bibinfo
  {pages} {332--332} (\bibinfo {year} {1958})}\BibitemShut {NoStop}%
\bibitem [{\citenamefont {Yu}\ and\ \citenamefont
  {Fan}(2009)}]{fan2009isolation}%
  \BibitemOpen
  \bibfield  {author} {\bibinfo {author} {\bibfnamefont {Z.}~\bibnamefont
  {Yu}}\ and\ \bibinfo {author} {\bibfnamefont {S.}~\bibnamefont {Fan}},\
  }\bibfield  {title} {\enquote {\bibinfo {title} {Complete optical isolation
  created by indirect interband photonic transitions},}\ }\href@noop {}
  {\bibfield  {journal} {\bibinfo  {journal} {Nat. Photon.}\ }\textbf {\bibinfo
  {volume} {3}},\ \bibinfo {pages} {91--94} (\bibinfo {year}
  {2009})}\BibitemShut {NoStop}%
\bibitem [{\citenamefont {Bahabad}\ \emph {et~al.}(2010)\citenamefont
  {Bahabad}, \citenamefont {Murnane},\ and\ \citenamefont
  {Kapteyn}}]{bahabad2010quasi}%
  \BibitemOpen
  \bibfield  {author} {\bibinfo {author} {\bibfnamefont {A.}~\bibnamefont
  {Bahabad}}, \bibinfo {author} {\bibfnamefont {M.~M.}\ \bibnamefont
  {Murnane}}, \ and\ \bibinfo {author} {\bibfnamefont {H.~C.}\ \bibnamefont
  {Kapteyn}},\ }\bibfield  {title} {\enquote {\bibinfo {title}
  {Quasi-phase-matching of momentum and energy in nonlinear optical
  processes},}\ }\href@noop {} {\bibfield  {journal} {\bibinfo  {journal} {Nat.
  Photonics}\ }\textbf {\bibinfo {volume} {4}},\ \bibinfo {pages} {570}
  (\bibinfo {year} {2010})}\BibitemShut {NoStop}%
\bibitem [{\citenamefont {Taravati}\ and\ \citenamefont
  {Caloz}(2017)}]{taravati2017nrmodulated}%
  \BibitemOpen
  \bibfield  {author} {\bibinfo {author} {\bibfnamefont {S.}~\bibnamefont
  {Taravati}}\ and\ \bibinfo {author} {\bibfnamefont {C.}~\bibnamefont
  {Caloz}},\ }\bibfield  {title} {\enquote {\bibinfo {title} {Nonreciprocal
  electromagnetic scattering from a periodically space-time modulated slab and
  application to a quasisonic isolator},}\ }\href@noop {} {\bibfield  {journal}
  {\bibinfo  {journal} {Phys. Rev. B}\ }\textbf {\bibinfo {volume} {65}},\
  \bibinfo {pages} {442--452} (\bibinfo {year} {2017})}\BibitemShut {NoStop}%
\bibitem [{\citenamefont {Chamanara}\ \emph {et~al.}(2017)\citenamefont
  {Chamanara}, \citenamefont {Taravati}, \citenamefont {Deck-L{\'e}ger},\ and\
  \citenamefont {Caloz}}]{chamanara2017optical}%
  \BibitemOpen
  \bibfield  {author} {\bibinfo {author} {\bibfnamefont {N.}~\bibnamefont
  {Chamanara}}, \bibinfo {author} {\bibfnamefont {S.}~\bibnamefont {Taravati}},
  \bibinfo {author} {\bibfnamefont {Z.-L.}\ \bibnamefont {Deck-L{\'e}ger}}, \
  and\ \bibinfo {author} {\bibfnamefont {C.}~\bibnamefont {Caloz}},\ }\bibfield
   {title} {\enquote {\bibinfo {title} {Optical isolation based on space-time
  engineered asymmetric photonic band gaps},}\ }\href@noop {} {\bibfield
  {journal} {\bibinfo  {journal} {Phys. Rev. B}\ }\textbf {\bibinfo {volume}
  {96}},\ \bibinfo {pages} {155409} (\bibinfo {year} {2017})}\BibitemShut
  {NoStop}%
\bibitem [{\citenamefont {Caloz}\ \emph {et~al.}(2018)\citenamefont {Caloz},
  \citenamefont {Al{\`u}}, \citenamefont {Tretyakov}, \citenamefont {Sounas},
  \citenamefont {Achouri},\ and\ \citenamefont
  {Deck-L{\'e}ger}}]{caloz2018nonreciprocity}%
  \BibitemOpen
  \bibfield  {author} {\bibinfo {author} {\bibfnamefont {C.}~\bibnamefont
  {Caloz}}, \bibinfo {author} {\bibfnamefont {A.}~\bibnamefont {Al{\`u}}},
  \bibinfo {author} {\bibfnamefont {S.}~\bibnamefont {Tretyakov}}, \bibinfo
  {author} {\bibfnamefont {D.}~\bibnamefont {Sounas}}, \bibinfo {author}
  {\bibfnamefont {K.}~\bibnamefont {Achouri}}, \ and\ \bibinfo {author}
  {\bibfnamefont {Z.-L.}\ \bibnamefont {Deck-L{\'e}ger}},\ }\bibfield  {title}
  {\enquote {\bibinfo {title} {Electromagnetic nonreciprocity},}\ }\href@noop
  {} {\bibfield  {journal} {\bibinfo  {journal} {Phys. Rev. Appl.}\ }\textbf
  {\bibinfo {volume} {10}},\ \bibinfo {pages} {047001} (\bibinfo {year}
  {2018})}\BibitemShut {NoStop}%
\bibitem [{\citenamefont {Caloz}\ and\ \citenamefont
  {Al{\`u}}(2018)}]{caloz2018editorial}%
  \BibitemOpen
  \bibfield  {author} {\bibinfo {author} {\bibfnamefont {C.}~\bibnamefont
  {Caloz}}\ and\ \bibinfo {author} {\bibfnamefont {A.}~\bibnamefont
  {Al{\`u}}},\ }\bibfield  {title} {\enquote {\bibinfo {title} {Guest editorial
  special cluster on magnetless nonreciprocity in electromagnetics},}\
  }\href@noop {} {\bibfield  {journal} {\bibinfo  {journal} {IEEE Antennas
  Wirel. Propag. Lett.}\ }\textbf {\bibinfo {volume} {17}},\ \bibinfo {pages}
  {1931--1937} (\bibinfo {year} {2018})}\BibitemShut {NoStop}%
\bibitem [{\citenamefont {Cassedy}\ and\ \citenamefont
  {Oliner}(1963)}]{cassedy1963dispersion}%
  \BibitemOpen
  \bibfield  {author} {\bibinfo {author} {\bibfnamefont {E.~S.}\ \bibnamefont
  {Cassedy}}\ and\ \bibinfo {author} {\bibfnamefont {A.~A.}\ \bibnamefont
  {Oliner}},\ }\bibfield  {title} {\enquote {\bibinfo {title} {Dispersion
  relations in time-space periodic media: {P}art {I}-{S}table interactions},}\
  }\href@noop {} {\bibfield  {journal} {\bibinfo  {journal} {Proc. IEEE}\
  }\textbf {\bibinfo {volume} {51}},\ \bibinfo {pages} {1342--1359} (\bibinfo
  {year} {1963})}\BibitemShut {NoStop}%
\bibitem [{\citenamefont {Cassedy}(1967)}]{cassedy1967}%
  \BibitemOpen
  \bibfield  {author} {\bibinfo {author} {\bibfnamefont {E.~S.}\ \bibnamefont
  {Cassedy}},\ }\bibfield  {title} {\enquote {\bibinfo {title} {Dispersion
  relations in time-space periodic media: {P}art {II}-{U}nstable
  interactions},}\ }\href@noop {} {\bibfield  {journal} {\bibinfo  {journal}
  {Proc. IEEE}\ }\textbf {\bibinfo {volume} {55}},\ \bibinfo {pages}
  {1154--1168} (\bibinfo {year} {1967})}\BibitemShut {NoStop}%
\bibitem [{\citenamefont {Biancalana}\ \emph {et~al.}(2007)\citenamefont
  {Biancalana}, \citenamefont {Amann}, \citenamefont {Uskov},\ and\
  \citenamefont {O’Reilly}}]{biancalana2007dynamics}%
  \BibitemOpen
  \bibfield  {author} {\bibinfo {author} {\bibfnamefont {F.}~\bibnamefont
  {Biancalana}}, \bibinfo {author} {\bibfnamefont {A.}~\bibnamefont {Amann}},
  \bibinfo {author} {\bibfnamefont {A.~V.}\ \bibnamefont {Uskov}}, \ and\
  \bibinfo {author} {\bibfnamefont {E.~P.}\ \bibnamefont {O’Reilly}},\
  }\bibfield  {title} {\enquote {\bibinfo {title} {Dynamics of light
  propagation in spatiotemporal dielectric structures},}\ }\href@noop {}
  {\bibfield  {journal} {\bibinfo  {journal} {Phys. Rev. E}\ }\textbf {\bibinfo
  {volume} {75}},\ \bibinfo {pages} {046607} (\bibinfo {year}
  {2007})}\BibitemShut {NoStop}%
\bibitem [{\citenamefont {Jackson}(1999)}]{jackson1999classical}%
  \BibitemOpen
  \bibfield  {author} {\bibinfo {author} {\bibfnamefont {J.~D.}\ \bibnamefont
  {Jackson}},\ }\href@noop {} {\emph {\bibinfo {title} {Classical
  electrodynamics}}}\ (\bibinfo  {publisher} {Wiley},\ \bibinfo {year}
  {1999})\BibitemShut {NoStop}%
\bibitem [{\citenamefont {Ostrovski{\u\i}}(1975)}]{ostrovskii1975}%
  \BibitemOpen
  \bibfield  {author} {\bibinfo {author} {\bibfnamefont {L.~A.}\ \bibnamefont
  {Ostrovski{\u\i}}},\ }\bibfield  {title} {\enquote {\bibinfo {title} {Some
  "moving boundaries paradoxes" in electrodynamics},}\ }\href@noop {}
  {\bibfield  {journal} {\bibinfo  {journal} {Sov. Phys. Usp.}\ }\textbf
  {\bibinfo {volume} {18}},\ \bibinfo {pages} {452} (\bibinfo {year}
  {1975})}\BibitemShut {NoStop}%
\bibitem [{\citenamefont {Shui}\ \emph {et~al.}(2014)\citenamefont {Shui},
  \citenamefont {Yue}, \citenamefont {Liu}, \citenamefont {Liu},\ and\
  \citenamefont {Guo}}]{shui2014one}%
  \BibitemOpen
  \bibfield  {author} {\bibinfo {author} {\bibfnamefont {L.-Q.}\ \bibnamefont
  {Shui}}, \bibinfo {author} {\bibfnamefont {Z.-F.}\ \bibnamefont {Yue}},
  \bibinfo {author} {\bibfnamefont {Y.-S.}\ \bibnamefont {Liu}}, \bibinfo
  {author} {\bibfnamefont {Q.-Chang.}\ \bibnamefont {Liu}}, \ and\ \bibinfo
  {author} {\bibfnamefont {J.-J.}\ \bibnamefont {Guo}},\ }\bibfield  {title}
  {\enquote {\bibinfo {title} {One-dimensional linear elastic waves at moving
  property interface},}\ }\href@noop {} {\bibfield  {journal} {\bibinfo
  {journal} {Wave Motion}\ }\textbf {\bibinfo {volume} {51}},\ \bibinfo {pages}
  {1179--1192} (\bibinfo {year} {2014})}\BibitemShut {NoStop}%
\bibitem [{\citenamefont {Deck-L{\'e}ger}\ and\ \citenamefont
  {Caloz}(2018)}]{deck2018interluminal}%
  \BibitemOpen
  \bibfield  {author} {\bibinfo {author} {\bibfnamefont {Z.-L.}\ \bibnamefont
  {Deck-L{\'e}ger}}\ and\ \bibinfo {author} {\bibfnamefont {C.}~\bibnamefont
  {Caloz}},\ }\bibfield  {title} {\enquote {\bibinfo {title} {Scattering at
  interluminal interface},}\ }in\ \href@noop {} {\emph {\bibinfo {booktitle}
  {2019 IEEE AP-S Int. Antennas Propag.}}}\ (\bibinfo {organization} {IEEE},\
  \bibinfo {year} {2018})\ pp.\ \bibinfo {pages} {165--166}\BibitemShut
  {NoStop}%
\bibitem [{\citenamefont {Lorentz}(1937)}]{lorentz1937electromagnetic}%
  \BibitemOpen
  \bibfield  {author} {\bibinfo {author} {\bibfnamefont {H.}~\bibnamefont
  {Lorentz}},\ }\bibfield  {title} {\enquote {\bibinfo {title} {Electromagnetic
  phenomena in a system moving with any velocity smaller than that of light},}\
  }in\ \href@noop {} {\emph {\bibinfo {booktitle} {Collected Papers}}}\
  (\bibinfo  {publisher} {Springer},\ \bibinfo {year} {1937})\ pp.\ \bibinfo
  {pages} {172--197}\BibitemShut {NoStop}%
\bibitem [{\citenamefont {Pauli}(1981)}]{pauli1981theory}%
  \BibitemOpen
  \bibfield  {author} {\bibinfo {author} {\bibfnamefont {W.}~\bibnamefont
  {Pauli}},\ }\href@noop {} {\emph {\bibinfo {title} {Theory of Relativity}}}\
  (\bibinfo  {publisher} {Courier Corporation},\ \bibinfo {year}
  {1981})\BibitemShut {NoStop}%
\bibitem [{\citenamefont {Einstein}(1905)}]{einstein1905elektrodynamik}%
  \BibitemOpen
  \bibfield  {author} {\bibinfo {author} {\bibfnamefont {A.}~\bibnamefont
  {Einstein}},\ }\bibfield  {title} {\enquote {\bibinfo {title} {Zur
  elektrodynamik bewegter {K}\"{o}rper},}\ }\href@noop {} {\bibfield  {journal}
  {\bibinfo  {journal} {Ann. Physik}\ }\textbf {\bibinfo {volume} {322}},\
  \bibinfo {pages} {891--921} (\bibinfo {year} {1905})},\ \bibinfo {note}
  {reprinted in \cite{einstein1952}}\BibitemShut {NoStop}%
\bibitem [{\citenamefont {Laue}(1907)}]{laue1907}%
  \BibitemOpen
  \bibfield  {author} {\bibinfo {author} {\bibfnamefont {M.}~\bibnamefont
  {Laue}},\ }\bibfield  {title} {\enquote {\bibinfo {title} {Die mitf\"{u}hrung
  des lichtes durch bewegte k\"{o}per nach dem relativit\"{a}ts-prinzip},}\
  }\href@noop {} {\bibfield  {journal} {\bibinfo  {journal} {Ann. d. Phys}\
  }\textbf {\bibinfo {volume} {29}},\ \bibinfo {pages} {077405--1:4} (\bibinfo
  {year} {1907})}\BibitemShut {NoStop}%
\bibitem [{\citenamefont {Fizeau}(1851)}]{fizeau1851hypotheses}%
  \BibitemOpen
  \bibfield  {author} {\bibinfo {author} {\bibfnamefont {H.}~\bibnamefont
  {Fizeau}},\ }\bibfield  {title} {\enquote {\bibinfo {title} {Sur les
  hypoth{\`e}ses relatives {\`a} l'{\'e}ther lumineux, et sur une
  exp{\'e}rience qui parait d{\'e}montrer que le mouvement des corps change la
  vitesse avec laquelle la lumi{\`e}re se propage dans leur int{\'e}rieur},}\
  }\href@noop {} {\bibfield  {journal} {\bibinfo  {journal} {C.R. Acad. Sci.}\
  }\textbf {\bibinfo {volume} {33}},\ \bibinfo {pages} {349} (\bibinfo {year}
  {1851})}\BibitemShut {NoStop}%
\bibitem [{\citenamefont {Kong}(2008)}]{kong2008theory}%
  \BibitemOpen
  \bibfield  {author} {\bibinfo {author} {\bibfnamefont {J.~A.}\ \bibnamefont
  {Kong}},\ }\href@noop {} {\emph {\bibinfo {title} {Electromagnetic Wave
  Theory}}},\ \bibinfo {edition} {seventh}\ ed.\ (\bibinfo  {publisher} {EMW
  Publication},\ \bibinfo {year} {2008})\BibitemShut {NoStop}%
\bibitem [{\citenamefont {Deck-L{\'e}ger}\ \emph {et~al.}(2018)\citenamefont
  {Deck-L{\'e}ger}, \citenamefont {Akbarzadeh},\ and\ \citenamefont
  {Caloz}}]{deck2017superluminal}%
  \BibitemOpen
  \bibfield  {author} {\bibinfo {author} {\bibfnamefont {Z.-L.}\ \bibnamefont
  {Deck-L{\'e}ger}}, \bibinfo {author} {\bibfnamefont {A.}~\bibnamefont
  {Akbarzadeh}}, \ and\ \bibinfo {author} {\bibfnamefont {C.}~\bibnamefont
  {Caloz}},\ }\bibfield  {title} {\enquote {\bibinfo {title} {Wave deflection
  and shifted refocusing in a medium modulated by a superluminal rectangular
  pulse},}\ }\href@noop {} {\bibfield  {journal} {\bibinfo  {journal} {Phys.
  Rev. B}\ }\textbf {\bibinfo {volume} {97}},\ \bibinfo {pages} {104305}
  (\bibinfo {year} {2018})}\BibitemShut {NoStop}%
\bibitem [{\citenamefont {Morgenthaler}(1958)}]{morgenthaler1958}%
  \BibitemOpen
  \bibfield  {author} {\bibinfo {author} {\bibfnamefont {F.~R.}\ \bibnamefont
  {Morgenthaler}},\ }\bibfield  {title} {\enquote {\bibinfo {title} {Velocity
  modulation of electromagnetic waves},}\ }\href@noop {} {\bibfield  {journal}
  {\bibinfo  {journal} {IEEE Trans. Microw. Theory Techn.}\ }\textbf {\bibinfo
  {volume} {6}},\ \bibinfo {pages} {167--172} (\bibinfo {year}
  {1958})}\BibitemShut {NoStop}%
\bibitem [{\citenamefont {Felsen}\ and\ \citenamefont
  {Whitman}(1970)}]{felsen1970wave}%
  \BibitemOpen
  \bibfield  {author} {\bibinfo {author} {\bibfnamefont {L.}~\bibnamefont
  {Felsen}}\ and\ \bibinfo {author} {\bibfnamefont {G.~M.}\ \bibnamefont
  {Whitman}},\ }\bibfield  {title} {\enquote {\bibinfo {title} {Wave
  propagation in time-varying media},}\ }\href@noop {} {\bibfield  {journal}
  {\bibinfo  {journal} {IEEE Trans. Antennas. Propag.}\ }\textbf {\bibinfo
  {volume} {18}},\ \bibinfo {pages} {242--253} (\bibinfo {year}
  {1970})}\BibitemShut {NoStop}%
\bibitem [{\citenamefont {Stokes}(2009)}]{stokes_2009}%
  \BibitemOpen
  \bibfield  {author} {\bibinfo {author} {\bibfnamefont {G.}~\bibnamefont
  {Stokes}},\ }\enquote {\bibinfo {title} {On the perfect blackness of the
  central spot in {N}ewton's rings, and on the verification of {F}resnel's
  formula for the intensities of reflected and reflacted rays},}\ in\
  \href@noop {} {\emph {\bibinfo {booktitle} {Mathematical and Physical
  Papers}}},\ \bibinfo {series} {Cambridge Library Collection - Mathematics},
  Vol.~\bibinfo {volume} {2}\ (\bibinfo  {publisher} {Cambridge University
  Press},\ \bibinfo {year} {2009})\ p.\ \bibinfo {pages} {89–103}\BibitemShut
  {NoStop}%
\bibitem [{\citenamefont {Fink}(1992)}]{fink1992time}%
  \BibitemOpen
  \bibfield  {author} {\bibinfo {author} {\bibfnamefont {M.}~\bibnamefont
  {Fink}},\ }\bibfield  {title} {\enquote {\bibinfo {title} {Time reversal of
  ultrasonic fields. {I}. {B}asic principles},}\ }\href@noop {} {\bibfield
  {journal} {\bibinfo  {journal} {IEEE Trans. Ultrason. Ferroelectrics. Freq.
  Contr.}\ }\textbf {\bibinfo {volume} {39}},\ \bibinfo {pages} {555--566}
  (\bibinfo {year} {1992})}\BibitemShut {NoStop}%
\bibitem [{\citenamefont {Born}\ and\ \citenamefont
  {Wolf}(1980)}]{born1980principles}%
  \BibitemOpen
  \bibfield  {author} {\bibinfo {author} {\bibfnamefont {M.}~\bibnamefont
  {Born}}\ and\ \bibinfo {author} {\bibfnamefont {E.}~\bibnamefont {Wolf}},\
  }\href@noop {} {\emph {\bibinfo {title} {Principles of optics}}}\ (\bibinfo
  {publisher} {Pergamon},\ \bibinfo {year} {1980})\BibitemShut {NoStop}%
\bibitem [{\citenamefont {Smith}\ and\ \citenamefont
  {Schurig}(2003)}]{smith2003electromagnetic}%
  \BibitemOpen
  \bibfield  {author} {\bibinfo {author} {\bibfnamefont {D.~R.}\ \bibnamefont
  {Smith}}\ and\ \bibinfo {author} {\bibfnamefont {D.}~\bibnamefont
  {Schurig}},\ }\bibfield  {title} {\enquote {\bibinfo {title} {Electromagnetic
  wave propagation in media with indefinite permittivity and permeability
  tensors},}\ }\href@noop {} {\bibfield  {journal} {\bibinfo  {journal} {Phys.
  Rev. Lett.}\ }\textbf {\bibinfo {volume} {90}},\ \bibinfo {pages}
  {077405--1:4} (\bibinfo {year} {2003})}\BibitemShut {NoStop}%
\bibitem [{\citenamefont
  {Seshadri}(1977{\natexlab{a}})}]{seshadri1977asymptotic1}%
  \BibitemOpen
  \bibfield  {author} {\bibinfo {author} {\bibfnamefont {S.}~\bibnamefont
  {Seshadri}},\ }\bibfield  {title} {\enquote {\bibinfo {title} {Asymptotic
  theory of mode coupling in a space-time periodic medium-{P}art {I}: {S}table
  interactions},}\ }\href@noop {} {\bibfield  {journal} {\bibinfo  {journal}
  {Proc. IEEE}\ }\textbf {\bibinfo {volume} {65}},\ \bibinfo {pages}
  {996--1004} (\bibinfo {year} {1977}{\natexlab{a}})}\BibitemShut {NoStop}%
\bibitem [{\citenamefont {Elnaggar}\ and\ \citenamefont
  {Milford}(2018)}]{elnaggar2018controlling}%
  \BibitemOpen
  \bibfield  {author} {\bibinfo {author} {\bibfnamefont {S.}~\bibnamefont
  {Elnaggar}}\ and\ \bibinfo {author} {\bibfnamefont {G.}~\bibnamefont
  {Milford}},\ }\bibfield  {title} {\enquote {\bibinfo {title} {Controlling
  nonreciprocity using enhanced {B}rillouin scattering},}\ }\href@noop {}
  {\bibfield  {journal} {\bibinfo  {journal} {IEEE Trans. Ant. Prop.}\ }\textbf
  {\bibinfo {volume} {66}},\ \bibinfo {pages} {3500--3511} (\bibinfo {year}
  {2018})}\BibitemShut {NoStop}%
\bibitem [{\citenamefont
  {Seshadri}(1977{\natexlab{b}})}]{seshadri1977asymptotic2}%
  \BibitemOpen
  \bibfield  {author} {\bibinfo {author} {\bibfnamefont {S.}~\bibnamefont
  {Seshadri}},\ }\bibfield  {title} {\enquote {\bibinfo {title} {Asymptotic
  theory of mode coupling in a space-time periodic medium-{P}art {II}:
  {U}nstable interactions},}\ }\href@noop {} {\bibfield  {journal} {\bibinfo
  {journal} {Proc. IEEE}\ }\textbf {\bibinfo {volume} {65}},\ \bibinfo {pages}
  {1459--1469} (\bibinfo {year} {1977}{\natexlab{b}})}\BibitemShut {NoStop}%
\bibitem [{\citenamefont {Abel{\`e}s}(1950)}]{abeles1950theorie}%
  \BibitemOpen
  \bibfield  {author} {\bibinfo {author} {\bibfnamefont {F.}~\bibnamefont
  {Abel{\`e}s}},\ }\bibfield  {title} {\enquote {\bibinfo {title} {La
  th{\'e}orie g{\'e}n{\'e}rale des couches minces},}\ }\href@noop {} {\bibfield
   {journal} {\bibinfo  {journal} {J. Ph. Radium}\ }\textbf {\bibinfo {volume}
  {11}},\ \bibinfo {pages} {307--309} (\bibinfo {year} {1950})}\BibitemShut
  {NoStop}%
\bibitem [{\citenamefont {Longhi}(2010)}]{longhi2010pt}%
  \BibitemOpen
  \bibfield  {author} {\bibinfo {author} {\bibfnamefont {S.}~\bibnamefont
  {Longhi}},\ }\bibfield  {title} {\enquote {\bibinfo {title}
  {$\mathcal{PT}$-symmetric laser absorber},}\ }\href@noop {} {\bibfield
  {journal} {\bibinfo  {journal} {Phys. Rev. A}\ }\textbf {\bibinfo {volume}
  {82}},\ \bibinfo {pages} {031801} (\bibinfo {year} {2010})}\BibitemShut
  {NoStop}%
\bibitem [{\citenamefont {Lorentz}\ \emph {et~al.}(1952)\citenamefont
  {Lorentz}, \citenamefont {Einstein}, \citenamefont {Minkowski}, \citenamefont
  {Weyl},\ and\ \citenamefont {Sommerfeld}}]{einstein1952}%
  \BibitemOpen
  \bibfield  {author} {\bibinfo {author} {\bibfnamefont {H.~A.}\ \bibnamefont
  {Lorentz}}, \bibinfo {author} {\bibfnamefont {A.}~\bibnamefont {Einstein}},
  \bibinfo {author} {\bibfnamefont {H.}~\bibnamefont {Minkowski}}, \bibinfo
  {author} {\bibfnamefont {H.}~\bibnamefont {Weyl}}, \ and\ \bibinfo {author}
  {\bibfnamefont {A.}~\bibnamefont {Sommerfeld}},\ }\href@noop {} {\emph
  {\bibinfo {title} {The principle of relativity: a collection of original
  memoirs on the special and general theory of relativity}}}\ (\bibinfo
  {publisher} {Dover Publications},\ \bibinfo {year} {1952})\BibitemShut
  {NoStop}%
\end{thebibliography}%

\end{document}